\documentclass[preprint, 3p,12pt,sort&compress]{elsarticle}
\journal{Journal}
\usepackage[utf8]{inputenc}
\usepackage{amsmath}
\usepackage[scr=boondox,  % heavily sloped
            ]   % slightly sloped
           {mathalpha}
\usepackage{booktabs}
\usepackage{empheq}
\usepackage{pifont}
\usepackage{amssymb}
\usepackage{graphicx}
\DeclareMathAlphabet      
{\mathbfit}{OML}{cmm}{b}{it}
\usepackage{multicol}
\usepackage{mathrsfs}
\usepackage{multirow}
\usepackage{pdfpages}
\usepackage[mathscr]{euscript}
\DeclareSymbolFont{rsfs}{U}{rsfs}{m}{n}
\DeclareSymbolFontAlphabet{\mathscrsfs}{rsfs}
\usepackage{comment}
\usepackage{longtable}
\usepackage{tabularx}
\usepackage{tabularray}
\graphicspath{ {./images/} }
\usepackage{float,booktabs,array}
\usepackage[pagewise]{lineno}
\usepackage[ruled,vlined,linesnumbered,algo2e,resetcount]{algorithm2e}
\usepackage{algorithm}
\usepackage{algcompatible}
\usepackage{lineno}
\usepackage{xcolor}
\usepackage{tikz,pgf} 
\usepackage{color, soul}
\usepackage{siunitx}
\usepackage{subcaption}
\usepackage[labelsep=period]{caption}
\usepackage[colorlinks]{hyperref}
\usepackage{enumitem}
\usepackage{setspace}
\usepackage{lipsum}
\usepackage{arydshln}

\newcolumntype{Y}{>{\centering\arraybackslash}X}
\usepackage[flushleft]{threeparttable}
\usepackage{xltabular}
\usepackage{diagbox}
\usepackage[bottom]{footmisc}
\usetikzlibrary{arrows.meta}
\definecolor{green}{rgb}{0.0,0.50,0.0}
\tikzset{>={Straight Barb[angle'=80, scale=1.1]}}
\hypersetup{
     colorlinks=true,
     linkcolor=blue,
     filecolor=blue,
     citecolor = blue,      
     urlcolor=blue,
     pdfauthor=author
     }

\usepackage{nomencl}
\usepackage{etoolbox}
\usepackage{pdflscape}
\usepackage{framed}
\usepackage{rotating}
\usepackage{adjustbox}
\usepackage{etoolbox}
\usepackage{framed}

\newcommand{\cmark}{\ding{51}} 
\newcommand{\xmark}{\ding{55}}

\makenomenclature
\renewcommand{\nomgroup}[1]{%
  \ifstrequal{#1}{A}{\item[\textbf{Acronyms}]\addvspace{6pt}}{%
  \ifstrequal{#1}{S}{\addvspace{15pt}\item[\textbf{List of Symbols}]}{}}%
}
\usepackage{multicol}
\usepackage{hyperref}
\usepackage{framed}
\renewcommand*\nompreamble{\begin{multicols}{2}}
\renewcommand*\nompostamble{\end{multicols}}

\usepackage{tikz,xcolor,hyperref}

\definecolor{lime}{HTML}{A6CE39}
\DeclareRobustCommand{\orcidicon}{%
	\begin{tikzpicture}
	\draw[lime, fill=lime] (0,0) 
	circle [radius=0.14] 
	node[white] {{\fontfamily{qag}\selectfont \tiny ID}};
	\draw[white, fill=white] (-0.0625,0.095) 
	circle [radius=0.007];
	\end{tikzpicture}
	\hspace{-2mm}
}

\foreach \x in {A, ..., Z}{%
	\expandafter\xdef\csname orcid\x\endcsname{\noexpand\href{https://orcid.org/\csname orcidauthor\x\endcsname}{\noexpand\orcidicon}}
}

\begin{document}
\begin{frontmatter}

\title{Deep Generative Models in Condition and Structural Health Monitoring: Opportunities, Limitations and Future Outlook}

\author[1]{Xin Yang\corref{correspondingauthor}\orcidA}
\ead{xin.yang@kuleuven.be}
\author[1,2]{Chen Fang\orcidB}
\author[1,3]{Yunlai Liao\orcidC}
\author[4]{Jian Yang\orcidD}
\author[1,5,6]{Konstantinos Gryllias\orcidE}
\author[1]{Dimitrios Chronopoulos\corref{correspondingauthor}\orcidF}
\ead{dimitrios.chronopoulos@kuleuven.be}

\address[1]{Department of Mechanical Engineering $\&$ Division of Mechatronic System Dynamics (LMSD), KU Leuven, Belgium}

\address[2]{\quad Department of Infrastructure Engineering, The University of Melbourne, Parkville, VIC 3010, Australia}

\address[3]{\quad School of Aerospace Engineering, Xiamen University, Xiamen, Fujian 361005, PR China}

\address[4]{Department of Mechanical, Materials and Manufacturing Engineering, University of Nottingham Ningbo China, Ningbo, 315100, PR China}

\address[5]{Flanders Make@KU Leuven, Belgium}

\address[6]{Leuven.AI - KU Leuven Institute for AI, B-3000 Leuven, Belgium}

\cortext[correspondingauthor]{Corresponding author}

\begin{abstract}
Condition and structural health monitoring (CM/SHM) is a pivotal component of predictive maintenance (PdM) strategies across diverse industrial sectors, including mechanical rotating machinery, aircraft structures, wind turbines, and civil infrastructures. Conventional deep learning models, while effective for fault diagnosis and anomaly detection through automatic feature learning from sensor data, often struggle with operational variability, imbalanced or scarce fault datasets, and multimodal sensory data from complex systems. Deep generative models (DGMs) including deep autoregressive models, variational autoencoders, generative adversarial networks, diffusion-based models, and emerging large language models, offer transformative capabilities by synthesizing high-fidelity data samples, reconstructing latent system states, and modeling complex multimodal data streams. This review systematically examines state-of-the-art DGM applications in CM/SHM across the four main industrial systems mentioned above, emphasizing their roles in addressing key challenges: data generation, domain adaptation and generalization, multimodal data fusion, and downstream fault diagnosis and anomaly detection tasks, with rigorous comparison among signal processing, conventional machine learning or deep learning models, and DGMs. Lastly, we discuss current limitations of DGMs, including challenges of explainable and trustworthy models, computational inefficiencies for edge deployment, and the need for parameter-efficient fine-tuning strategies. Future research directions can focus on zero-shot and few-shot learning, robust multimodal data generation, hybrid architectures integrating DGMs with physics knowledge, and reinforcement learning with DGMs to enhance robustness and accuracy in industrial scenarios.

\end{abstract}


\end{frontmatter}

% \linenumbers
\nomenclature[A]{PdM}{Predictive maintenance}
\nomenclature[A]{CM}{Condition monitoring}
\nomenclature[A]{FD\&AD}{Fault diagnosis and anomaly detection}
\nomenclature[A]{SHM}{Structural health monitoring}
\nomenclature[A]{NDT}{Non-destructive testing}
\nomenclature[A]{DNNs}{Deep neural networks}
\nomenclature[A]{GenAI}{Generative artificial intelligence}
\nomenclature[A]{DGMs}{Deep generative model}
\nomenclature[A]{DAR}{Deep autoregressive}
\nomenclature[A]{VAE}{Variational autoencoder}
\nomenclature[A]{GAN}{Generative adversarial network}
\nomenclature[A]{LLM}{Large language model}
\nomenclature[A]{NLL}{Negative log-likelihood}
\nomenclature[A]{ELBO}{Evidence Lower Bound}
\nomenclature[A]{DDPMs}{Denoising diffusion probabilistic models}
\nomenclature[A]{GPT-4}{Generative pre-trained transformer 4}
\nomenclature[A]{GMMs}{Gaussian mixture models}
\nomenclature[A]{HMMs}{Hidden Markov models}
\nomenclature[A]{FID}{Fréchet inception distance}
\nomenclature[A]{DDIMs}{Denoising diffusion implicit models}
\nomenclature[A]{SMOTE}{Synthetic minority over-sampling technique}
\nomenclature[A]{SVR}{Support vector regression}
\nomenclature[A]{RNN}{Recurrent neural networks}
\nomenclature[A]{LSTM}{Long short-term memory}
\nomenclature[A]{GRU}{Gated recurrent unit}
\nomenclature[A]{KNN}{K-nearest neighbors}
\nomenclature[A]{TL}{Transfer learning}
\nomenclature[A]{DA}{Domain adaptation}
\nomenclature[A]{SCADA}{Supervisory Control and Data Acquisition}
\nomenclature[A]{LoRA}{Low-Rank Adaptation}
\nomenclature[A]{PEFT}{Parameter-efficient fine-tuning}
\nomenclature[A]{GenXAI}{Explainable Generative AI}
\nomenclature[A]{SHAP}{SHapley Additive exPlanations}
\nomenclature[A]{RAG}{Retrieval-augmented generation}
\nomenclature[A]{PTQ}{Post-training quantization}
\nomenclature[A]{QLoRA}{Quantized LoRA}
\nomenclature[A]{PINNs}{Physics-informed neural networks}
\nomenclature[A]{RL}{Reinforcement learning}

\nomenclature[S]{$T$}{Time steps}
\nomenclature[S]{$A$}{Amplitude}
\nomenclature[S]{$\mathcal{D}_{KL}$}{Kullback–Leibler divergence}
\nomenclature[S]{$\mathcal{X}$}{Target distribution}
\nomenclature[S]{$\mathbfit{x}$}{Target sample}
\nomenclature[S]{$\mathcal{Z}$}{Tractable distribution}
\nomenclature[S]{$\mathbfit{z}$}{Latent vector}
\nomenclature[S]{$\mathcal{G}$}{Mapping generator}
\nomenclature[S]{$\mathbb{R}$}{The set of real numbers}
\nomenclature[S]{$\theta$}{Model parameters}
\nomenclature[S]{$\mathbb{E}$}{Expectation}
\nomenclature[S]{$n$}{Number of data samples}
\nomenclature[S]{$p$}{Probability distribution}
\nomenclature[S]{$\mathscrsfs{D}$}{Training dataset}
\nomenclature[S]{$\mathcal{L}$}{Loss function}
\nomenclature[S]{$G$}{GAN generator}
\nomenclature[S]{$D$}{GAN discriminator}
\nomenclature[S]{$q(\cdot)$}{Forward transition probability}
\nomenclature[S]{$p(\cdot)$}{Backward transition probability}
\nomenclature[S]{$\mathcal{N}$}{Gaussian distribution}
\nomenclature[S]{$\beta_t$}{Noise scale}
\nomenclature[S]{$\mathbf{I}$}{The identity matrix}
\nomenclature[S]{$\mu$}{Mean value}
\nomenclature[S]{$\Sigma$}{Covariance matrix}
\nomenclature[S]{$\mathbf{C}$}{Constant}
\nomenclature[S]{$\mathbf{W}$}{Weight matrices}
\nomenclature[S]{$h$}{Hidden state}
\nomenclature[S]{$\varepsilon$}{Gaussian noise variable}
\nomenclature[S]{$\mathbf{Q}$}{Query matrices}
\nomenclature[S]{$\mathbf{K}$}{Key matrices}
\nomenclature[S]{$\mathbf{V}$}{Value matrices}
\nomenclature[S]{$\Delta \text{W}$}{Weight update matrix}

\begin{table*}[!htp]
\begin{framed}
\printnomenclature
\end{framed}
\end{table*}

\begin{spacing}{1.4}
\tableofcontents
\end{spacing}

\section{Introduction}
\label{Sec:1}
\subsection{Background}
Safety and reliability are of great importance for industrial operational systems, driving the need for predictive strategies to mitigate risks and optimize performance. Predictive maintenance (PdM) has emerged as a pivotal strategy, leveraging intelligent data analysis and monitoring techniques to prevent potential failures, minimize downtime, and prolong equipment lifespans \cite{NUNES202353, 10440027, huxiaosong2021}. As a core component of PdM, condition monitoring (CM), also termed condition-based maintenance \cite{AHMEDMURTAZA2024102935}, is a real-time process involving sensor-based technologies to track equipment parameters such as vibration, temperature and acoustics in order to assess system health and detect early anomalies. Therefore, CM also involves fault diagnosis and anomaly detection (FD\&AD) tasks using real-time data and enables immediate corrective actions to prevent failures. Specifically, in rotating machinery systems, CM stands out by identifying subtle deviations from normal operating profiles, detecting incipient faults (e.g., bearing wear, gear degradation, misalignment due to improper installation, asymmetric loads). These insights are pivotal for rotating systems, where unexpected failures can lead to safety hazards, production losses, or environmental impacts. For static structures, such as pipelines, bridges, or fixed machinery, structural health monitoring (SHM) and non-destructive testing (NDT) are frequently used to assess structural integrity without disrupting operations \cite{YANG2024111597, AliMardanshahi2025}. NDT techniques, including ultrasonic testing and thermography, assess structural safety without compromising functionality by identifying defects such as cracks or corrosion. SHM expands on this by deploying permanent sensor networks to continuously monitor static systems for gradual degradation.

For intelligent CM/SHM systems, deep neural networks (DNNs) are widely utilized for FD\&AD tasks. Representative frameworks, such as convolutional neural networks \cite{CHEN2019106272}, recurrent neural networks \cite{CHEN2020106683, 2022_Yang}, and more recent Transformer-based networks \cite{WU2023439}, have demonstrated effectiveness in FD\&AD tasks by extracting useful features from heterogeneous data modalities, including images, acoustic signals, and time series sensor data. However, limitations and challenges still exist for applying DNNs within industrial scenarios:
\begin{itemize}
    \item \textbf{Scarce or missing data:} Prevalent DNNs are limited by their dependence on large, high-quality labeled datasets to achieve robust classification or prediction performance. Industrial environments frequently struggle to collect sufficient labeled data, or the presence of missing data due to sensor failure, communication interruption, and environmental interference \cite{VANDREVEN2024132711, ZHANG2025110663};  
    \item \textbf{Domain mismatch:} Sensor data collected under normal and abnormal conditions often exhibits severe class imbalance, as industrial systems predominantly operate within nominal states during routine cycles. Even when fault data are available, they may display uneven distribution across distinct fault categories, compromising model training and generalization \cite{ZHAO2024122807}.  
    \item \textbf{Multimodal data complexity:} Industrial sensor data originate from heterogeneous sources, such as vibration time series, thermal imaging, acoustic signals, and emission sensors. These multimodal datasets require advanced techniques to address challenges in feature alignment, temporal synchronization, and cross-modal fusion \cite{ZHANG2025110663}.  
    \item \textbf{Generalization limitations:} Current FD\&AD methods often lack adaptability and stability when deployed across varying operational environments, equipment configurations, or operating conditions, restricting their practical applicability in dynamic industrial scenarios \cite{ZHANG2025124378, hu2024evaluationimprovement}.
\end{itemize}

As a promising solution, more recently, deep generative models, under the umbrella of generative artificial intelligence (GenAI), have attracted broad interest in addressing the challenges in the CM/SHM field.

\subsection{Deep generative models}  
Deep generative models (DGMs) are deep neural networks with multiple hidden layers trained to approximate complicated, high-dimensional probability distributions of training samples \cite{ruthotto2021introduction, Bond_Taylor_2022}. These models have become the dominant paradigm in GenAI, driven by advances in stochastic optimization and the availability of massive training datasets \cite{manduchi2025chall, ZHANG2025125059, GM2020100285}. The term \textit{deep} indicates the use of deep neural networks as parametric models in the approximation process \cite{Tomczak2022}. To anchor the view in industrial health monitoring, this review focuses on five prominent DGM frameworks: deep autoregressive (DAR) models, variational autoencoders (VAEs), generative adversarial networks (GANs), diffusion-based models, and large language models (LLMs). Other DGM families, such as normalizing flow models \cite{Kobyzev_2021}, energy-based models \cite{NEURIPS2019_378a063b}, and score-based models \cite{song2021scorebasedgenerative}, often integrate with diffusion-based or LLM-based models and are more prevalent in the computer science community. To maintain consistency and avoid misalignment, this review will focus on the aforementioned models, which have already shown promising applications in industrial CM and SHM systems.

Comprehensive scientometric analysis was performed among multiple databases, including Web of Science, Scopus, IEEE Xplore and Google Scholar, to identify relevant publications up to the end of 2025. Considering that GenAI represents an emerging and rapidly developing research hotspot, numerous studies are available as preprints before being sent to formal peer review. Therefore, the search scope was extended to include repositories such as arXiv, SSRN to capture the most recent trend. The search strategy combined terms related to generative modeling and industrial applications, using keywords such as \textit{deep generative model}, \textit{generative adversarial network}, \textit{condition monitoring}, and \textit{structural health monitoring}. To ensure completeness, variations and synonyms of these terms (e.g., \textit{fault diagnosis}, \textit{predictive maintenance}, and \textit{data-driven modeling}) were also incorporated. In this way, we provided a broad overview of the evolving scope of generative models applied to industrial health monitoring tasks and systems.

\begin{figure}[H]
    \centering
    \makebox[1\textwidth][c]{
    \includegraphics[width=1.2\linewidth]{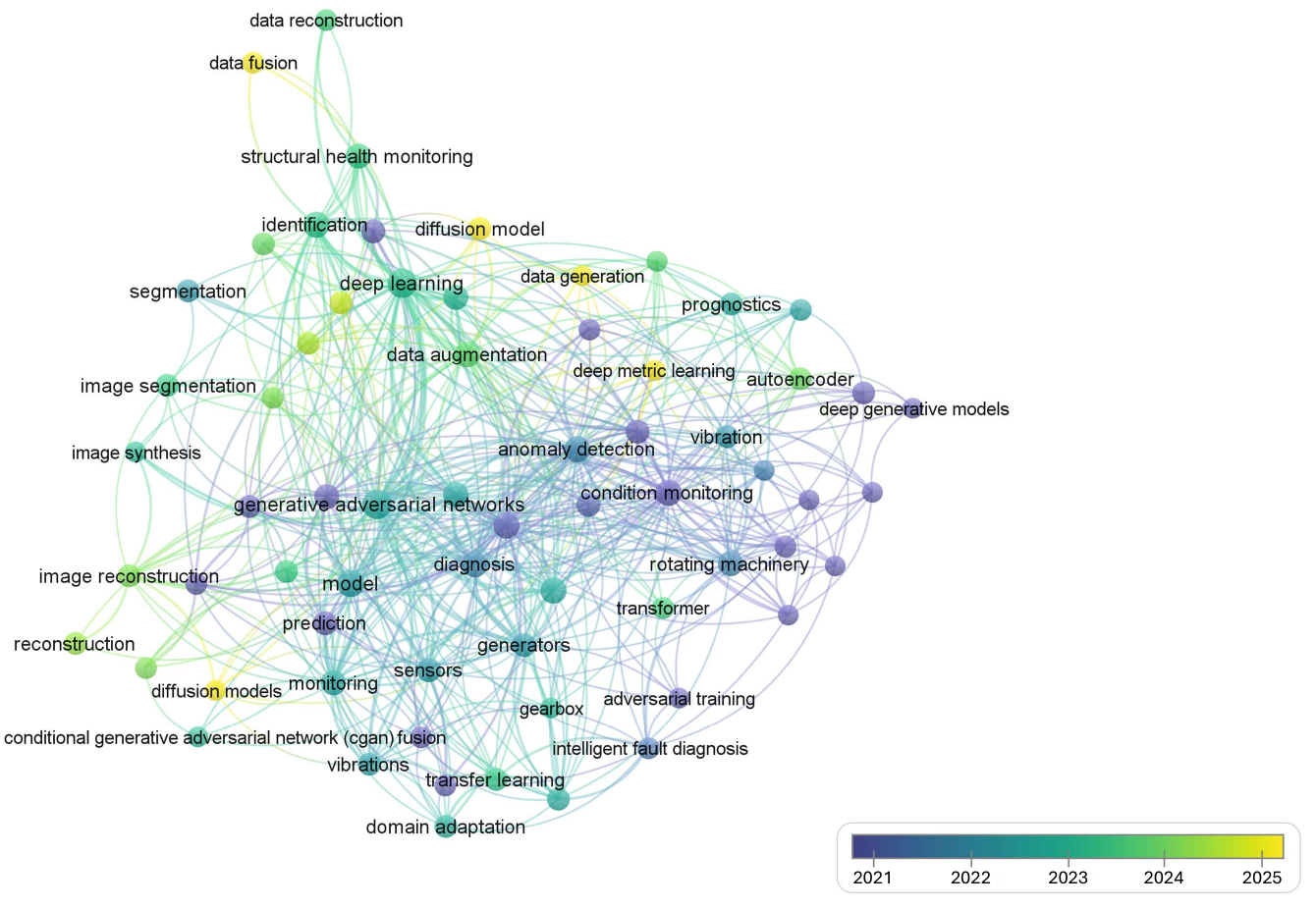}}
    \caption{The cluster to indicate the co-occurrence among keywords (condition monitoring, structural health monitoring, deep generative models, and etc.).}
    \label{fig1}
\end{figure}

To further elaborate on the co-occurrence among the keywords, we create a map of the cluster of keywords in Fig.~\ref{fig1} using VOSviewer developed by the Center for Science and Technology Studies at Leiden University \cite{van2010software}. The visualization reveals that the generative model exhibits a strong semantic relation with terms such as condition monitoring, fault diagnosis, and data augmentation for vibration signals. However, recent trends indicate a shift toward applications in image reconstruction, synthetic data generation, diffusion model, and etc. This trend suggests that the deployment of DGMs in industrial CM/SHM remains in its nascent stage, with promising opportunities for future industrial scenarios.

\subsection{Motivations}

In response to emerging challenges and research trends, this review systematically summarizes DGMs in CM/SHM applications, addressing downstream tasks by identifying faults and detecting possible anomalies. This review provides three key contributions: (i) a thorough summary of state-of-the-art DGMs; (ii) an exploration of their implementations across four technical domains, including data augmentation\&imputation, domain adaptation\&generalization, multimodal data fusion, and FD\&AD tasks in CM/SHM systems; and (iii) a forward-looking perspective on current limitations and research opportunities. This work arrives at a critical moment in the evolution of CM/SHM, where these systems are transitioning from effective signal processing techniques to intelligent, interactive, AI-driven architectures. The emergence of GenAI positions DGMs as transformative tools in this paradigm shift. Building on cross-disciplinary advances from the computer vision to PdM communities, this review establishes a technical framework to address the following questions.

\begin{itemize}
    \item \textit{What DGMs have been previously used to support research related to CM/SHM systems?}
    
    Previous studies on using DGM in CM/SHM systems mainly focus on DAR, VAEs, and GANs, which are dedicated to data augmentation, domain adaptation and generalization. Among them, i) DAR models are widely used in time series augmentation; ii) VAEs focus on data augmentation through latent space modeling; and iii) GANs enhance unbalanced or missing data by synthesizing limited fault data through adversarial training. Diffusion models and LLMs are recent emerging trends. Diffusion-based models are mainly used for signal denoising and data augmentation, while the application of LLMs is still limited to fault diagnosis and damage detection based on pre-trained large models and fine-tuned on customized datasets.

    \item \textit{Why DGMs have great potential to enhance or even replace signal processing-based methods in specific application scenarios?}

    Unlike signal processing techniques that typically rely on hand-crafted feature extraction methods such as Fourier transform or wavelet transform, DGM can automatically learn hierarchical representations from multisource and multimodal data. This enables DGM to perform end-to-end learning  without the need for multi-step processing and feature extraction. However, in resource-constrained environments, or when interpretability and computational efficiency are critical, signal processing techniques may still be preferred.
    
    \item \textit{What are the remaining challenges and future research directions in using DGMs for CM/SHM system?}
    
    The high computational cost, difficulty in creating high-quality synthetic data, and interpretability challenges remain the main limiting factors for applying DGM in the field of CM/SHM. Future research can explore physics-based generative networks to improve the consistency and fidelity of generated signals, leverage multimodal data generation techniques, and combine reinforcement learning with LLM to simulate extreme failure scenarios (e.g., catastrophic equipment failures or rare structural failures), thereby improving the model robustness of the entire industrial ecosystem.
\end{itemize}

The remainder of this paper is organized as follows: Section \ref{Sec:2} presents the theoretical background and recent state-of-the-art DGMs. Section \ref{Sec:3} reviews the tasks and specific industrial applications. Section \ref{Sec:4} discusses existing challenges and future research directions. Section \ref{Sec:5} concludes the review.

\section{Preliminaries of DGMs}
\label{Sec:2}
\subsection{Theoretical preliminaries}
Mathematically, the primary goal of DGMs involves learning representations of intractable probability distributions $\mathcal{X}$ over $\mathbb{R}^m$, where $m$ is typically large and the distribution exhibits complex characteristics. This learning process uses a finite but substantial collection of independent and identically distributed (i.i.d.) samples from $\mathcal{X}$, referred to as training data. Unlike conventional statistical inference that seeks explicit probability expressions, the main objective of DGMs is to construct a generator $\mathcal{G}$ \cite{ruthotto2021introduction}:
\begin{equation}
    \mathcal{G}:\mathbb{R}^k \xrightarrow{}\mathbb{R}^m
\end{equation}
where $\mathcal{G}$ maps data from a tractable distribution $\mathcal{Z}$ supported on $\mathbb{R}^k$ into data resembling the target distribution $\mathcal{X}$. Formally, let us assume that for each $\mathbfit{x} \sim \mathcal{X}$ , there exists at least one latent vector $\mathbfit{z} \sim \mathcal{Z}$ satisfying $\mathcal{G}(\mathbfit{z}) \approx \mathbfit{x}$. The transformed distribution $\mathcal{G}(\mathcal{Z})$ thus approximates $\mathcal{X}$, which can be illustrated in Fig.~\ref{fig2}.

\begin{figure}[H]
\begin{center}
\begin{tikzpicture}

\draw[fill=blue!20, draw=none, shift={(0.2, 0.7)},scale=0.5]
  (0, 0) to[out=20, in=140] (1.5, -0.2) to [out=60, in=160]
  (5, 0.5) to[out=130, in=60] cycle;

\shade[thin, left color=green!10, right color=blue!30, draw=none,
  shift={(0.2, 0.7)},scale=1]
  (0, 0) to[out=10, in=140] (3.3, -0.8) to [out=60, in=190] (5, 0.5)
    to[out=130, in=60] cycle;

\draw[fill=blue!20, draw=none, shift={(6.2, 1.2)},scale=1] 
  (2, -0.5) to[out=5, in=120] (4.5, -1.5) to [out=60, in=160]
  (7, 0) to[out=130, in=60] cycle;

\fill[inner color=blue!20,outer color=white] (10.5,-3) circle (8mm);

\draw[thick,<->,black] 
  (9.3, 1) to [out=150, in=30] node[midway, above, xshift=2pt, yshift=-2pt] {$\text{Approximation}$} (3.8, 1); 

\draw[thick,->,black] 
  (10.5, -2) to  node[midway, right, xshift=2pt, yshift=-2pt] {} (10.5, -0.5); 

\draw[text=black] (10.5,-3) node  {\large $\mathbfit{z}$};
\draw[text=black] (10.5,1.3) node {\large $\mathcal{G}_{\bf{\theta}}$($\cal Z$)};
\draw[text=black] (2.5,1.3) node  {\large ${\cal X}$ };
\end{tikzpicture}

\caption{A deep generative model $\mathcal{G}_{\bf{\theta}}$ is trained to map samples from a tractable distribution $\mathcal{Z}$ to the more complicated distribution $\mathcal{G}_{\bf{\theta}}(\cal Z)$, which approximates the true distribution $\cal X$. Finding an objective function that minimizes the discrepancy between the generated samples and the original samples is the key obstacle to train the DGM.}
\label{fig2}
\end{center}
\end{figure}

\subsection{State-of-the-art DGMs}
\label{section2.2}
\subsubsection{Deep autoregressive models}
Deep autoregressive models belong to a class of DGMs that generate new data samples by iteratively invoking the model based on past observations \cite{urain2024dgmsrobotics, ren2025aigcindustrial}. This makes them particularly suitable for sequential data modeling tasks such as time series forecasting. Given a dataset $\mathscrsfs{D}=\{\mathbfit{x}_i\}_{i=1}^n$ with each $\mathbfit{x}_i \in \mathbb{R}^m$, the training objective for a DAR model minimizes the negative log-likelihood (NLL) of the data under the model, computed via the chain rule of probability \cite{Bond_Taylor_2022}:
\begin{equation}
    \mathcal{L}_{DAR}=-\mathbb{E}_{\mathbfit{x} \sim p_{\text{data}}}\Big[\log p_{\text{model}}(\mathbfit{x}_{i}|\mathbfit{x}_{1},...,\mathbfit{x}_{i-1}; \theta)\Big]
\end{equation}
where $p_{\text{model}}(\mathbfit{x}_{i}|\mathbfit{x}_{1},...,\mathbfit{x}_{i-1}; \theta)$ is the conditional probability density for the $i$-th time step given previous steps, parameterized by $\theta$. $\mathbb{E}$ denotes the expectation over all samples. In practice, the model learns to maximize the likelihood of each subsequent value in a sequence, given its previous values.

\subsubsection{Variational autoencoders}
VAEs represent a typical class of DGMs specializing in data compression and reconstruction \cite{Bond_Taylor_2022}. Their architecture consists of two networks: an encoder and a decoder. The encoder maps high-dimensional inputs to a lower-dimensional latent space that captures essential data features, while the decoder reconstructs approximations of the original input from these latent representations. This framework achieves effective dimensionality reduction by identifying latent variables that encode the intrinsic structure of data. Formally, a VAE is a probabilistic model that processes an input vector $\mathbfit{x} \in \mathcal{X}$ by introducing a latent variable $\mathbfit{z}\in \mathbb{R}^k$ with a predefined prior distribution (e.g., standard Gaussian distribution). Given a dataset $\mathscrsfs{D}=\{\mathbfit{x}_i\}_{i=1}^n$ with each $\mathbfit{x}_i \in \mathbb{R}^m$, the model optimizes the log-likelihood of the data under the generative process:
\begin{equation}
    \log p_{\theta}(\mathbfit{x}_1,...,\mathbfit{x}_n)=\log \prod_{i=1}^{n} p_{\theta}(\mathbfit{x}_i)
\end{equation}

To tractably maximize this log-likelihood, VAEs optimize the Evidence Lower Bound (ELBO) for each data point $\mathbfit{x}_i$, defined as \cite{SHI2023101940}:
\begin{equation}
\begin{split}
\log p_{\theta}(\mathrm{\mathbfit{x}}_i)   &= \log \int p_{\theta}(\mathbfit{x}_i|\mathbfit{z})p(\mathbfit{z})\frac{q(\mathbfit{z}|\mathbfit{x}_i)}{q(\mathbfit{z}|\mathbfit{x}_i)}\mathrm{d}\mathbfit{z} \\
                                &= \log \mathbb{E}_{\mathbfit{z} \sim q_{\phi}(\mathbfit{z}|\mathbfit{x}_i)} \Big[p_{\theta}(\mathbfit{x}_i|\mathbfit{z})\frac{p(\mathbfit{z})}{q(\mathbfit{z}|\mathbfit{x}_i)}\Big]
                                \\
                                &\geq \underbrace{\mathbb{E}_{q_{\phi}(\mathbfit{z}|\mathbfit{x}_i)}\Big[\log p_{\theta}(\mathbfit{x}_i|\mathbfit{z})\Big]}_{\text{reconstruction error}}-\underbrace{\mathcal{D}_{KL}\Big(q_{\phi}(\mathbfit{z}|\mathbfit{x}_i)||p_{\theta}(\mathbfit{z})\Big)}_{\text{regularsation}} \\
                                &= \mathcal{L}_{\text{VAE}}(\mathbfit{x}_i,\theta,\phi)
\end{split}
\label{eq_2}
\end{equation}
where $\theta$ denotes the parameters of the decoder $p_{\theta}(\mathbfit{x}|\mathbfit{z})$ reconstructing data from latent variables. $\phi$ represents the parameters of the encoder $q_{\phi}(\mathbfit{z}|\mathbfit{x})$ approximating the posterior distribution over latent variables. $\mathcal{D}_{KL}$ stands for the Kullback–Leibler divergence. By balancing accurate reconstruction with latent space regularization, VAEs efficiently learn compressed representations while maintaining the ability to generate coherent outputs from the latent space.

\subsubsection{Generative adversarial networks}
GANs \cite{NIPS2014_f033ed80} are a class of DGMs that employ a unique adversarial training process. A GAN consists of two competing networks: a generator (denoted $G$) and a discriminator (denoted $D$). The generator tries to create synthetic samples (e.g., thermal images, vibration acceleration signals) given a random noise, while the discriminator attempts to distinguish between whether the generated samples are real or fake \cite{wen2024gear}. This adversarial process leads the generator to progressively refine its output samples, ultimately producing realistic high-fidelity outputs. The objective function governing this process is formulated as a minimax optimization problem \cite{he2024llmsmeetmultimodalgeneration}:
\begin{equation}
    \operatorname*{min}_{\theta_{G}} \operatorname*{max}_{\theta_{D}} \mathcal{L}_{\text{GAN}}=\operatorname*{min}_{\theta_{G}} \operatorname*{max}_{\theta_{D}} \mathbb{E}_{\mathbfit{x}\sim{p_{\text{data}}}}\bigg[\log D(\mathbfit{x};\theta_{D})\bigg]+\mathbb{E}_{\mathbfit{z}\sim p_{\text{model}}}\bigg[\log \Big(1-D\big(G(\mathbfit{z};\theta_{G});\theta_{D}\big)\Big)\bigg]
\end{equation}
where $\mathcal{L}_{\text{GAN}}$ represents the adversarial loss of GAN, and $\mathbfit{x}$ denotes the data sampled from $p_{\mathrm{data}}$. $p_{\mathrm{model}}$ is the probabilistic distribution in latent space (typically Gaussian).
 
\subsubsection{Diffusion models}
Denoising diffusion probabilistic models (DDPMs) \cite{ho2020denoisingdiffusion}, commonly referred to as diffusion models, form the foundation of the diffusion-based GenAI framework. These models represent a class of DGMs that have demonstrated exceptional performance in image generation tasks. The core of the DDPMs is to construct two Markov chains: a forward diffusion process and its corresponding backward denoising. The forward diffusion process incrementally corrupts an initial data sample $\mathbfit{x}_0 \sim q(\mathbfit{x}_0)$ over $T$ discrete steps by introducing additive Gaussian noise, until it is transformed into a pure noise sample $\mathbfit{x}_T$. Reversely, the backward denoising, as it indicates, iteratively reverses this transformation to reconstruct the original data distribution from pure noise. Formally, the forward process defines a Markov chain $\{\mathbfit{x}_{t}\}_{t=0}^T$ governed by the forward transition probability $q(\cdot)$ \cite{xiao2022tacklinggenerative, ho2020denoisingdiffusion}:
\begin{equation}
\begin{split}
    q(\mathbfit{x}_t|\mathbfit{x}_{t-1})&=\mathcal{N}(\mathbfit{x}_t;\sqrt{1-\beta_t}\mathbfit{x}_{t-1},\beta_t\mathbf{I}) \\
    q(\mathbfit{x}_{1:T}|\mathbfit{x}_0) &= \prod_{t=1}^T q(\mathbfit{x}_t|\mathbfit{x}_{t-1})
\end{split}
\end{equation}
where $\beta_t \in (0,1)$ denotes the time-dependent noise scale at step $t$, $\mathbf{I}$ is the identity matrix, and $\mathcal{N}$ represents the Gaussian distribution. The backward denoising denoted as $p_{\theta}(\cdot)$ iteratively reconstructs the original data distribution $q(\mathbfit{x}_0)$ from pure noise $\mathbfit{x}_T$. Specifically, the backward process defines another Markov chain governed by \cite{ho2020denoisingdiffusion}:
\begin{equation}
\begin{split}
    p_{\theta}(\mathbfit{x}_{t-1}|\mathbfit{x}_{t}) &= \mathcal{N}(\mathbfit{x}_{t-1};\mu_{\theta}(\mathbfit{x}_t,t),\Sigma_{\theta}(\mathbfit{x}_t,t)) \\
    p_{\theta}(\mathbfit{x}_{0:T}) &= p_{\theta}(\mathbfit{x}_T) \prod_{t=1}^{T} p_{\theta}(\mathbfit{x}_{t-1}|\mathbfit{x}_{t})
\end{split}
\end{equation}
where the mean $\mu_{\theta}$ and covariance $\Sigma_{\theta}$ are learned by a neural network (typically using a U-Net architecture), parameterized by $\theta$. The loss function of training is to maximize the likelihood $p_{\theta}(\mathbfit{x}_0) = \int p_{\theta}(\mathbfit{x}_{0:T}) \text{d} \mathbfit{x}_{1:T}$ by maximizing the ELBO, $\mathcal{L}_{\text{Diff}} \leq \log p_{\theta}(\mathbfit{x}_{0})$. The ELBO can be written as matching the true denoising distribution $q(\mathbfit{x}_{t-1} | \mathbfit{x}_{t})$ with the parameterized denoising model $p_{\theta}(\mathbfit{x}_{t-1} | \mathbfit{x}_{t})$ by \cite{song2022ddim}:
\begin{equation}
    \mathcal{L}_{\text{Diff}} = - \sum_{t \geq 1} \mathbb{E}_{q(\mathbfit{x}_t)}\bigg[\mathcal{D}_{KL} \Big(q(\mathbfit{x}_{t-1} | \mathbfit{x}_{t}) || p_{\theta}(\mathbfit{x}_{t-1} | \mathbfit{x}_{t})  \Big) \bigg] + \text{C}
\end{equation}
where $\text{C}$ is a constant. The forward diffusion and backward denoising processes, and key characteristics of the diffusion model are also summarized in Table~\ref{table_1}.

\subsubsection{Large language models}

LLMs have become a dominant paradigm for addressing various engineering challenges. These models are particularly notable for their ability to process diverse inputs, such as text prompts, questions, or visual images through prompt engineering. Representative LLMs such as the generative pre-trained transformer 4 (GPT-4) \footnote{https://openai.com/index/gpt-4/}, Llama 4 \footnote{https://www.llama.com/}, Claude \footnote{https://claude.ai/}, and DeepSeek \footnote{https://www.deepseek.com/}, all utilize a Transformer-based architecture pre-trained on massive datasets. This architecture integrates attention mechanisms to capture both local and global dependencies between tokens \cite{vaswani2023attentionneed}. Mathematically, LLMs function as autoregressive models that predict the next token conditioned on previous tokens, which can be expressed as the following conditional probability:

\begin{equation}
    p(\mathbfit{x}_n|\mathbfit{x}_{1:{n-1}})=\text{softmax}(\mathbf{W}h_{n-1})
\end{equation}
where $p(\mathbfit{x}_n|\mathbfit{x}_{1:{n-1}})$ represents the transition probability of the next token given all previous tokens up to the $n-1$ position. $\mathbf{W}$ denotes the learnable weight matrices, $h_{n-1}$ is the hidden state derived from the Transformer by processing the previous input sequence $\mathbfit{x}_{1:{n-1}}$, and $\text{softmax}$ refers to the softmax activation function that converts logits into a probability distribution for token selection. This indicates the model will output the token with the highest probability after the softmax function. A comparison of the model architectures and specific characteristics of the DGMs presented is illustrated in Table \ref{table_1}.

\begin{table}[H]
\footnotesize
\caption{Comparison of basic architectures of DAR, VAE, GAN, Diffusion model, and LLM}
\label{table_1}
\renewcommand{\arraystretch}{1.3}
\makebox[\textwidth][c]{   
\begin{tabularx}{1.2\textwidth}{@{\extracolsep{\fill}} m{0.6cm} m{6.5cm} m{3.8cm} m{3.8cm}}
\toprule

\textbf{Models} & 
\multicolumn{1}{>{\centering\arraybackslash}m{6.5cm}}{\textbf{Base architectures}} & 
\textbf{Pros} & 
\textbf{Cons} \\
\midrule\midrule[.1em]

DAR & 
  \begin{tabular}[c]{@{}l@{}}
    \includegraphics[width=0.4\textwidth]{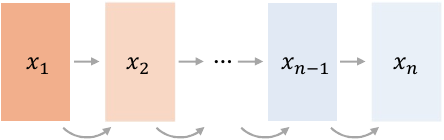} \\ 
  \end{tabular} &
    \begin{itemize}[itemsep=0pt, label={-}, leftmargin=*]
       \item Sequential generation given previous inputs
       \item Specialize for time series or flow data \cite{Yanfei2020}
    \end{itemize} & 
    \begin{itemize}[itemsep=0pt, label={-}, leftmargin=*] 
           \item Slow to train and converge
       \item Struggle to model complex, non-autoregressive distributions \cite{ye2025autoregression}
       \end{itemize} \\
    
\cline{2-4} 
\addlinespace[1.5ex]

VAE & 
  \begin{tabular}[c]{@{}l@{}}
    \includegraphics[width=0.4\textwidth]{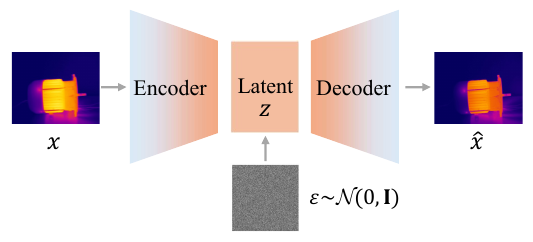} \\ 
  \end{tabular}
    & \begin{itemize}[itemsep=0pt, label={-}, leftmargin=*]
       \item Probabilistic encoder-decoder architecture \cite{Kingma_2019}
        \item Good to learn latent representation
        \item Stable training process
        \end{itemize}
    &   \begin{itemize}[itemsep=0pt, label={-}, leftmargin=*]     
        \item Sensitive to a limited dataset
        \item May suffer from posterior collapse \cite{Lucas2019UnderstandingPC}
        \end{itemize} \\
    
\cline{2-4} 
\addlinespace[1.5ex]

GAN & 
  \begin{tabular}[c]{@{}l@{}}
    \includegraphics[width=0.4\textwidth]{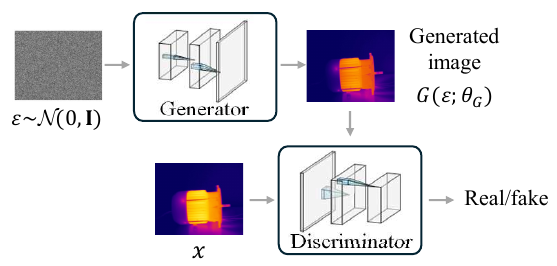} \\ 
  \end{tabular} 
    & \begin{itemize}[itemsep=0pt, label={-}, leftmargin=*]
        \item Adversarial generator-discriminator framework
        \item Produces high resolution synthetic samples \cite{xiao2022tacklinggenerative}
    \end{itemize} & 
    \begin{itemize}[itemsep=0pt, label={-}, leftmargin=*]
        \item Prone to training instability \cite{ZHANG2025125059}
        \item The training loss can be slow to converge \cite{9934291}
        \end{itemize} \\
    
\cline{2-4}
\addlinespace[1.5ex]

Diffusion model &  
  \begin{tabular}[c]{@{}l@{}}
    \includegraphics[width=0.5\textwidth]{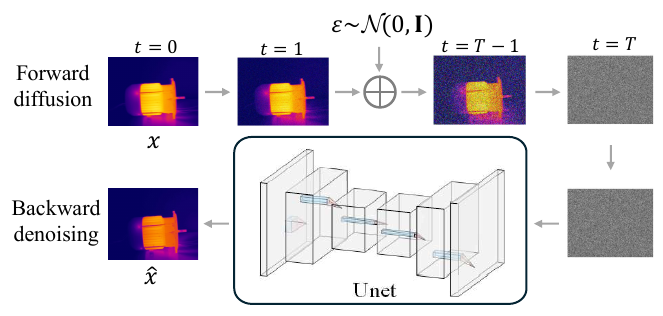} \\ 
  \end{tabular} 
    & \begin{itemize}[itemsep=0pt, label={-}, leftmargin=*]
        \item Forward diffusion and backward denoising process \cite{ho2020denoisingdiffusion}
        \item Stable training via gradually removing noise
        \item High-fidelity sample generation \cite{urain2024dgmsrobotics}
        \end{itemize}
    & \begin{itemize}[itemsep=0pt, label={-}, leftmargin=*]
         \item Large number of denoising steps to cause training complexity \cite{song2022ddim}
         \item Hard to control over the generation process
    \end{itemize}
    \\
    
\cline{2-4}
\addlinespace[1.5ex]

LLM & \begin{tabular}[c]{@{}l@{}}
    \includegraphics[width=0.5\textwidth]{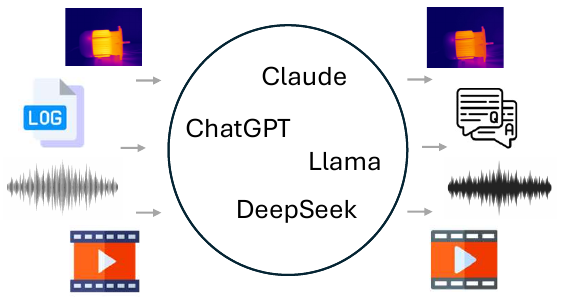} \\
  \end{tabular}  & \begin{itemize}[itemsep=0pt, label={-}, leftmargin=*]
        \item Transformer-based architecture
        \item Pre-trained on large image text paired datasets
        \item Multimodal fusion via attention mechanisms \cite{vaswani2023attentionneed} \end{itemize}
         & \begin{itemize}[itemsep=0pt, label={-}, leftmargin=*]
         \item Large computation resources both in the training and inference stages 
         \item Large high-quality training data
    \end{itemize}\\
    \bottomrule
\end{tabularx}
}
\scriptsize{$^*$ The thermal images of the motor illustrated in the table were retrieved from \cite{NajafiDataset}}.\\
\end{table}

\subsection{DGMs vs Classical generative models}
Section \ref{section2.2} outlines the basic theory of state-of-the-art DGMs. Although generative models have developed rapidly in recent years, some classic methods remain useful and conceptually significant for CM/SHM. Representative classical generative methods include Gaussian mixture models (GMMs) \cite{WEN2024107562}, hidden Markov models (HMMs) \cite{CORACA2023109025}, and autoregressive time-series models such as ARIMA. These are often referred to as \textit{statistical} or \textit{classical} generative models. Classical generative models are parametrically compact: a small number of parameters, low computational cost, strong interpretability, and good performance on small datasets as they can explicitly encode domain-specific structures such as linear dynamics or known statistical dependencies. However, their representational capability is limited, as they are often too shallow to capture the highly nonlinear, high-dimensional, or multimodal patterns present in multi-sensor data.

It is worth noting that the autoregressive models span a wide range of models from these classical autoregressive models through deep autoregressive networks to the latest GPT-style autoregressive large models (i.e., LLMs) \cite{naveed2024comprehensiveoverviewlargelanguage}. On that basis, we compare classical generative models and DGMs with respect to model capacity, data and computational requirements, inference speed, and interpretability, as shown in Table~\ref{table_2}.

\begin{table}[H]
\footnotesize
\caption{Comparison between classical generative models and DGMs}
\renewcommand{\arraystretch}{1.5}
\makebox[\textwidth][c]{   
\begin{tabularx}{1\textwidth}{@{\extracolsep{\fill}}  m{3cm} m{5cm} m{6.5cm}}
\toprule

\multirow{2}{3cm}{\textbf{Metric}}         & \textbf{Classical generative models} & \textbf{Deep generative models} \\
        & (GMM, HMM, ARIMA, etc.) & (DAR, VAE, GAN, Diffusion, LLMs, etc.) \\

\midrule\midrule[.1em]
Training objective  & Maximum likelihood, Expectation–Maximization, least squares & ELBO (VAE), adversarial loss(GAN), denoising score (Diffusion model) \\
Data types & Low dimension data, time series & High-dimensional complex data: images, audio,  multimodal sensor data \\
Model complexity & Typical range: $10^2-10^4$ \cite{3408318}& Typical range:$10^4-10^{9+}$ (VAEs/GANs $\approx 10^6-10^8$ \cite{hibat2024framework}, transformers/LLMs $\geq 10^8$ \cite{zhang2024scalingmeetsllmfinetuning})\\
Evaluation metrics & Log-likelihood, RMSE, prediction interval coverage &  NLL, ELBO, Fréchet inception distance, task-specific scores (F1-Score, AUC, etc.)\\
Inference speed & $\leq 10 \mathrm{ms}$, typically on CPU \cite{Zhu_2022} & Normally on GPU; Diffusion/LLM inference: $\geq100 \mathrm{ms}$ \cite{NEURIPS2024_9ad996b5} \\
Interpretability & Highly interpretable & Low/moderate interpretability due to a large amount of model parameters \\
\bottomrule
\end{tabularx}}
\label{table_2}
\end{table}

\section{DGMs application in CM/SHM}
\label{Sec:3}
The previous section introduced the latest DGMs in detail and compared the advantages and disadvantages between classical generative models and DGMs. This section systematically investigates DGMs in different tasks and various industrial systems. Section~\ref{Section3.1} describes the specific tasks performed by DGMs, and Section~\ref{Section3.2} concludes the industrial systems applied in the CM/SHM communities.
\begin{figure}[H]
    \centering
    \makebox[1\textwidth][c]{
    \includegraphics[width=1\linewidth]{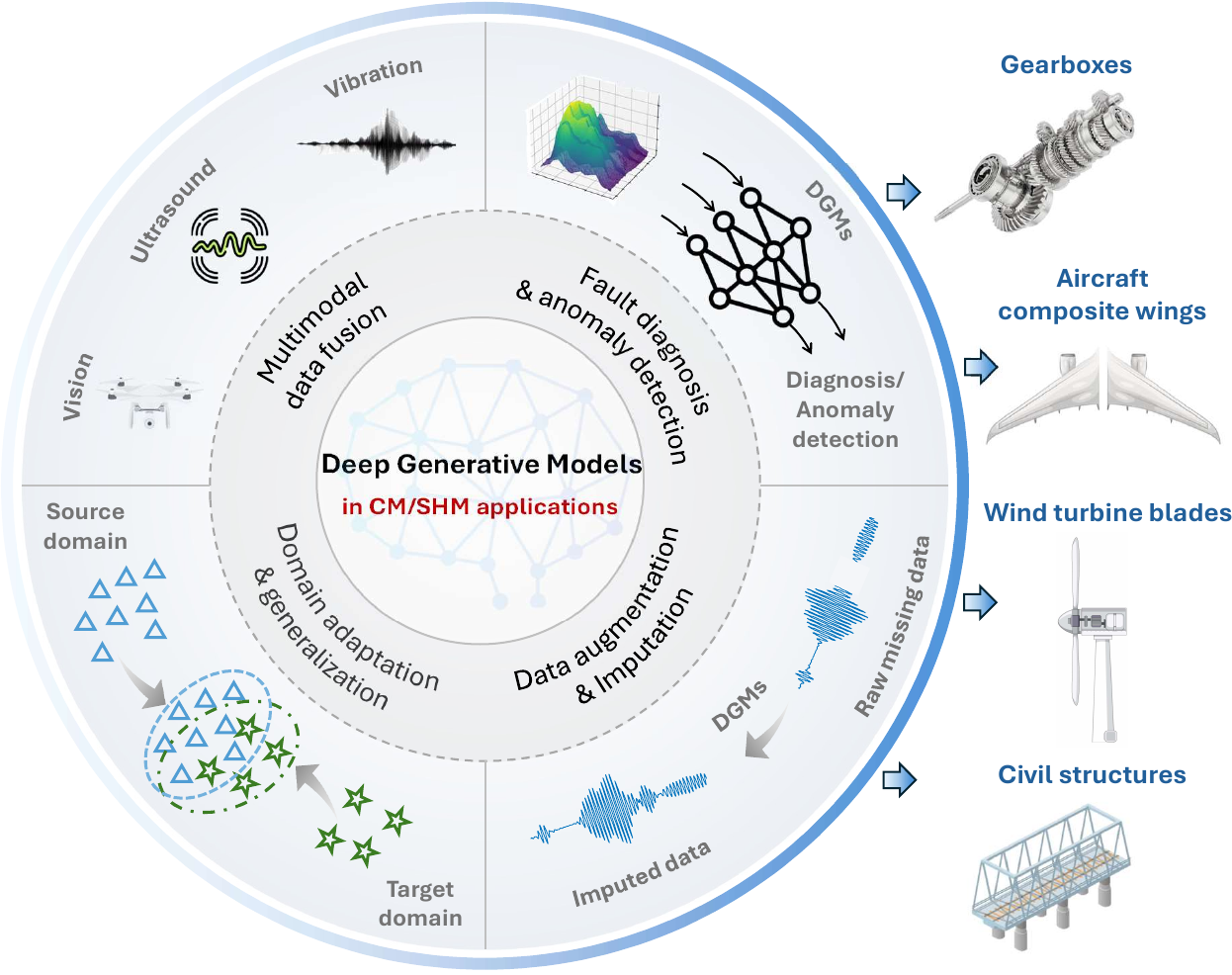}}
    \caption{Evolving applications of deep generative models in CM/SHM: from focused tasks to integrated industrial systems.}
    \label{fig3}
\end{figure}

\subsection{Applications in specific tasks}
\label{Section3.1}
\subsubsection{Performance metrics}
\label{Section3.1.1}
Before introducing the DGM application to specific tasks, the quantitative metrics are given in Table~\ref{table_3} to fully evaluate the performance of the DGM. These metrics are widely implemented in various subsequent tasks. For instance, the mean absolute error (MAE) can be used to evaluate the quality of synthesized data compared to real data in data augmentation tasks, as well as in fault diagnosis and anomaly detection tasks. The Pearson correlation coefficient (PCC) is used to evaluate the similarity between real and synthetic data in the augmentation task. 

\begin{table}[H]
\footnotesize
\caption{Summary table for performance metrics}
\renewcommand{\arraystretch}{1.7}
\makebox[\textwidth][c]{   
\begin{tabularx}{1\textwidth}{@{\extracolsep{\fill}}  m{6.7cm} m{10cm}}
\toprule

\textbf{Performance metrics}         & \textbf{Mathematic expressions} \\

\midrule\midrule[.1em]
Mean absolute error (MAE)      & $ \frac{1}{N} \sum_{i=1}^N |x_i-\widetilde{x}_i|$  \\
Mean absolute percentage error (MAPE)   & $ \frac{1}{N} \sum_{i=1}^N \frac{|x_i-\widetilde{x}_i|}{x_i} \times 100$  \\
Root mean square error (RMSE)  &  $\sqrt{\frac{1}{N} \sum_{i=1}^N (x_i-\widetilde{x}_i)^2}$ \\
Accuracy (ACC)                      &  $1-\frac{\sqrt{\sum_i^N (x-\widetilde{x_i})^2}}{\sqrt{\sum_i^N x_i^2}}$  \\
Correlation coefficient ($\text{R}^2$)    & $1-\frac{\sum_i^N (x_i-\widetilde{x}_i)^2}{\sum_i^N(x_i-\bar{x})^2}$   \\
Root mean square (RMS)                    & $\sqrt{\frac{1}{N}\sum_{i=1}^N x_i^2}$ \\
 Pearson correlation coefficient (PCC)   & $\frac{\sum_{i=1}^{N}\left(x_{i}-\overline{x}\right)\left(\widetilde{x_{i}}-\overline{\widetilde{x}}\right)}{\sqrt{\sum_{i=1}^{N}\left(x_{i}-\overline{x}\right)^{2}} \sqrt{\sum_{i=1}^{N}\left(\widetilde{x_{i}}-\overline{\widetilde{x}}\right)^{2}}}$ \\
Crest factor                        & $\text{Peak}(x)/\text{RMS}$ \\

Maximum mean discrepancy (MMD)            & $\left[\frac{1}{m^{2}} \sum_{i, j=1}^{m} k\left(x_{i}, x_{j}\right)-\frac{2}{m n} \sum_{i=1}^{m} \sum_{j=1}^{n} k\left(x_{i}, \widetilde{x}_{j}\right)
+\frac{1}{n^{2}} \sum_{i, j=1}^{n} k\left(\widetilde{x}_{i}, \widetilde{x}_{j}\right)\right]^{\frac{1}{2}}$ \\ 
KL Divergence ($\mathcal{D}_{KL}$)        &  $\sum_{i=1}^N P(x_i)\log \frac{P(x_i)}{Q(\widetilde{x}_j)}$ \\
Structural similarity index measure (SSIM) &  $\frac{\left(2 \mu_{x} \mu_{\widetilde{x}}+c_{1}\right)\left(2 \sigma_{x \widetilde{x}}+c_{2}\right)}{\left(\mu_{x}^{2}+\mu_{\widetilde{x}}^{2}+c_{1}\right)\left(\sigma_{x}^{2}+\sigma_{\widetilde{x}}^{2}+c_{2}\right)}$  \\
Signal-to-noise ratio (SNR)            &  $20 \log_{10}(A_{signal}/A_{noise})$ \\
Earth mover’s distance (EMD) & $\inf _{\gamma \in \Pi(P, Q)} \mathbb{E}_{(x, \widetilde{x}) \sim \gamma}[d(x, \widetilde{x})]$ \\
Classification metrics & Accuracy, Precision, Recall, F1-score, confusion matrix \\
\bottomrule
\end{tabularx}}
\label{table_3}
\vspace{0.2cm}
\scriptsize{\textbf{Note}: 
$x$, $\bar{x}$, $P(x_i)$, $\sigma$ denote the real data, its averaged value, probability distribution and standard deviation, respectively; $\widetilde{x}$ and $Q(\widetilde{x})$ represent synthetic data, generated, estimated or imputed data, and its probability distribution. 
}
\end{table}

\subsubsection{DGMs for data generation}
\label{Section3.1.2}
\begin{figure}[H]
    \centering
    \makebox[1\textwidth][c]{
    \includegraphics[width=1\linewidth]{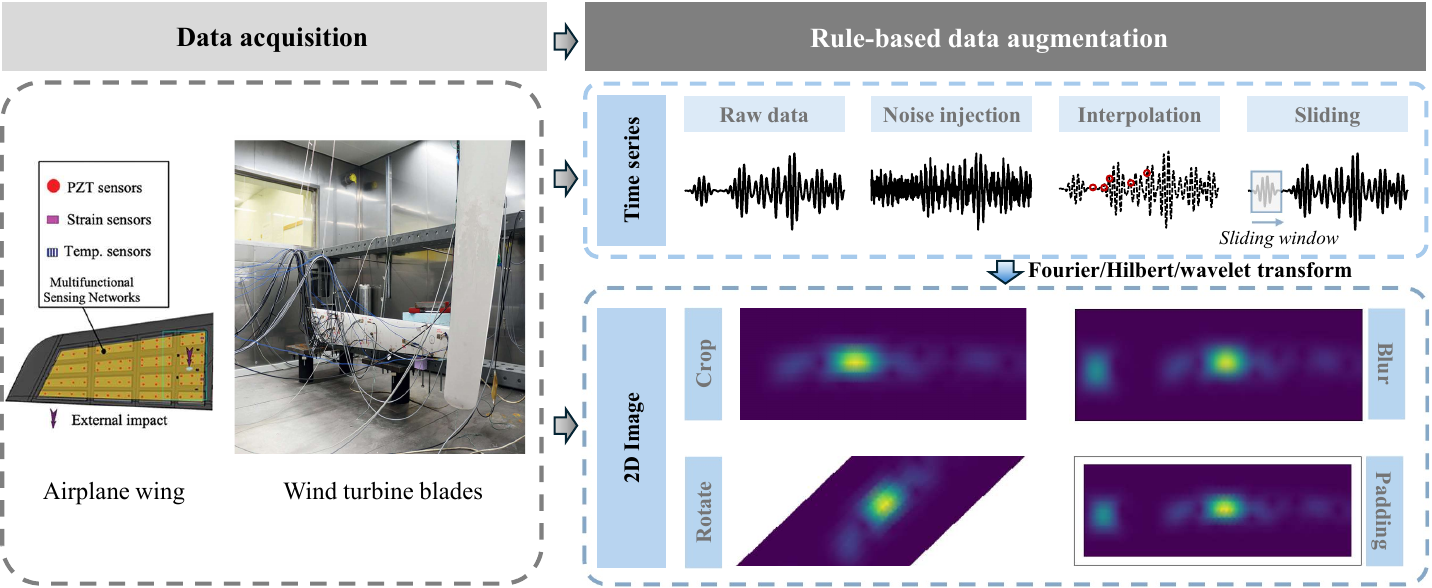}}
    \caption{Data transformation-based augmentation approach. In CM/SHM applications, both time series and 2D image formats are utilized. The 2D representations are typically generated from 1D time series through signal transformation techniques such as the short-time Fourier transform and the wavelet transform. The airplane wing and wind turbine blade figures are reused from Refs.~\cite{Qing03042022} and \cite{Yaowen2021}, respectively.}
    \label{fig4}
\end{figure}

Data generation serves as an umbrella concept with data augmentation, reconstruction and imputation, which differ mainly in how and why the data are generated. Data augmentation can improve model robustness by generating new samples from existing data, while data reconstruction tries to restore or rebuild signals from partial or compressed data. In contrast, data imputation focuses on filling in missing or corrupted values to complete the dataset. Following the data augmentation methods defined in \cite{Ju_2024}, these techniques can be categorized into three primary groups: the synthetic minority over-sampling technique (SMOTE), data transformation-based augmentation, and generative model-based approaches. SMOTE-based methods specifically address class imbalance by synthetically generating minority class samples (e.g., fault or defect samples in CM/SHM applications). Whereas data transformation-based approaches enhance dataset diversity through operations such as noise injection or image manipulations (e.g., cropping, rotating, flipping, and padding). Fig.~\ref{fig4} is an illustration for both 1D time-series and 2D image augmentation in CM/SHM applications. The third category comprises generative model-based approaches, which in this work refer to DGM-based approaches.

DGM-based data augmentation methods have emerged as indispensable tools to address challenges in scenarios where real-world data is scarce and noisy. These models enable robust supplementation and augmentation of existing datasets by synthesizing high-quality data that mirrors the statistical properties of real data samples, especially when acquiring experimental operation data is often prohibitively costly and labor-intensive \cite{CHEN2024128167, Zhao3649447}. Thus, by adopting DGM-based data augmentation, industrial sectors are expected to mitigate data scarcity and address imbalance issues inherent to industrial settings, thus enriching training datasets for downstream FD\&AD models while preserving the integrity of original data distributions \cite{LIU2024124511, LIU2021107488}. Beyond data augmentation, DGMs also play a critical role in \textit{data imputation}, a solution to resolve missing data points in industrial systems \cite{ZHANG2025110663}. Missing data may seriously compromise the diagnostic or detective performance of DGM-based models \cite{ma2021identifiable}. The inherent ability of DGMs to learn complex data distributions makes them well-suited for data imputation tasks. Instead of simply filling in missing values with fixed estimates, DGMs can leverage the underlying patterns and correlations present in the data to generate plausible values for missing entries. Table~\ref{table_4} summarizes the existing literature on applying DGMs to data augmentation and imputation tasks. We comprehensively evaluate signal processing, conventional ML-based approaches (e.g., support vector regression (SVR), recurrent neural networks (RNN), long short-term memory (LSTM), gated recurrent unit (GRU)), and DGM-based methods using the performance metrics outlined in Table~\ref{table_4}. SMOTE-based approaches are classified as conventional ML, as SMOTE generates synthetic data through linear interpolation between minority class samples and their k-nearest neighbors (KNN) \cite{elreedy2024theoretical}. 

Existing studies explicitly focused on signal processing, conventional ML, and DGM are listed in \textbf{Refs}. Works comparing the performance improvements across these approaches using evaluation metrics in Table \ref{table_4} are detailed in \textbf{Quantitative comparison}. From the table, it is found: (i) most works focus exclusively on data augmentation or imputation without cross-comparative analysis; (ii) in quantitative comparisons, DDPM and GAN-based models consistently outperform conventional ML methods (e.g., LSTM, SMOTE) across metrics such as MMD and EMD; (iii) cross comparisons between signal processing methods, conventional DNNs, and DGMs remain scarce. This is largely because signal processing techniques are typically used as preprocessing steps, while DGMs are generally regarded as complementary tools for data generation that do not explicitly incorporate physical knowledge.

Moreover, diffusion-based models are typically large-scale architectures containing billions of parameters. Recent studies have examined the relationship between the training time and data generation quality in diffusion-based models \cite{HU2025113122, SHU2025111783, YANG2025112996}. As discussed in these works, the number of diffusion steps is a key factor determining the final data generation quality, which typically ranges from 500 to 1000 steps \cite{YANG2025112996}. This observation has motivated the development of denoising diffusion implicit models (DDIMs) \cite{song2022ddim}, which aim to accelerate the sampling process while maintaining generation fidelity. In practice, the training time for a typical diffusion model often exceeds 10 hours even when using commercial-grade GPUs.

\begin{table}[H]
  \centering
  \caption{Quantitative evaluation of signal processing, ML/DNNs, and DGMs across various tasks}
  \label{table_4}
  \scriptsize
  \renewcommand{\arraystretch}{1.3}
  \makebox[\textwidth][c]{ 
  \begin{tabularx}{1.2\textwidth}{@{\extracolsep{\fill}} 
      m{1.8cm} m{5.5cm} m{5.5cm} m{5.5cm}}
    \toprule
    & \textbf{Signal processing} & \textbf{Conventional ML/DNNs} & \textbf{DGMs} \\
    \midrule
    \diagbox[width=1.9cm]{Tasks}{Models} & Data transformation-based/Least-squares & SMOTE/CNN/RNN/LSTM  & DAR/VAE/GAN/Diffusion \\
    \midrule\midrule[.1em]

    \multirow{6}{2cm}{\textbf{Data Augmentation}} 
    & \textbf{Refs}: Data transformation-based \cite{9259076, NGUYEN2022114172, 9673114, OH202072}
    & \textbf{Refs}: SMOTE-based \cite{JANSSEN2020115483, 8815712, 8737894}
    & \textbf{Refs}: DAR-based \cite{CHEN2023116063, ZHANG2022109175, djemili2024wind, Chen10081265}, VAEs \cite{LI2025110027, li2023machine}, GANs \cite{LI2025112062,Xiaoming2021,10058512, WANG2025110760, ZHUANG2024111186, WANG2023106872, LIU2022110888, WAN2025109614, LI2022101552, LI20242958, ZHANG2024114795, SHEN2024109159, Sun10504387, WEN2024111663,XU2023107063, GUO2025110854}, Diffusion-based \cite{ZHENG2025119685, SHEN2024109299, DENG2025116595, SHU2025111783, 10750268, MUELLER2024107696, ZHAO2025109520, YU2024110343, Wang0309714, HerTerng, YANG2025110312, YI2024111481, WANG2024110394}      \\
    & \textbf{Quantitative comparison}:  &  &  \\
    & \multicolumn{3}{l}{-- \cite{GUO2025115951} DDPM-based method surpasses GAN-based and autoencoder model in augmentation task with metric: MMD $\downarrow (\leq 1.046)$.}   \\
    & \multicolumn{3}{l}{-- \cite{Fu_2023} Different GAN-based variants for wavelet transformed image augmentation: MMD$\downarrow$, EMD$\downarrow$, $\mathcal{D}_{KL}\downarrow$, SSIM$(>0.9)$.}   \\
    &  \multicolumn{3}{l}{-- \cite{10136604} Comparison of LSTM, VAE, GAN-based data augmentation: Improved diagnosis performance ACC $\uparrow$ (confusion matrix).}   \\
    &  \multicolumn{3}{l}{-- \cite{10239215} SMOTE and GAN-based data augmentation performance comparison:  classification accuracy $\uparrow$ (F1-score, confusion matrix).} \\
    \cmidrule(lr){2-4}

    \multirow{10}{2cm}{\textbf{Data Imputation}} 
    & \textbf{Refs}: Least-squares \cite{van2015time}, Bayesian \cite{QiAngWang2022}
    & \textbf{Refs}: KNN \cite{chirici2016meta}, Principal component analysis (PCA) \cite{9851940}, RNN \cite{YE2024}
    & \textbf{Refs}: DAR-based \cite{ZHANG2022108718, WANG2023110703}, VAEs \cite{Ahang_2024}, GANs \cite{QU2020106610, ZHENG2025119694,GAO2023105277,Huachen2022}, Diffusion-based \cite{NEURIPS2024_cb1ba6a4} \\
    & \textbf{Quantitative comparison}:  &    &  \\
    & \multicolumn{3}{l}{-- \cite{LI2021109377} The proposed LSTM-based methods outperform random forest (RF), light gradient boosting machine, and multiple}  \\
    & \multicolumn{3}{l}{\hspace{0.9cm} linear regression for dams missing data imputation task: $\text{R}^2(\geq0.838)$, MAE $\downarrow$, RMSE $\downarrow$.}  \\
    & \multicolumn{3}{l}{-- \cite{REN2021107734} The proposed Bayesian matrix learning approach outperforms PCA in SHM data imputation: higher ACC ($\geq$0.95).}  \\
    & \multicolumn{3}{l}{-- \cite{tan2024missing} Comparison among KNN, RF, matrix factorization, RNN, their proposed model: MAE (7.77), RMSE (10.92).}  \\
    & \multicolumn{3}{l}{-- \cite{song2024missing} Dam missing value imputation among GRU, LSTM, and their proposed model: MAE (0.593), MAPE (0.076), RMSE (0.731).}  \\
    & \multicolumn{3}{l}{-- \cite{HOU2022111206} GAN surpasses SVR, radial basis function network, and RNN in data imputation: lower MSE (0.0237), higher $\text{R}^2$(0.9946).}  \\
    & \multicolumn{3}{l}{-- \cite{GAO2022112095} The proposed GAN-based method surpasses LSTM and SVR in bridge data imputation: ACC $\uparrow$ ($\geq$0.9), lower execution time.}  \\
    & \multicolumn{3}{l}{-- \cite{WANG2025110973} The method outperforms the VAE and other GAN variants with: lower RMSE ($\leq 6.1$), MAE ($\leq 4.86$) values, and higher $\text{R}^2$.}  \\
    \cmidrule(lr){2-4}

    \multirow{6}{2cm}{\textbf{Domain Adaptation}} 
    & \textbf{Refs}: Not applicable
    & \textbf{Refs}: Gaussian mixture model \cite{gardner2022domain}, CNN-based models \cite{chen2023deep}
    & \textbf{Refs}: GANs \cite{SOLEIMANIBABAKAMALI2023110404, LULECI2023110370, LULECI2023106146, Lou9789138, ir202307, WANG2023455, GE2024111236, CACERESCASTELLANOS20237746, Sun10504387}, Diffusion-based \cite{ZHANG2025107031} \\
    & \textbf{Quantitative comparison}:  &   &  \\
    & \multicolumn{3}{l}{-- \cite{GARDNER2020106550} Comparison among different DA methods on four SHM scenarios: ACC $\uparrow$, F1-score (1.000).} \\
    & \multicolumn{3}{l}{-- \cite{WANG2025110662} Improved diagnostic performance of proposed dynamic collaborative adversarial domain adaptation network in unsupervised} \\
    & \multicolumn{3}{l}{\hspace{0.9cm} DA task: recall $\uparrow$, precision $\uparrow$, accuracy $\uparrow$, F1-score $\uparrow$ (classification metrics).} \\
    & \multicolumn{3}{l}{-- \cite{XU2024112396} Multi-source DA comparison among diffusion and DNN models: improved classification performance (confusion matrix, accuracy$\uparrow$).} \\
    \cmidrule(lr){2-4}

    \multirow{9}{2cm}{\textbf{Domain Generalization}} 
    & \textbf{Refs}: Not applicable
    & \textbf{Refs}: Mete-learning techniques \cite{finn2017modelagnostiCMetalearningfastadaptation}
    & \textbf{Refs}: GANs \cite{GUO2025110854}, Diffusion-based model \cite{LIU2025116989}  \\
    & \textbf{Quantitative comparison}:  \\
    & \multicolumn{3}{l}{-- \cite{Zhuang9759507} The proposed adversarial domain generalization framework that outperforms LSTM, CNN-based, temporal CNN, and} \\
    & \multicolumn{3}{l}{\hspace{0.9CM} multiscale CNN with evaluation metrics on unknown target domain: RMSE$\downarrow$ (0.1376), MAE$\downarrow$ (0.1117).} \\
    &  \multicolumn{3}{l}{-- \cite{10091197} Meta-learning-based DG model outperforms domain adversarial neural network, correlation alignment, meta-learning DG, } \\
    &  \multicolumn{3}{l}{\hspace{0.9CM} aggregation with improved metrics: accuracy ($\geq 83.11$), enhanced classification performance with confusion matrix.} \\
    &  \multicolumn{3}{l}{-- \cite{10688397} The proposed similarity-based meta-representation learning method for DG task outperforms several baseline methods} \\
    &  \multicolumn{3}{l}{\hspace{0.9CM} including neural network, hierarchical neural network, model-agnostic meta-learning (MAML), and hierarchical MAML} \\
    &  \multicolumn{3}{l}{\hspace{0.9CM} with improved metric: ACC $(15.44\% \rightarrow{31.62\%})$.} \\
    \bottomrule
  \end{tabularx}
}
\end{table}

\subsubsection{DGMs for domain adaptation and generalization}
\label{Section3.1.3}
\begin{figure}[H]
    \centering
    \makebox[1\textwidth][c]{
    \includegraphics[width=1\linewidth]{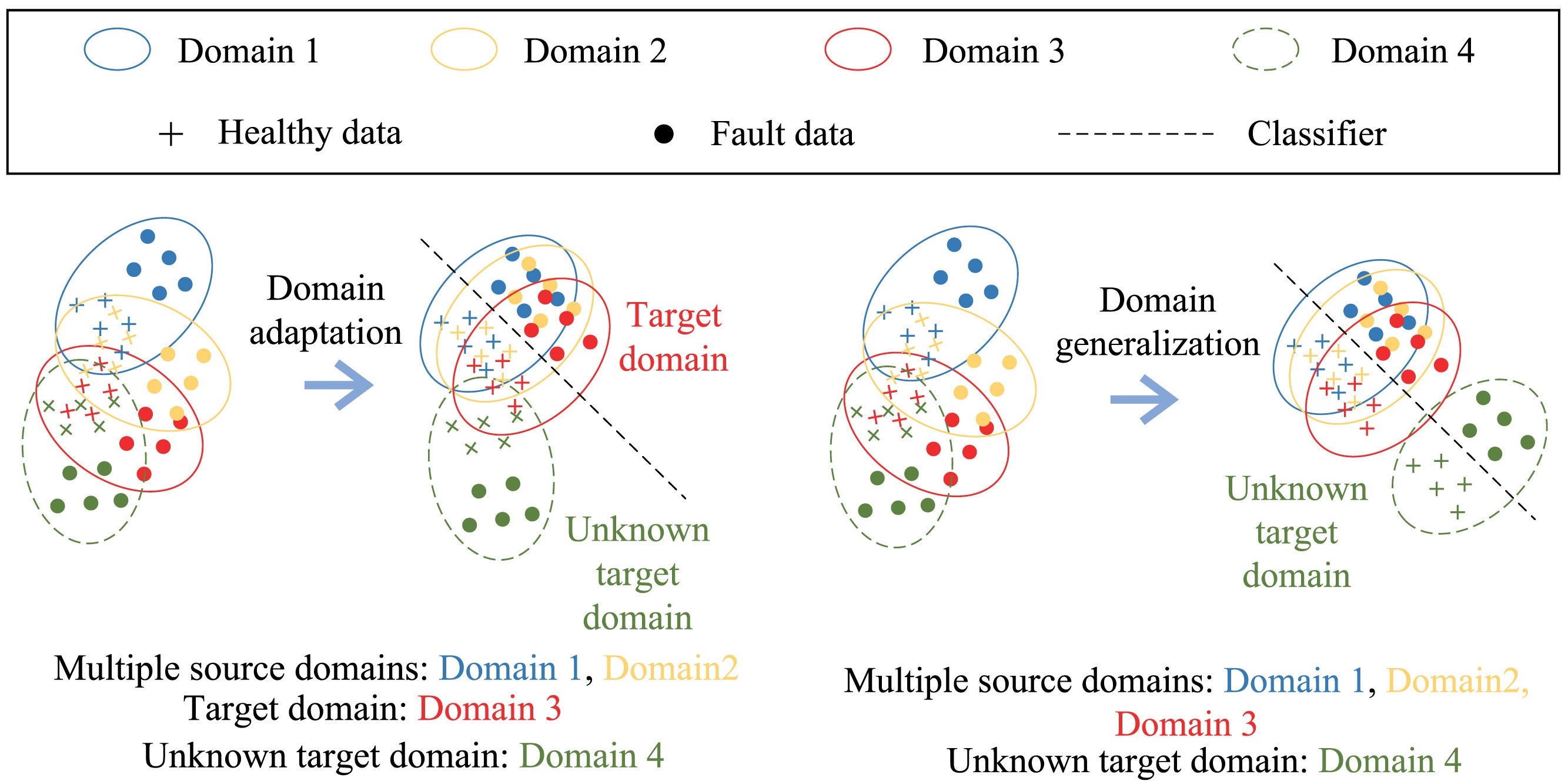}}
    \caption{An illustration of domain adaptation and domain generalization \cite{WAN2025109614}. Domain adaptation focuses on minimizing data distribution discrepancies between source and target domains, whereas domain generalization seeks to transfer knowledge from multiple source domains to an unseen target domain without utilizing any target domain data.}
    \label{fig5}
\end{figure}

Transfer learning (TL) aims to leverage knowledge from a pre-trained model on a source task to improve performance on a different yet related target domain or task \cite{wang2022generalizing, YANG2024115387}. Within this framework, \textit{domain adaptation} (DA) and \textit{domain generalization} (DG) emerge as two pivotal subfields. DA focuses on minimizing data distribution discrepancies between the source and target domains \cite{XU2024102650, Kouw_2021}, while DG seeks to transfer knowledge from multiple source domains to an unseen target domain without utilizing any data from the target domain. Hence, the main difference between DA and DG lies in data accessibility during training: DA can utilize target domain data during training (even when such data are scarce or limited), while DG relies solely on source domain knowledge. This makes DG particularly suitable for evaluating the robustness of the model in completely unseen working scenarios where the target domain data remain inaccessible \cite{WAN2025109614}. Fig.~\ref{fig5} illustrates the main difference between DA and DG concepts. 

Following the framework established in the previous subsection, Table \ref{table_4} outlines the applications of DGMs for DA and DG within the TL framework. Apparently, signal processing methods are not applicable to either DA or DG, as they belong to transfer learning, which is indeed one category of ML methods. Nevertheless, recent work mainly focuses on GAN-based models for DA, including: (i) undamaged-to-damaged data transition under scarce damage data \cite{SOLEIMANIBABAKAMALI2023110404, LULECI2023110370, LULECI2023106146}; (ii) adaptation between simulation and experimental data \cite{GE2024111236, Lou9789138, WANG2025110662}, and (iii) adaptation among different rotating machine types, speeds, or operating conditions \cite{ir202307, WANG2023455, Sun10504387}). 

A smaller proportion of studies \cite{GUO2025110854, LIU2025116989, 10688397, 10091197, Zhuang9759507} address DG under unknown or unseen working conditions. Meta-learning \cite{10688397, 10091197} stands out as a complementary method to DG by training a model to adapt quickly to new tasks or environments by leveraging knowledge gained from multiple related tasks. This is particularly relevant because many DGM-based DG approaches rely primarily on statistical learning from available data, without explicitly incorporating physical knowledge of the system. Since DG is highly desirable in industrial settings, where target domain data is unavailable, there is an emerging trend toward the use of diffusion model-based DG for CM/SHM systems \cite{GUO2025115951}. It is noted that certain citations span multiple application categories. For instance, \cite{GUO2025110854, Sun10504387, GUO2025115951} proposed methodologies that combine data augmentation with domain generalization to handle unknown target domains. As shown in Table \ref{table_4}, these approaches inherently span both categories since they both involve augmenting scarce data while achieving data generalization to unseen domains.

\subsubsection{DGMs for multimodal data fusion}
\label{section3.1.4}
In the context of multimodal data fusion in condition monitoring, the application of DGMs remains in its early stages. Notably, diffusion-based and LLM-based DGMs exhibit unique promise largely because of their Transformer-based architectures, which inherently enable the dynamic fusion of heterogeneous data types, including text, visual images, and audio inputs (known as \textit{multimodal data}), into a unified latent space. The attention mechanism \cite{vaswani2023attentionneed} within these Transformer frameworks serves as a foundation for cross-modal fusion. By dynamically computing attention scores across modalities, this mechanism enables context-aware weighting of input features, allowing the model to prioritize statistically salient or semantically correlated signals during fusion. These capabilities are particularly important for diffusion-based and LLM-based DGMs, as the attention mechanism not only enhances representational power but also ensures the coherence of model outputs with the joint distribution of multimodal inputs. 

\begin{figure}[H]
    \centering
    \makebox[1\textwidth][c]{
    \includegraphics[width=0.9\linewidth]{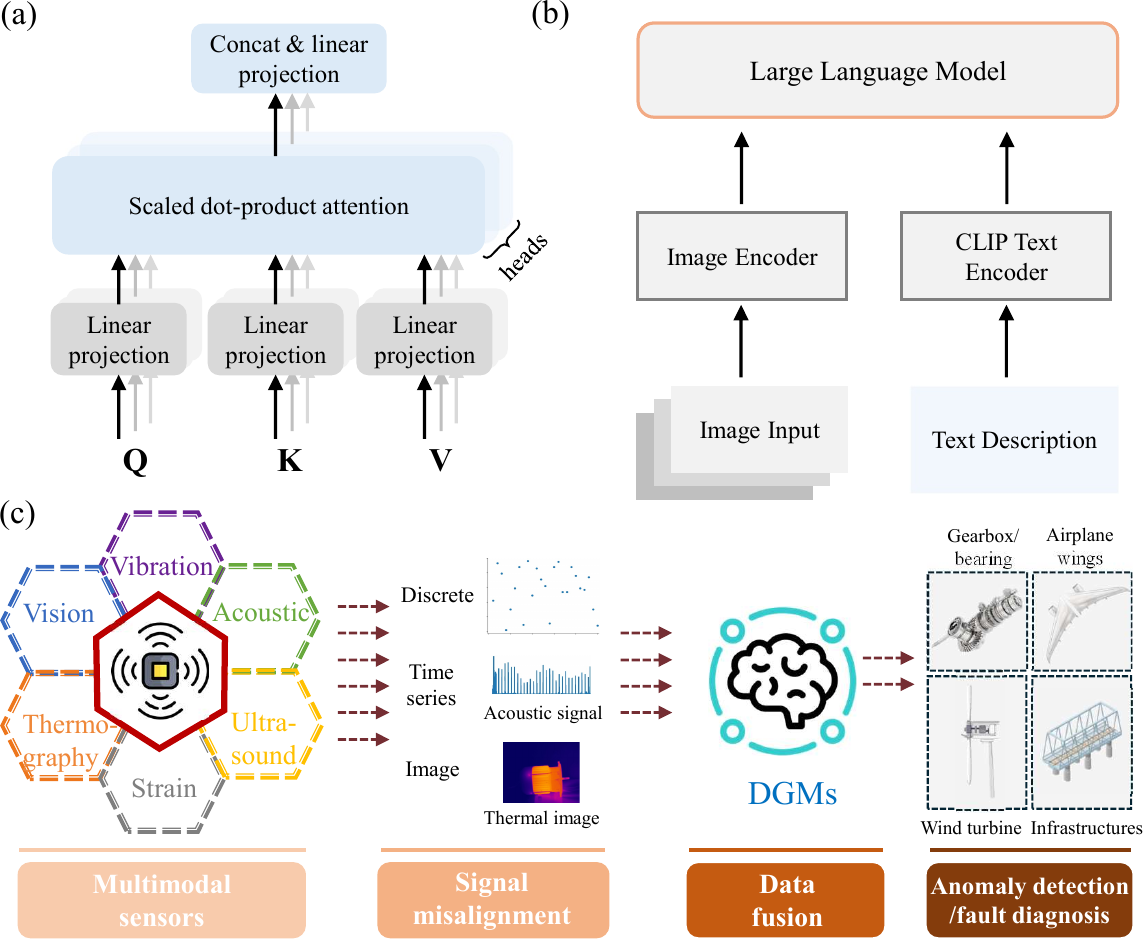}}
    \caption{DGMs application in multimodal data fusion. (a) Multihead attention mechanism in Transformer-based networks; (b) LLM-based image and text prompt modal fusion; (c) DGMs apply multimodal data fusion in condition monitoring.}
    \label{fig6}
\end{figure}

Fig.~\ref{fig6}(a) provides an illustrative scheme of the multi-head attention mechanism in the Transformer network, which is a fundamental component in both Diffusion-based (e.g., Stable Diffusion \cite{rombach2022highresolution}) and LLM-based models. Fig.~\ref{fig6}(b) illustrates the implementation of cross-modal fusion in LLMs, where paired image and text inputs are jointly processed. In this framework, the image (or visual) inputs are sent to a feature extraction module via an image encoder to obtain latent representations, which are subsequently projected into key ($\mathbf{K}$) and query ($\mathbf{Q}$) matrices. Meanwhile, the text caption is encoded by a text encoder (e.g., CLIP text encoder \cite{radford2021learning}) into a value ($\mathbf{V}$) matrix, as shown in Fig.~\ref{fig6}(a). In this way, the LLM computes attention scores by enabling the dynamic fusion of image and text modalities. Fig.~\ref{fig6}(c) depicts the main flow of multimodal data fusion based on DGM. Various data formats, such as discrete data, continuous time series, or images, can be obtained from multimodal heterogeneous sensors (such as vibration accelerometers, stereo cameras, and acoustic emission sensors). The DGMs can operate on and dynamically fuse this information to facilitate downstream FD\&AD tasks. Table \ref{table_5} concludes the DGM application for multimodal data fusion.

In Table~\ref{table_5}, we categorize diffusion-based and LLM-based models derived from Transformers as DGMs, while other Transformer-based networks are classified as DNNs. Within multimodal data fusion, diffusion and LLM-based models constitute a significant proportion of DGMs. Conversely, works using Kalman filtering for multimodal fusion are classified as signal processing methods. Although Kalman filtering is a popular and effective approach for real-time multimodal fusion in monitoring scenarios, it struggles to handle large volumes of heterogeneous data streams. This limitation indicates the potential for applying diffusion and LLM-based models to future multimodal data fusion techniques.

\begin{table}[H]
  \centering
  \caption{Quantitative evaluation of signal processing, ML/DNNs, and DGMs across various tasks}
  \label{table_5}
  \scriptsize
  \renewcommand{\arraystretch}{1.3}
  \makebox[\textwidth][c]{ 
  \begin{tabularx}{1.2\textwidth}{@{\extracolsep{\fill}} 
      m{1.8cm} m{5.5cm} m{5.5cm} m{5.5cm}}
    \toprule
    & \textbf{Signal processing} & \textbf{Conventional ML/DNNs} & \textbf{DGMs} \\
    \midrule
    \diagbox[width=1.9cm]{Tasks}{Models} & Kalman filtering/Least-squares & CNN/RNN/LSTM/Transformer  & Diffusion/LLM \\
    \midrule\midrule[.1em]

    \multirow{4}{2cm}{\textbf{Multimodal Data Fusion}} 
    & \textbf{Refs}: Kalman filtering \cite{PAL2024111482, ZHU2023116573, SHEN2023100442}
    & \textbf{Refs}: LSTM-based \cite{Jichen2025}, Transformer \cite{10922758, FALCHI2024111382}   
    & \textbf{Refs}: Diffusion-based \cite{YAO2024111397}, LLM-based \cite{electronics13244912, ZHANG2025124378, TAO2025112127, PANG2024110312, qaid2024fdllmlargelanguagemodel} \\
    & \textbf{Quantitative comparison}: \\
    & \multicolumn{3}{l}{-- \cite{CHEN2025111813} A multi-scale Transformer network was proposed with acceleration feature fusion achieving improved performance comparing} \\
    & \multicolumn{3}{l}{\hspace{0.9cm} to CNNs and sigle-scale Transformer network: Accuracy (96.12\%), confusion matrix.} \\
    & \multicolumn{3}{l}{-- \cite{YingHan2025} The proposed Swin-Transformer-based model achieves highest diagnostic accuracy compared to ResNet18, EfficienceV2,} \\
    & \multicolumn{3}{l}{\hspace{0.9cm} and Vgg16 networks by fusing infrared thermal and visible light images: Accuracy $\uparrow$, confusion matrix.}\\
    \cmidrule(lr){2-4}

    \multirow{5}{2cm}{\textbf{Fault Diagnosis \&Anomaly Detection}} 
    & \textbf{Refs}: Not applicable 
    & \textbf{Refs}: CNN-based \cite{liu2022data, JANA2022108723, RUAN2023101877}, Transformer \cite{li2022structural, 9625031, 10922758, FALCHI2024111382, WANG2024111067, JIN2022379}   
    & \textbf{Refs}: Diffusion-based, LLM-based \cite{ZHANG2025124378, TAO2025112127, PANG2024110312, qaid2024fdllmlargelanguagemodel} \\
    & \textbf{Quantitative comparison}: \\
    & \multicolumn{3}{l}{-- \cite{10922758} A two-stage encoder Transformer was proposed with 50\% improvement in accuracy comparing to extreme gradient boosting.} \\
    & \multicolumn{3}{l}{-- \cite{FALCHI2024111382} The temporal fusion Transformer network achieves higher accuracy compared to the autoregressive model: MAE$\uparrow$, MSE$\uparrow$, MAPE$\uparrow$.}  \\
    & \multicolumn{3}{l}{-- \cite{10103569} The proposed model outperforms ShuffleNet, ResNet, DenseNet models with improved: average accuracy ($93.1\%$).}  \\
    \bottomrule
  \end{tabularx}
}
\end{table}

\subsubsection{DGMs for fault diagnosis and anomaly detection}
For downstream FD\&AD tasks, it should be clarified that LLM-based models play a fundamentally different role compared with generative models such as VAE, GAN, or Diffusion models. In VAE- or GAN-based frameworks, DGMs are typically employed for signal-level tasks, such as data generation, feature extraction, or anomaly detection based on reconstruction error. These models primarily operate on the statistical distribution of sensor signals. In contrast, LLM-based approaches are not used as data generators for sensor-level signals. Instead, they function as pretrained representation learning and reasoning modules that are fine-tuned to adapt to diagnostic tasks. The core advantage of LLMs lies in their large-scale pretraining on diverse multimodal data, enabling them to perform contextual reasoning, cross-modal understanding, and decision support.

\begin{figure}[H]
    \centering
    \makebox[1\textwidth][c]{
    \includegraphics[width=0.8\linewidth]{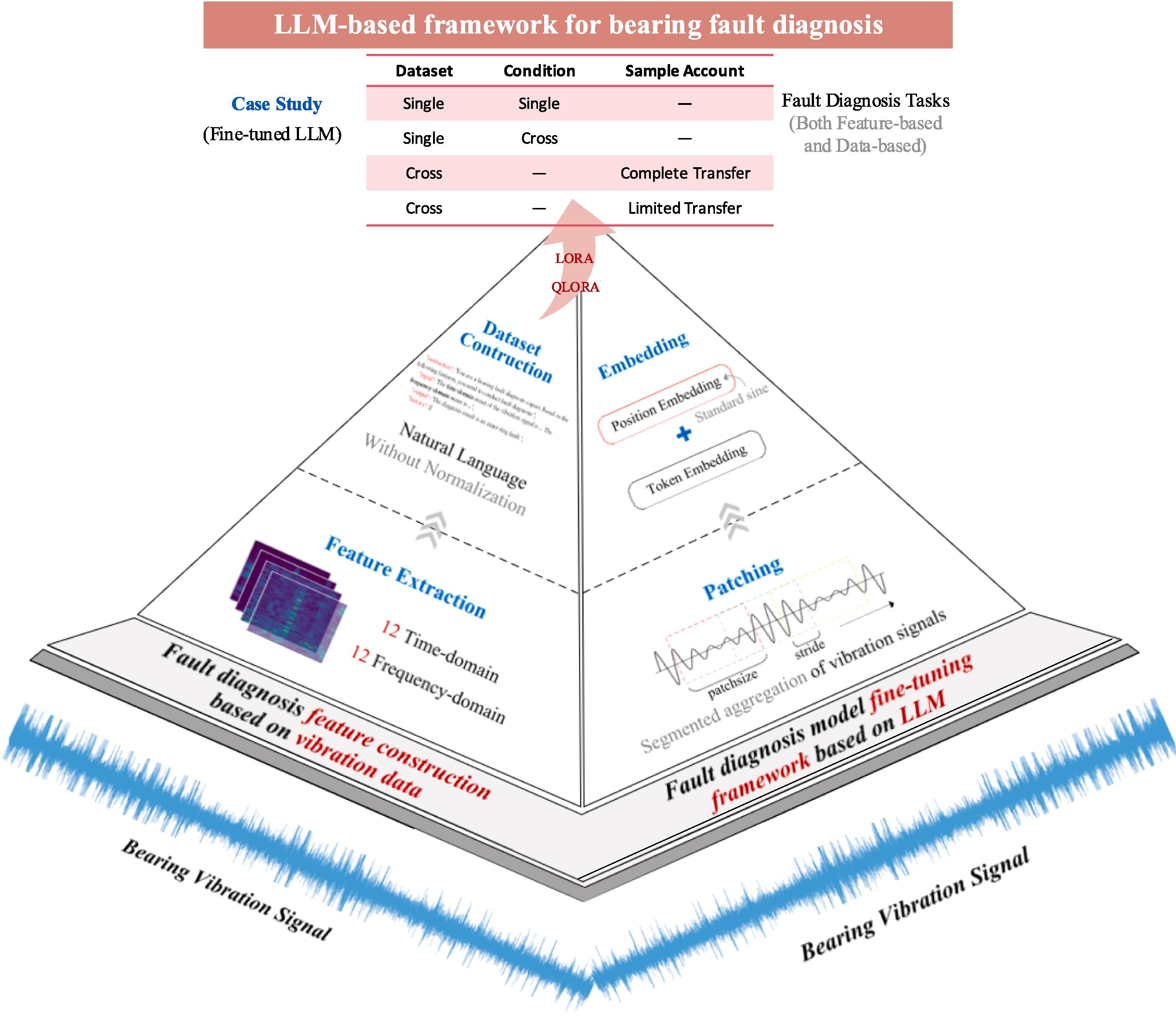}}
    \caption{LLMs for bearing fault diagnosis \cite{TAO2025112127}. The core idea for applying LLMs in bearing fault diagnosis is to pair vibration data and text descriptions and fine-tune LLMs for direct fault diagnosis.}
    \label{fig7}
\end{figure}

The basic idea behind applying LLM-based models in FD\&AD tasks is that the Transformer architecture in LLMs integrates a large number of model parameters that are pre-trained on large text-image datasets. Then the pre-trained LLMs can be adapted to bearing or structural FD\&AD tasks by fine-tuning the customized dataset. As shown in Fig.~\ref{fig7}, the authors conducted fault diagnosis by using Low-Rank Adaptation (LoRA) fine-tuned GPT-2 model based on bearing vibration datasets. In Table \ref{table_5}, we summarize the LLM-based FD\&AD tasks presented for the wind turbine system \cite{ZHANG2025124378}, and the bearing/gearbox \cite{TAO2025112127, PANG2024110312, qaid2024fdllmlargelanguagemodel}. Most of the publications for LLM-based FD\&AD tasks focused on bearing/gearbox and were also based on LoRA fine-tuning techniques. We also detailed the different types of fine-tuning techniques in the following Section \ref{section4.1.1}. 

It is noted that the fine-tuning efficiency and model size of LLMs largely determine their final performance in FD\&AD tasks. Zheng et al.~\cite{ZHENG2024110382} compared the diagnostic accuracy of Llama-2 models of different sizes (7B vs 13B), and reported that the 7B model can achieve satisfactory results, while larger models are generally preferred for tasks requiring higher precision or involving more complex datasets. Additional studies discussing the impact of fine-tuning strategies and model scale on LLM performance for FD\&AD can be found in Refs.~\cite{chen2025large, j2024finetuningllmenterprise, 10774330, LIN2025103208}.

\subsection{Applications in industrial systems}
\label{Section3.2}
\begin{figure}[H]
    \centering
    \makebox[1\textwidth][c]{
    \includegraphics[width=1.2\linewidth]{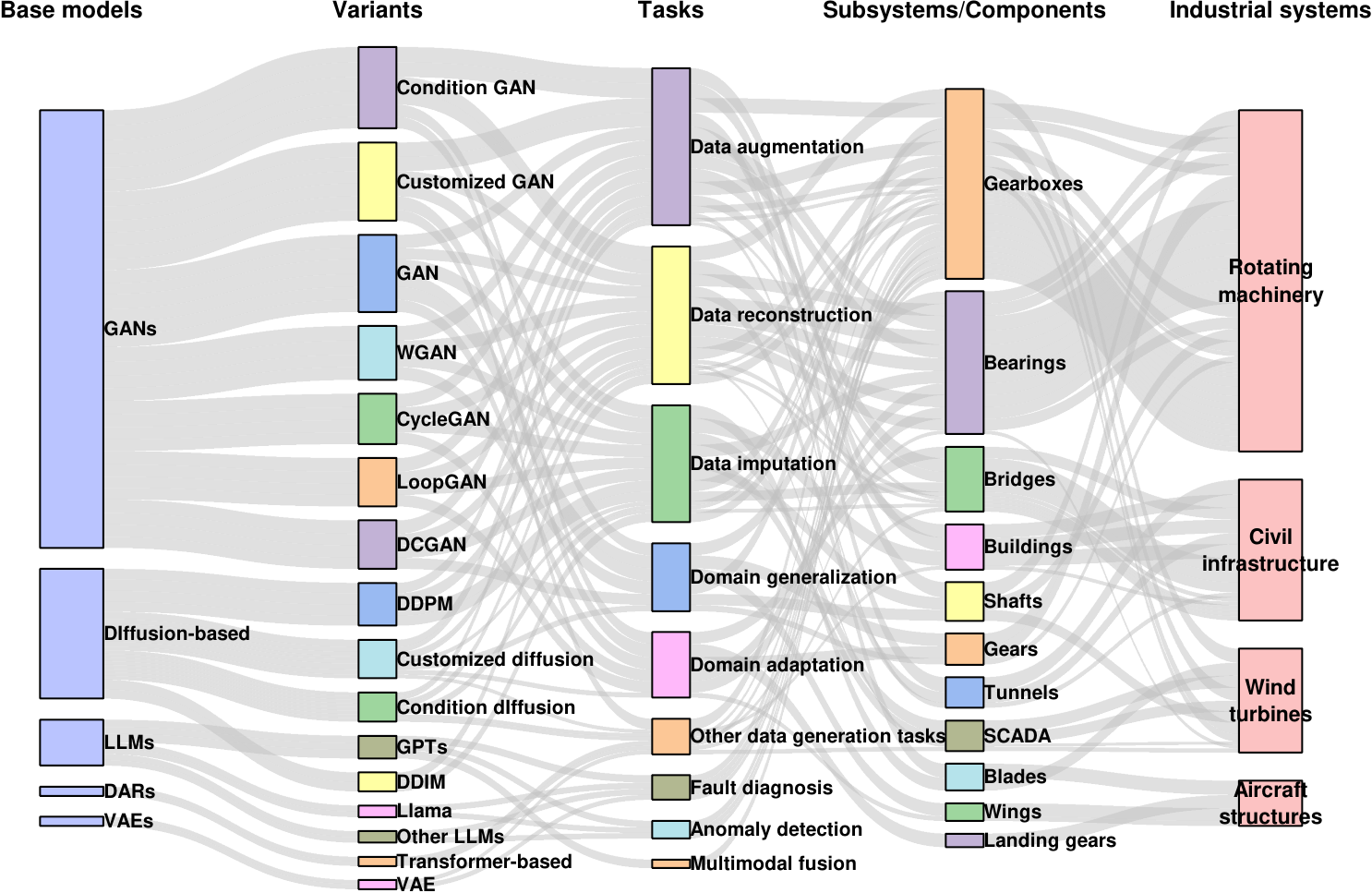}}
    \caption{The Sankey diagram illustrates how the presented DGMs perform various tasks and how they are applied in industrial CM/SHM.}
    \label{figSankey}
\end{figure}

In the previous subsection, we summarize the main tasks performed by DGMs. To illustrate how these tasks transfer into practical use, this section describes DGM applications across subsystems and components in four major industrial sectors: rotating machinery, aircraft structures, wind turbines, and civil infrastructure. Fig.~\ref{figSankey} maps base DGM families and their variants to the tasks they perform and the subsystems they enable in CM/SHM. From the figure, it can be observed that GAN and its variants dominate in data generation task and most of tasks are applied to rotating machinery and civil infrastructure sectors. The diffusion-based models and LLMs remain in the early stages of industrial practice, as they require substantial computational resources and large datasets.

\subsubsection{Rotating machinery}
\label{Sec3.2.1}
Rotating machinery constitutes a fundamental and critical component of numerous mechanical systems, primarily serving to generate, transmit, or regulate rotational motion and torque \cite{LEI2013108}. A representative example is the gearbox, which is designed to adjust rotational speed and output power to specified levels. Within the gearbox, gears transmit torque through precise meshing, while bearings support power transmission by reducing friction and stabilizing the rotating shafts. These components, however, are highly susceptible to unexpected failures: bearings are likely to experience inner race, outer race, or rolling element faults \cite{WANG2025110760}, whereas gears are prone to surface fatigue (pitting, case crushing and spalling) and even tooth breakage \cite{SINGHRAGHAV2023104020}. Continuous monitoring is therefore essential for timely fault detection and reliable failure prediction.

As shown in Table~\ref{table_6}, DGMs have been widely applied to vibration signals, supporting a variety of tasks discussed in Section~\ref{Section3.1}. At the component or subsystem level, gearboxes and bearings are the most representative components in rotating machinery. GAN and diffusion-based models are commonly used for data generation, domain adaptation, and generalization tasks. Recently, LLM-based approaches for multimodal fusion and fault diagnosis have emerged as a growing trend. The text prompting capability enabled by LLMs allows the recording and interpretation of metadata, including operational logs and state transitions, which cannot be directly processed by other DGMs. This capability provides new opportunities for knowledge-enhanced condition monitoring.

\subsubsection{Aircraft structures}
Aircraft SHM focuses on maintaining the structural integrity of load-bearing elements such as wings, fuselage panels, and landing gear structures. The composite wings and fuselage structures in commercial aircraft are subject to extreme environmental conditions from sub-zero temperatures at cruise altitudes to intense solar heating and thermal cycling on the ground. Continuous monitoring and maintenance in such an environment is exceedingly difficult. Typical sensing approaches primarily rely on NDT techniques, including ultrasonic sensing and visual inspection, to detect potential damage such as delamination or fiber breakage. DGMs assist in aircraft SHM by generating realistic signals and features for scarce experimental dataset as shown in Table~\ref{table_6}, enabling cross-modal data synthesis, and producing conditioned damage progression data for downstream diagnostic and prognostic purposes \cite{YANG2025112996}.

It should be noted that engine health monitoring is an extensive topic treated under energy-based studies and will not be discussed in this work.

\subsubsection{Wind turbines}
Wind turbines represent a crucial source of renewable energy. In their operational state monitoring, Supervisory Control and Data Acquisition (SCADA) systems continuously record a wide range of parameters in real time, including wind speed, gear oil temperature, lubricant pressure, generator output power, and rotor speed. At the system level, SCADA data consists of comprehensive state monitoring parameters, aligning closely with the CM community. It should be noted that the rotating components such as the gearbox in wind turbines are classified as DGM-based applications in rotating machinery as discussed in Section \ref{Sec3.2.1}.

At the component level, blade inspection constitutes a key focus area within the SHM community. The turbine blades are constantly subjected to environmental stress from wind, rain, sand, and other particles, leaving them extremely susceptible to wear and tear \cite{KHAZAEE20221568}. Due to these harsh conditions, detecting blade damage using vibration or vision-based inspection becomes a critical task in wind turbine health monitoring.

\subsubsection{Civil infrastructures}
Civil structure monitoring addresses the safety, serviceability, and longevity of large-scale infrastructure systems such as bridges, tunnels, structures, dams and etc. At the system level, SHM emphasizes long-term sensing and decision-making to manage aging, extreme loadings, and changing operational demands.

At the component level, subsystems of interest include bridges, tunnels, dams, and buildings. Common modalities span point sensors (strain gauges, accelerometers), distributed fiber optic sensing systems, NDT techniques (guided waves, acoustic emission), and remote imaging (UAV photography, LiDAR). DGMs complement these capabilities by enabling multimodal fusion, robust data imputation, synthetic generation of damage scenarios (such as cracks, corrosion), conditional/physics-aware simulation for digital twin calibration, domain adaptation across sites and sensor layouts, virtual sensing to estimate unmeasured fields, etc. Although UAV-based inspections are increasingly important for dams and large-span structures, vibration and strain-based data generation tasks remain prevalent across civil applications. The mapped DGM tasks and component-level applications are summarized in Table~\ref{table_6}.

\clearpage
\begin{landscape}
% \color{red}
\footnotesize
\renewcommand{\arraystretch}{0.6}

\begin{longtable}{@{\extracolsep{\fill}} 
>{\centering\arraybackslash}m{1.5cm} m{2cm} m{3cm} m{3cm} m{3cm} m{8cm}}

\caption{Summary table for DGMs in industrial systems}
\label{table_6} \\

\toprule
Industrial sectors & Subsystems  &  DGMs & Data types & Specific tasks & Strengths and limits \\
\midrule\midrule[.1em]
\endfirsthead

\caption[]{Summary table for DGMs in industrial systems (continued)} \\
\toprule
Industrial sectors & Subsystems & DGMs & Data types & Specific tasks & Strength and limits \\
\midrule
\endhead

\bottomrule
\endfoot

\bottomrule
\endlastfoot

 &  & VAE \cite{NarayanaPichika, BOOYSE2020106612, WANG2022234} & Vibration signals; Acoustic signals & Fault diagnosis & \begin{itemize}[leftmargin=*, label={}]
\item[\cmark] Learns compact latent representations for effective fault diagnosis from noisy sensor data
\item[\xmark] Reconstruction-based diagnosis struggles with subtle faults and generalization
\end{itemize} \\

&  & GAN \cite{WANG2025110760, ZHUANG2024111186, WANG2023106872, ZHANG2024114795, LIU2022110888, WAN2025109614, LI2022101552, LI20242958, SHEN2024109159, GUO2025110854, Sun10504387, WEN2024111663, XU2023107063,app13053136, 10018473} & Vibration signals; Acoustic signals; Thermal images & Data generation & \begin{itemize}[leftmargin=*, label={}]
\item[\cmark] Effectively synthesizes realistic fault signals to alleviate data imbalance
\item[\xmark] Training instability and limited physical interpretability, especially with scarce or noisy data
\end{itemize} \\

&  & GAN \cite{Lou9789138, WANG2025110662,Zhuang9759507,GUO2025110854,ir202307, WANG2023455, Sun10504387, 10892083} &Vibration signals; Acoustic signals & Domain adaptation and generalization & \begin{itemize}[leftmargin=*, label={}]
\item[\cmark] Adversarial alignment reduces cross-domain distribution mismatch
\item[\xmark] Severe domain shifts and unstable training can cause overfitting to synthetic domains
\end{itemize}\\

\multirow{1}{1.5cm}{Rotating machinery} & \multirow{1}{2cm}{Gearboxes, Bearings} & Diffusion \cite{MUELLER2024107696, GUO2025115951, ZHAO2025109520, YU2024110343, Wang0309714, HerTerng, YANG2025110312, YI2024111481, CHEN2024110101} & Vibration signals; Acoustic signals & Data generation & \begin{itemize}[leftmargin=*, label={}]
\item[\cmark] Produces high-fidelity, diverse fault samples with stable training, alleviating data imbalance
\item[\xmark] Slow iterative sampling and heavy computation hinder real-time use and large-scale augmentation
\end{itemize}\\

&  & Diffusion \cite{XU2024112396, GUO2025115951} & Vibration signals & Domain adaptation and generalization & \begin{itemize}[leftmargin=*, label={}]
\item[\cmark] Condition-guided diffusion aligns multi-domain distributions under unknown working conditions
\item[\xmark] Domain adaptation requires careful conditioning design and high computation
\end{itemize}\\

&  & LLM \cite{TAO2025112127, PANG2024110312, qaid2024fdllmlargelanguagemodel} & Vibration signals; Text prompt & Multimodal fusion & \begin{itemize}[leftmargin=*, label={}]
\item[\cmark] Multimodal fusion by integrating heterogeneous sensor data, text, and expert knowledge
\item[\xmark] Fusion reliability depends on prompt design and data quality
\end{itemize}\\

&  & LLM \cite{TAO2025112127, PANG2024110312, qaid2024fdllmlargelanguagemodel} & Vibration signals; Text prompt & Fault diagnosis, anomaly detection & \begin{itemize}[leftmargin=*, label={}]
\item[\cmark] Supports knowledge-driven fault diagnosis and anomaly detection under sparse, incomplete data
\item[\xmark] Lacks physical grounding and needs careful fine-tuning for customized dataset
\end{itemize}\\

\cmidrule{2-6}
 & Blades & GAN \cite{LEE2025112301} & Loading data & Data generation & \begin{itemize}[leftmargin=*, label={}]
\item[\cmark] Generates realistic blade fault signal data, mitigating scarcity of aircraft propeller damage samples
\item[\xmark] Training instability and limited physics-awareness may produce unrealistic blade damage patterns
\end{itemize}\\

\multirow{2}{1.5cm}{Aircrafts} & Landing gears & GAN \cite{PAN2025110175} & Landing gear data & Data generation & \begin{itemize}[leftmargin=*, label={}]
\item[\cmark] Generates diverse landing gear load data, alleviating data scarcity
\item[\xmark] GAN samples may violate physical load constraints without explicit physics-informed regularization
\end{itemize}\\

& Wings, skins  & GAN \cite{10577252} & Visual images  & Data generation & \begin{itemize}[leftmargin=*, label={}]
\item[\cmark] Augments wing and skin defect data under limited samples
\item[\xmark] Synthetic data without considering material physics and damage mechanisms
\end{itemize}\\

&   & Diffusion \cite{CAO2025113155, XinY101049} & Ultrasonic signals  & Data generation & \begin{itemize}[leftmargin=*, label={}]
\item[\cmark] Generates high-fidelity composite wing and skin damage data
\item[\xmark] Intensive training and slow sampling hinder scalability for large aircraft structural datasets
\end{itemize}\\

\cmidrule{2-6}
 &  & VAE \cite{Liu_2023, 10153965} &  &  & \begin{itemize}[leftmargin=*, label={}]
\item[\cmark] Generates realistic SCADA data by capturing normal operational patterns
\item[\xmark] Posterior collapse may suppress rare fault features in generated SCADA data
\end{itemize}\\

& SCADA & GAN \cite{QU2020106610, CHATTERJEE2025103991} & SCADA data  & Data generation & \begin{itemize}[leftmargin=*, label={}]
\item[\cmark] Mitigates data imbalance by generating diverse SCADA samples
\item[\xmark] Training instability and mode collapse can distort temporal dependencies in generated SCADA data.
\end{itemize}\\

\multirow{1}{1.5cm}{Wind turbines} &  & Diffusion \cite{YAO2025127925, YAO2024111397} &  &  & \begin{itemize}[leftmargin=*, label={}]
\item[\cmark] Generates high-fidelity SCADA data, preserving temporal correlations and operational variability
\item[\xmark] High computational cost and slow sampling for large-scale SCADA data augmentation
\end{itemize}\\

& Blades & VAE \cite{YANG2024117689} & Vibration signals & Anomaly detection & \begin{itemize}[leftmargin=*, label={}]
\item[\cmark] Unsupervised anomaly detection under limited labeled blade fault data
\item[\xmark] Reconstruction errors may miss subtle damage and degrade under nonstationary conditions
\end{itemize}\\

&  & Diffusion \cite{10697302}& Visual images & Anomaly detection & \begin{itemize}[leftmargin=*, label={}]
\item[\cmark] Unsupervised damage detection by capturing complex blade signal distributions
\item[\xmark] Computation-intensive reconstruction and inference for real-time anomaly detection deployment
\end{itemize}\\

\cmidrule{2-6}
  & Bridges  & DAR \cite{CHEN2023116063} & Strain data & Data generation & \begin{itemize}[leftmargin=*, label={}]
\item[\cmark] Accurately reconstructs long-term bridge strain data by preserving thermal-induced dependencies
\item[\xmark] Limited generative diversity and weak extrapolation under unseen conditions
\end{itemize}\\

 & & VAE \cite{anaissi2023multi} & Vibration signals & Anomaly detection & \begin{itemize}[leftmargin=*, label={}]
\item[\cmark] Models normal bridge behavior for unsupervised anomaly detection under limited damage data
\item[\xmark] Reconstruction-based indicators may miss localized anomalies under strong environmental variability
\end{itemize}\\

 \multirow{1}{1.5cm}{Civil structures} & & GAN \cite{ZHENG2025119694,HOU2022111206, Xiaoming2021, GAO2023105277} &  Vibration signals; Strain data; Visual images & Data generation & \begin{itemize}[leftmargin=*, label={}]
\item[\cmark] Generates realistic bridge sensor data for effective missing data imputation
\item[\xmark] Training instability and mode collapse may distort long-term correlations and damage patterns.
\end{itemize}\\

&  & GAN \cite{GE2024111236} & Vibration signals & Domain adaptation and generalization & \begin{itemize}[leftmargin=*, label={}]
\item[\cmark] Adversarial alignment improves cross-bridge domain adaptation under varying conditions
\item[\xmark] Adaptation performance depends on physics priors and adversarial stability
\end{itemize}\\

 &  & Diffusion \cite{ZHENG2025119685, SHU2025111783, wei2025generation} & Vibration signals  & Data generation & \begin{itemize}[leftmargin=*, label={}]
\item[\cmark] Generates high-fidelity bridge vibration responses by preserving temporal correlations
\item[\xmark] Computationally intensive training and slow sampling for large-scale data generation
\end{itemize}\\

& Tunnels & GAN \cite{LI2025112062} &  Vibration signals  & Data generation & \begin{itemize}[leftmargin=*, label={}]
\item[\cmark] Rapidly generates realistic tunnel vibration data under compressive sensing constraints
\item[\xmark] Adversarial training instability may reduce fidelity for complex tunnel responses
\end{itemize}\\

& Buildings & GAN \cite{asami2022data} & Visual images & Data generation & \begin{itemize}[leftmargin=*, label={}]
\item[\cmark] Generates diverse damaged-building roof images to augment scarce post-disaster datasets
\item[\xmark] Synthetic damage patterns may lack generalization across diverse building types
\end{itemize}\\

&  & GAN \cite{SOLEIMANIBABAKAMALI2023110404, LULECI2023110370, LULECI2023106146} & Vibration signals & Domain adaptation and generalization & \begin{itemize}[leftmargin=*, label={}]
\item[\cmark] Zero-shot and cross-domain translation, improving generalization across building damage states
\item[\xmark] Domain shifts beyond training distributions can cause unrealistic translations
\end{itemize}\\

& Dams & GAN \cite{rs17203398} & Visual images & Anomaly detection & \begin{itemize}[leftmargin=*, label={}]
\item[\cmark] Anomaly segmentation by using infrared and visual images under complex surface conditions
\item[\xmark] Detection reliability depends on training data diversity
\end{itemize}\\
\bottomrule
\end{longtable}
\end{landscape}
\clearpage

\section{Challenges and future prospects}
\label{Sec:4}
This section summarizes the current limitations of DGMs in CM/SHM in three main aspects, as described in Section~\ref{section4.1}: the need for efficient fine-tuning strategies, interpretable and reliable models, and mitigating the computational inefficiency of edge deployment. The corresponding future prospects will be discussed in Section~\ref{section4.2}.

\subsection{Challenges and limitations}
\label{section4.1}
\subsubsection{Fine-tuning techniques}
\label{section4.1.1}
DGMs are generally trained on large datasets during the initial pre-training stage. To adapt these pre-trained DGMs to specific tasks or domains, fine-tuning techniques are usually adopted by further training on a smaller, custom dataset. By leveraging knowledge from pre-training while learning nuances of the target domain, fine-tuning bridges the performance gap between the source and target domains. This also indicates that fine-tuning is among the most widely used strategies for applying DGMs in transfer learning and domain adaptation across diverse applications. Fine-tuning techniques for DGMs consist primarily of two branches: 1) Full fine-tuning, and 2) parameter-efficient fine-tuning (PEFT) \cite{peft}. Full fine-tuning updates all parameters of the pre-trained model while preserving its original architecture, and the size of the model after fine-tuning remains the same as the pre-trained model. However, this method requires a large storage space for DGMs and is also not achievable for deploying the fine-tuned models for downstream applications. While PEFT methods selectively update a small subset of parameters, drastically reducing computational and storage overheads \cite{houlsby2019parameterefficient}. PEFT methods are further categorized into:

\begin{itemize}
    \item Prefix/Prompt-Tuning \cite{li2021prefixtuning}: Additional trainable prefix tokens added to the input or hidden layer (these prefixes are continuous pseudo tokens that do not correspond to real tokens), and only these prefix parameters are trained.
    \item Adapter-Tuning \cite{hu2023llmadapters}: Insert smaller neural network layers or modules into each layer of the pre-trained model. These newly inserted neural modules are called adapters. When fine-tuning downstream tasks, only these adapter parameters are trained.
    \item Low-Rank Adaptation \cite{hu2021lora}: Learning a low-rank matrix with small parameters to approximate the parameter update of the model weight matrix, only the low-rank matrix parameters are optimized during training.
\end{itemize}

\begin{figure}[H]
    \centering
    \makebox[1\textwidth][c]{
    \includegraphics[width=0.8\linewidth]{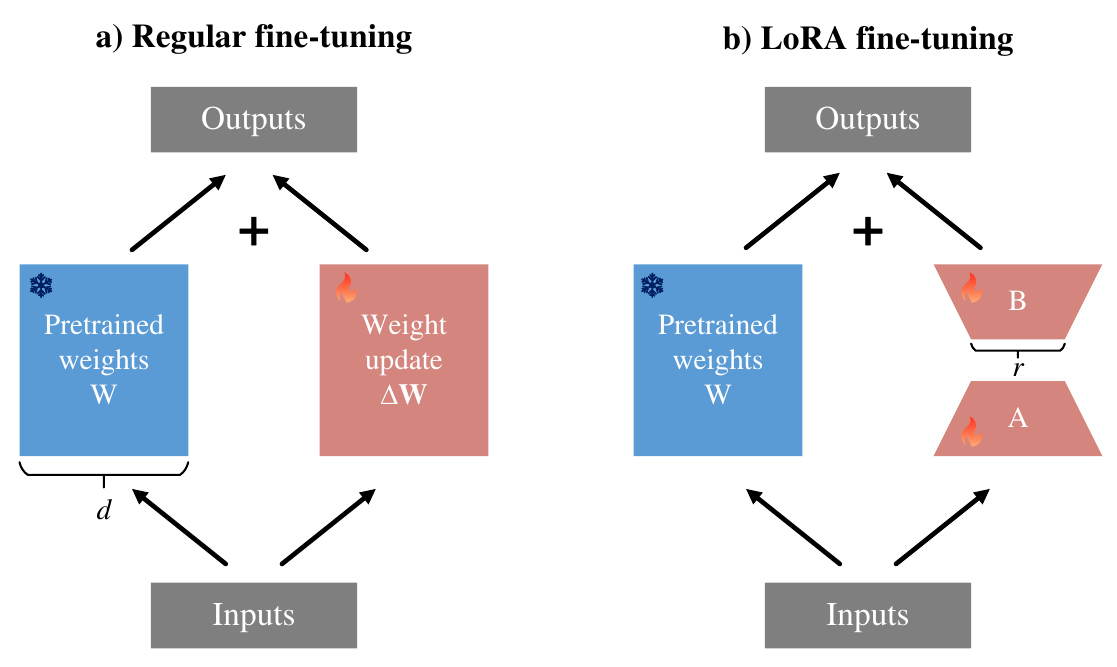}}
    \caption{LoRA fine-tuning techniques. Comparing with regular fine-tuning updating pre-trained weights $\text{W} \in \mathbb{R}^{d \times d}$, LoRA is performed by replacing the weight update matrix $\Delta \text{W}$ with two smaller matrices ($A \in \mathbb{R}^{d \times r}$ and $B \in \mathbb{R}^{r \times d}$, where $r \ll d$) through low-rank decomposition, which greatly reduces the number of trainable parameters.}
    \label{fig8}
\end{figure}

As illustrated in Fig.~\ref{fig8}, LoRA operates by decomposing model updates into low-rank matrices, enabling efficient adaptation without altering the original parameter count. 

In industrial applications, efficient fine-tuning enables pre-trained DGMs to generate domain-specific sensor data (such as vibration accelerations, infrared images, and acoustic emission signals) while maintaining their foundational understanding of complex patterns \cite{YANG2025112996}. For instance, a DGM pre-trained on diverse mechanical fault datasets can be fine-tuned with LoRA or adapters based on the Hugging Face PEFT package \cite{peft} using only a small number of labeled data from a particular wind turbine gearbox. This enables the model to detect subtle anomalies on-site without requiring a complete retraining of the model from scratch, which is a promising direction for real-time condition monitoring and possible deployment in industry.

\subsubsection{Explanability, interpretability, and trustworthiness}
Explainable models pose another obstacle in deploying DGMs for industrial generative tasks. Since DGMs combine generative modeling with DNNs, which are often characterized by the notion of black-box models, it is important to explain the generated results and interpret internal DGM mechanisms, which are collectively termed as Explainable Generative AI (GenXAI) \cite{XU2025102721}. Note that \textit{explainability} and \textit{interpretability} represent distinct but related concepts, where the former tries to explain and illustrate model decisions or outputs in human-understandable terms \cite{zhao2023explainability}, whereas the latter emphasizes understanding the internal processes governing model behavior \cite{RETZLAFF2024101243}. In contrast to traditional DNN models, the scale of modern DGMs, particularly LLMs, with billions of parameters and vast training datasets introduces huge challenges for GenXAI research.

\begin{figure}[H]
    \centering
    \makebox[1\textwidth][c]{
    \includegraphics[width=0.8\linewidth]{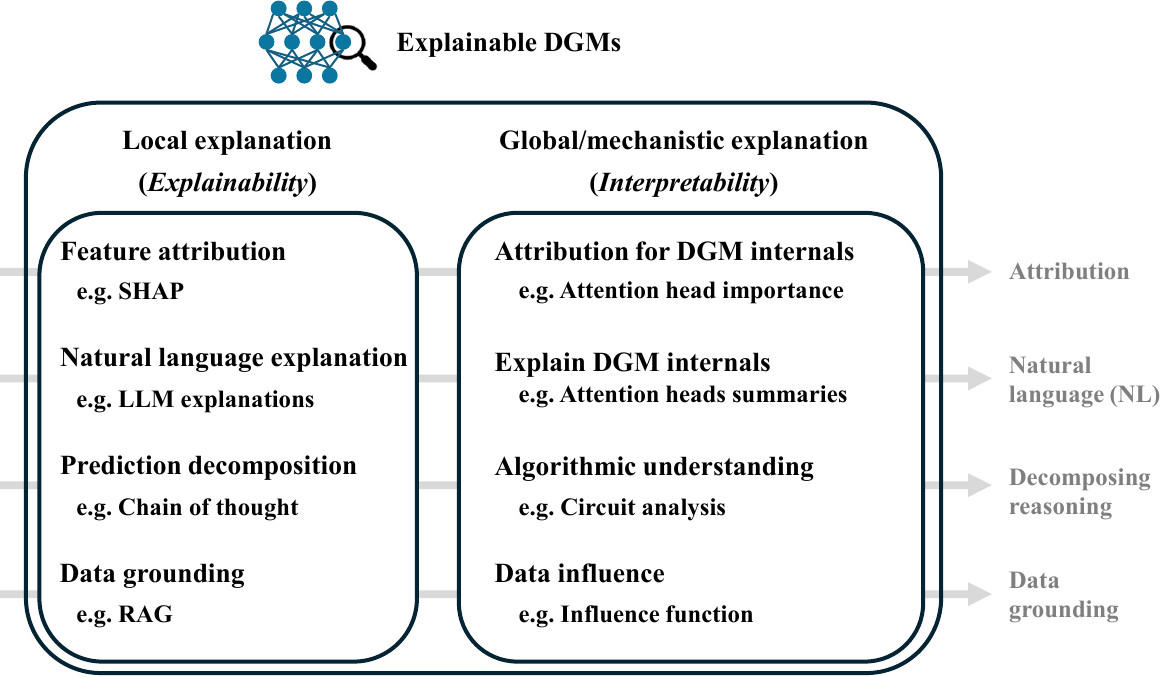}}
    \caption{Explainable DGMs can be classified into two categories: local explanation and global/mechanistic explanation. Local explanation refers to explaining the generated content or outputs from the DGM, and global/mechanistic explanation focuses on interpreting the DGM itself. The figure was adapted from Ref.~\cite{singh2024rethinking}.}
    \label{fig9}
\end{figure}

To date, different explainable skills have been developed in the realm of GenXAI, which can be primarily categorized into two types of methods: local explanation (seeks to explain the result) and global/mechanistic explanation (interprets its internal mechanism) \cite{singh2024rethinking}. Local approaches often adopt techniques from traditional DNN explainability, such as feature attribution methods (in this case, SHapley Additive exPlanations (SHAP) \cite{NIPS2017_7062}). More recent local explanation techniques use LLMs themselves to yield explanations, e.g., through post-hoc natural language explanations, asking an LLM to build explanations into its generation process, or through data grounding (e.g., Retrieval-Augmented Generation (RAG) \cite{moran2025interpretableDGMs}). As an alternative, the second direction is for global or mechanistic explanation, which also means to interpret the entire model. Similar techniques have also been developed, such as attribution of DGM internals, interpretation of model internals, algorithmic understanding, and data influence \cite{schneider2024GenXAI}. To conclude the aforementioned content, we elaborate on the difference between local and global explanation as shown in Fig.~\ref{fig9} from four well-built facets: attribution, natural language, decomposing reasoning, and data grounding.

Trustworthiness remains a critical concern, particularly for LLMs. While these models are good at generating accurate predictions by leveraging statistical patterns and semantic patterns from training data, their reliability is inherently tied to the limitations of DNN models \cite{liu2024trustworthyllms}. In LLMs, the models may get a good handle on grammar, syntax, and semantic relationships in language usage, but they may not necessarily map the words to real-world entities as humans do, which is crucial for ensuring the credibility and trustworthiness of the generated results. Instead, current LLMs can be prone to hallucinations, which result in the generation of content that deviates from facts or user input \cite{huang2024trustllm}. This discrepancy could pose a serious obstacle to practical implementation and cast doubts about the reliability of their diagnostic results.

\subsubsection{Quantization and deployment}
\begin{figure}[H]
    \centering
    \makebox[1\textwidth][c]{
    \includegraphics[width=1\linewidth]{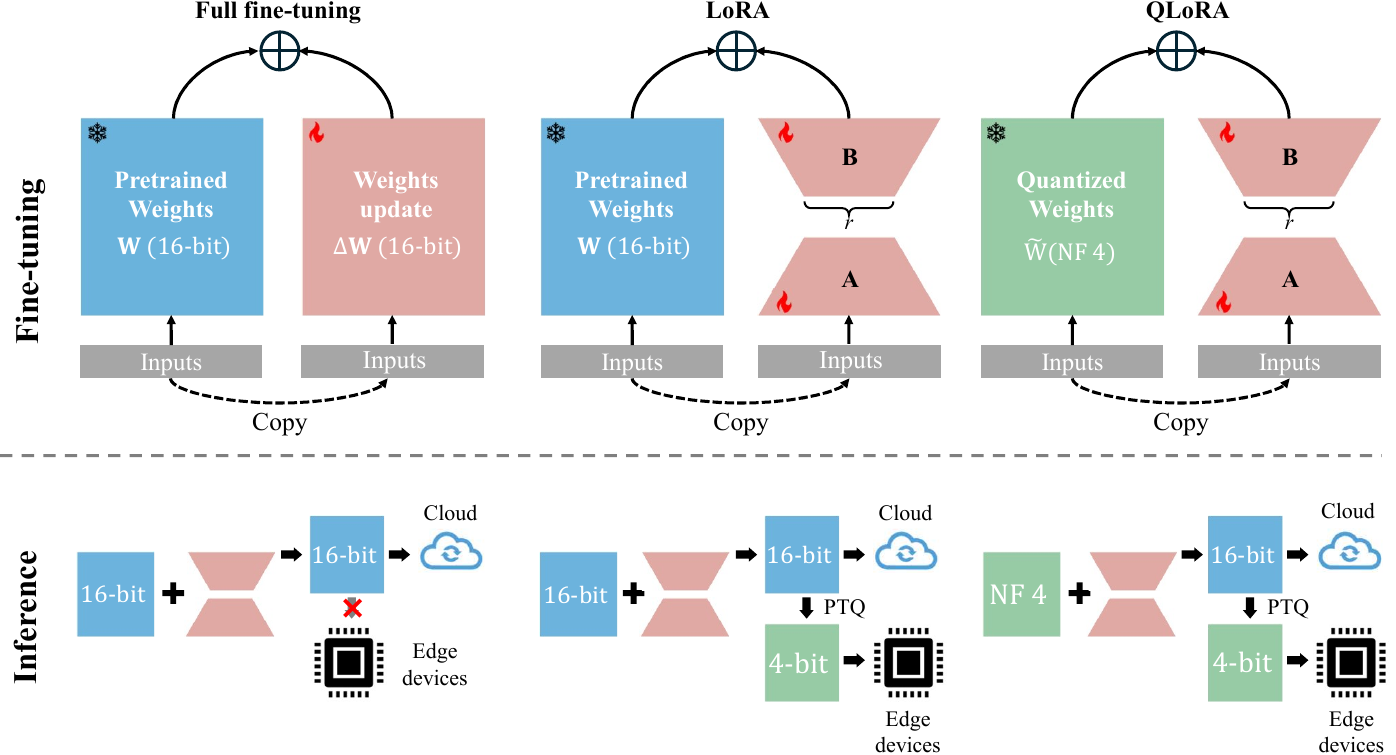}}
    \caption{An illustration of DGM quantization and deployment. Fully fine-tuned DGMs can be inferred in cloud environments but are not achievable on edge devices. LoRA fine-tuned DGMs, while deployable in the cloud using 16-bit floating-point precision, require quantization to 4-bit integers to reduce computational overhead for edge device compatibility. QLoRA addresses this limitation by quantizing the base pre-trained model to 4-bit NormalFloat (NF4), a data type that surpasses both 4-bit integers and 4-bit floats in accuracy while significantly minimizing memory consumption.}
    \label{fig10}
\end{figure}

Deployment involves transferring the algorithm from a development environment into a real-world operational setting where it can start monitoring equipment. Model quantization is an important step before deploying large DGMs on edge devices (e.g., mobile devices on-site) to enable real-time monitoring. The key issue is to eliminate the heavy computational burden caused by the large number of parameters of DGMs, in particular LLMs \cite{xu2023qalora}. At the very beginning, the fine-tuned DGMs need to be quantized so that 16-bit floating-point models are converted into a compact low-bit (4-bit or 8-bit) fixed-point format. This transformation needs to achieve a balance between the efficiency of fixed-point computation and the benefits of reduced memory usage. Other notable techniques include model pruning and knowledge distillation, both of which aim to reduce network size while preserving as much of the original model performance as possible \cite{li2024genqquantizationlowdata}. The main steps involved in this process include: 1) Efficient fine-tuning of DGMs on the custom dataset to adapt to specific application scenarios; 2) Performing post-training quantization (PTQ) after fine-tuning, and the DGMs are quantized to lower accuracy, that is, a 4/8-bit fixed-point model; 3) Suitable model pruning to deploy the fine-tuned model on edge devices. Quantized LoRA (QLoRA) \cite{dettmers2023qlora} is an efficient quantization method for LLM-based LoRA fine-tuning, as shown in Fig.~\ref{fig10}. Following these steps, the pre-trained large DGMs can be applied to commercial-grade chips so that their generative ability can be used in real-world condition monitoring scenarios.

\subsection{Future outlooks}
\label{section4.2}
\subsubsection{Zero-shot/few-shot/multimodal data generation} 
DGMs, particularly LLMs, are revolutionizing zero-shot, few-shot, and multimodal data generation paradigms by enabling robust and diverse generalization in data-scarce scenarios. Zero-shot learning \cite{wang2022generalizing} focuses on diagnosing or predicting unseen samples, which seamlessly refers to domain generalization in Section \ref{Section3.1.3}. Few-shot learning, on the other hand, aims to achieve high performance with only a small number of training samples, often employing meta-learning techniques \cite{hospedales2020metalearning, vanschoren2018metalearningsurvey}, which is a broader approach where models learn to learn, enabling them to adapt to new tasks quickly with minimal training data. It can be foreseen that DGMs can be pre-trained on large text-sensor data pairs and then can be exploited for zero-shot or few-shot reasoning for either generative tasks or diagnostic tasks. Multimodal generation will also be greatly enhanced with DGMs, especially LLMs \cite{he2024llmsmeetmultimodalgeneration}. It is interesting to note that in Section \ref{section3.1.4}, the multimodal data fusion aims for multimodal \textit{input} data, which herein DGMs similarly enable effective generation of multimodal \textit{output} data. In industrial systems, DGMs can generate diverse data streams, such as vibration spectra, thermal imaging, acoustic emissions, and even operational logs, which collectively indicate equipment health.
    
\subsubsection{Physics-informed DGMs}
Physics-informed neural networks (PINNs) \cite{raissi2017PINN} integrated with DGMs represent a promising direction for next-generation condition monitoring systems. By embedding the governing equations of the monitored system as regularization terms or loss functions, PINNs regulate the generated data to align with \textit{in-situ} physical dynamics. In particular, combining physics-aware principles or rules with LLMs creates new opportunities to simulate industrial phenomena, such as variable loading conditions, thermal gradients, and mechanical stresses. To date, no published research has explored the application of physics-informed LLMs for industrial condition monitoring. Future work might focus on hybrid architectures in which PINNs guide LLMs to generate physically consistent degradation trajectories (e.g., bearing wear or crack propagation) for robust predictive maintenance. Additionally, enhancing physics-informed latent spaces by directly linking generative features to measurable physical parameters (e.g., strain, temperature) would strengthen the interpretability and reliability of generated data.

\subsubsection{Reinforcement learning with DGMs}
Reinforcement learning (RL) has proved highly effective in computer vision for autonomous driving and robotics. The application of RL for real-time monitoring is still in its initial stages. RL is known for its autonomous and adaptive learning features, and has been instrumental in the training and optimization of large-scale models as well as in solving goal-directed decision-making problems \cite{LI2025110570}. However, conventional RL struggles with sparse rewards, high-dimensional state spaces, and safety-critical constraints \cite{OGUNFOWORA2023244}, but DGMs can address these challenges by modeling system dynamics, simulating environments, and generating synthetic failure samples. For instance, LLMs can learn latent representations of equipment degradation patterns from historical sensor data, allowing RL agents to train in simulated environments that capture rare fault modes or extreme operational conditions. This reduces reliance on real-world data while improving robustness to unseen anomalies. By integrating LLMs with RL agents, systems can achieve policy optimization through the way in which LLMs predict future states under different maintenance actions, enabling agents to explore strategies safely and efficiently. For partial observability issues, DGMs reconstruct missing sensor signals or infer hidden system states, enhancing RL decision-making under uncertainty. Future work may focus on hierarchical RL frameworks to coordinate long-term maintenance with short-term anomaly detection.

\section{Conclusions}
\label{Sec:5}
This work systematically reviews recent advances and emerging trends in deep generative models (DGMs) for industrial condition and structural health monitoring systems. We examine state-of-the-art DGMs, including deep autoregressive models, variational autoencoders, generative adversarial networks (GANs), diffusion-based models, and large language models (LLMs), and categorize their applications into four critical areas: (i) data generation, (ii) domain adaptation and generalization, (iii) multimodal data fusion, and (iv) fault diagnosis and anomaly detection (FD\&AD) tasks. The review addresses key challenges across diverse industrial domains, such as rotating machinery, aircraft structures, wind turbines, and civil infrastructure. Challenges still exist, including the need for efficient fine-tuning strategies, enhanced model interpretability, and optimization techniques such as quantization for edge device implementation. In short, the following main conclusions can be drawn from this review.
\begin{itemize}
    \item GAN-based approaches dominate in data generation task, whereas diffusion-based models and LLMs remain in the early stages of industrial practice.
    \item DGMs for domain generalization need to be further explored, especially under unseen scenarios.
    \item Multimodal data fusion in LLM can be sensibly integrated with real-time monitoring data in the condition monitoring process.
    \item For FD\&AD tasks, the LLM can fuse different data streams such as vibration, acoustics, and even operational logging texts, but effective fine-tuning such as LoRA still needs to be in place.
    \item For future perspectives, DGMs are still in their infancy for industrial condition and structural health monitoring; this review provides academic and industrial insights for both research and practical implementation.
\end{itemize}

\clearpage
\bibliography{Refs}

@article{MUELLER2024107696,
    title = {Attention-enhanced conditional-diffusion-based data synthesis for data augmentation in machine fault diagnosis},
    journal = {Engineering Applications of Artificial Intelligence},
    volume = {131},
    pages = {107696},
    year = {2024},
    issn = {0952-1976},
    doi = {https://doi.org/10.1016/j.engappai.2023.107696},
    author = {Philipp N. Mueller},
}

@article{WEN2024107562,
    title = {A new unsupervised health index estimation method for bearings early fault detection based on Gaussian mixture model},
    journal = {Engineering Applications of Artificial Intelligence},
    volume = {128},
    pages = {107562},
    year = {2024},
    issn = {0952-1976},
    doi = {https://doi.org/10.1016/j.engappai.2023.107562},
    author = {Long Wen and Guang Yang and Longxin Hu and Chunsheng Yang and Ke Feng},
}

@article{KHAZAEE20221568,
    title = {A comprehensive study on Structural Health Monitoring (SHM) of wind turbine blades by instrumenting tower using machine learning methods},
    journal = {Renewable Energy},
    volume = {199},
    pages = {1568-1579},
    year = {2022},
    issn = {0960-1481},
    doi = {https://doi.org/10.1016/j.renene.2022.09.032},
    author = {Meghdad Khazaee and Pierre Derian and Anthony Mouraud},
}

@inproceedings{wen2024gear,
    title={A gear health indicator based on f-anogan},
    author={Wen, Hao and Van Maele, Djordy and Poletto, Jean Carlos and De Baets, Patrick and Gryllias, Konstantinos},
    booktitle={PHM Society European Conference},
    volume={8},
    doi={https://doi.org/10.36001/phme.2024.v8i1.4046},
    pages={12--12},
    year={2024}
}

@misc{zhao2023explainability,
    title={Explainability for Large Language Models: A Survey}, 
    author={Haiyan Zhao and Hanjie Chen and Fan Yang and Ninghao Liu and Huiqi Deng and Hengyi Cai and Shuaiqiang Wang and Dawei Yin and Mengnan Du},
    year={2023},
    eprint={2309.01029},
    archivePrefix={arXiv},
    primaryClass={cs.CL},
    url={https://arxiv.org/abs/2309.01029}, 
}

@incollection{NIPS2017_7062,
    title = {A Unified Approach to Interpreting Model Predictions},
    author = {Lundberg, Scott M and Lee, Su-In},
    booktitle = {Advances in Neural Information Processing Systems 30},
    editor = {I. Guyon and U. V. Luxburg and S. Bengio and H. Wallach and R. Fergus and S. Vishwanathan and R. Garnett},
    pages = {4765--4774},
    year = {2017},
    publisher = {Curran Associates, Inc.},
    url = {http://papers.nips.cc/paper/7062-a-unified-approach-to-interpreting-model-predictions.pdf}
}

@misc{vanschoren2018metalearningsurvey,
    title={Meta-Learning: A Survey}, 
    author={Joaquin Vanschoren},
    year={2018},
    eprint={1810.03548},
    archivePrefix={arXiv},
    primaryClass={cs.LG},
    url={https://arxiv.org/abs/1810.03548}, 
}

@misc{hospedales2020metalearning,
    title={Meta-Learning in Neural Networks: A Survey}, 
    author={Timothy Hospedales and Antreas Antoniou and Paul Micaelli and Amos Storkey},
    year={2020},
    eprint={2004.05439},
    archivePrefix={arXiv},
    primaryClass={cs.LG},
    url={https://arxiv.org/abs/2004.05439}, 
}

@article{YU2024110343,
    title = {ReF-DDPM: A novel DDPM-based data augmentation method for imbalanced rolling bearing fault diagnosis},
    journal = {Reliability Engineering \& System Safety},
    volume = {251},
    pages = {110343},
    year = {2024},
    issn = {0951-8320},
    doi = {https://doi.org/10.1016/j.ress.2024.110343},
    author = {Tian Yu and Chaoshun Li and Jie Huang and Xiangqu Xiao and Xiaoyuan Zhang and Yuhong Li and Bitao Fu},
}

@inproceedings{NIPS2014_f033ed80,
    author = {Goodfellow, Ian J. and Pouget-Abadie, Jean and Mirza, Mehdi and Xu, Bing and Warde-Farley, David and Ozair, Sherjil and Courville, Aaron and Bengio, Yoshua},
    booktitle = {Advances in Neural Information Processing Systems},
    editor = {Z. Ghahramani and M. Welling and C. Cortes and N. Lawrence and K.Q. Weinberger},
    publisher = {Curran Associates, Inc.},
    title = {Generative Adversarial Nets},
    volume = {27},
    year = {2014}
}

@article{NUNES202353,
    title = {Challenges in predictive maintenance – {A} review},
    journal = {CIRP Journal of Manufacturing Science and Technology},
    volume = {40},
    pages = {53-67},
    year = {2023},
    issn = {1755-5817},
    doi = {https://doi.org/10.1016/j.cirpj.2022.11.004},
    author = {P. Nunes and J. Santos and E. Rocha},
}

@article{GE2024111236,
    title = {Domain adaptation for structural health monitoring via physics-informed and self-attention-enhanced generative adversarial learning},
    journal = {Mechanical Systems and Signal Processing},
    volume = {211},
    pages = {111236},
    year = {2024},
    issn = {0888-3270},
    doi = {https://doi.org/10.1016/j.ymssp.2024.111236},
    author = {Liangfu Ge and Ayan Sadhu},
}

@article{chen2023deep,
    title={Deep convolutional transfer learning-based structural damage detection with domain adaptation},
    author={Chen, Zuoyi and Wang, Chao and Wu, Jun and Deng, Chao and Wang, Yuanhang},
    journal={Applied intelligence},
    volume={53},
    number={5},
    pages={5085--5099},
    year={2023},
    publisher={Springer}
}

@article{gardner2022domain,
    title={Domain-adapted Gaussian mixture models for population-based structural health monitoring},
    author={Gardner, Paul and Bull, Lawrence A and Dervilis, Nikolaos and Worden, Keith},
    journal={Journal of Civil Structural Health Monitoring},
    volume={12},
    number={6},
    pages={1343--1353},
    year={2022},
    publisher={Springer}
}

@article{GARDNER2020106550,
    title = {On the application of domain adaptation in structural health monitoring},
    journal = {Mechanical Systems and Signal Processing},
    volume = {138},
    pages = {106550},
    year = {2020},
    issn = {0888-3270},
    doi = {https://doi.org/10.1016/j.ymssp.2019.106550},
    author = {P. Gardner and X. Liu and K. Worden},
}

@article{ZHENG2025119685,
    title = {Data augmentation of dynamic responses for structural health monitoring using denoising diffusion probabilistic models},
    journal = {Engineering Structures},
    volume = {328},
    pages = {119685},
    year = {2025},
    issn = {0141-0296},
    doi = {https://doi.org/10.1016/j.engstruct.2025.119685},
    author = {Wenhao Zheng and Jun Li and Hong Hao},
}

@article{ZHENG2025119694,
    title = {Missing data imputation for structural health monitoring using unsupervised domain adaptation and pretraining techniques},
    journal = {Engineering Structures},
    volume = {328},
    pages = {119694},
    year = {2025},
    issn = {0141-0296},
    doi = {https://doi.org/10.1016/j.engstruct.2025.119694},
    author = {Wenhao Zheng and Jun Li and Hong Hao and Gao Fan},
}

@article{song2024missing,
    title={Missing data imputation model for dam health monitoring based on mode decomposition and deep learning},
    author={Song, Jintao and Yang, Zhaodi and Li, Xinru},
    journal={Journal of Civil Structural Health Monitoring},
    volume={14},
    number={5},
    pages={1111--1124},
    year={2024},
    publisher={Springer}
}

@article{tan2024missing,
    title={Missing data imputation in tunnel monitoring with a spatio-temporal correlation fused machine learning model},
    author={Tan, Xuyan and Chen, Weizhong and Tan, Xianjun and Fan, Chengkai and Mao, Yuhao and Cheng, Ke and Du, Bowen},
    journal={Journal of Civil Structural Health Monitoring},
    pages={1--12},
    year={2024},
    publisher={Springer}
}

@inproceedings{Ahang_2024,
    title={Condition Monitoring with Incomplete Data: An Integrated Variational Autoencoder and Distance Metric Framework},
    url={http://dx.doi.org/10.1109/CASE59546.2024.10711553},
    DOI={10.1109/case59546.2024.10711553},
    booktitle={2024 IEEE 20th International Conference on Automation Science and Engineering (CASE)},
    publisher={IEEE},
    author={Ahang, Maryam and Abbasi, Mostafa and Charter, Todd and Najjaran, Homayoun},
    year={2024},
    month=aug, pages={3213–3218} 
}

@article{WANG2023110703,
    title = {A combined method of autoregressive model and matrix factorization for recovery and forecasting of cyclic structural health monitoring data},
    journal = {Mechanical Systems and Signal Processing},
    volume = {202},
    pages = {110703},
    year = {2023},
    issn = {0888-3270},
    doi = {https://doi.org/10.1016/j.ymssp.2023.110703},
    author = {Chunsheng Wang and Peijie Zhang},
}

@article{ZHANG2022108718,
    title = {Autoregressive matrix factorization for imputation and forecasting of spatiotemporal structural monitoring time series},
    journal = {Mechanical Systems and Signal Processing},
    volume = {169},
    pages = {108718},
    year = {2022},
    issn = {0888-3270},
    doi = {https://doi.org/10.1016/j.ymssp.2021.108718},
    author = {Peijie Zhang and Pu Ren and Yang Liu and Hao Sun},
}

@article{YAO2025127925,
    title = {A novel diffusion model with Shapley value analysis for anomaly detection and identification of wind turbine},
    journal = {Expert Systems with Applications},
    volume = {284},
    pages = {127925},
    year = {2025},
    issn = {0957-4174},
    doi = {https://doi.org/10.1016/j.eswa.2025.127925},
    author = {Qingtao Yao and Bohua Chen and Aijun Hu and Dong Zhen and Ling Xiang},
}

@article{CHATTERJEE2025103991,
    title = {Leveraging generative adversarial networks for data augmentation to improve fault detection in wind turbines with imbalanced data},
    journal = {Results in Engineering},
    volume = {25},
    pages = {103991},
    year = {2025},
    issn = {2590-1230},
    doi = {https://doi.org/10.1016/j.rineng.2025.103991},
    author = {Subhajit Chatterjee and Yung-Cheol Byun},
}

@article{CAO2025113155,
    title = {A conditional diffusion model-based data augmentation method for structural health monitoring of composites},
    journal = {Mechanical Systems and Signal Processing},
    volume = {238},
    pages = {113155},
    year = {2025},
    issn = {0888-3270},
    doi = {https://doi.org/10.1016/j.ymssp.2025.113155},
    author = {Licai Cao and Fei Gao and Tianxiao Zhang and Jin Cui and Anastasios P. Vassilopoulos},
}

@article{LEE2025112301,
    title = {Propeller fault-detection method for electric-propulsion aircraft using motor signals and generative model-based semi-supervised learning},
    journal = {Engineering Applications of Artificial Intelligence},
    volume = {162},
    pages = {112301},
    year = {2025},
    issn = {0952-1976},
    doi = {https://doi.org/10.1016/j.engappai.2025.112301},
    author = {Sanga Lee and Dohyeong Kim and Minkyun Noh and Shinkyu Jeong and Jikang Kong and Youngjun Yoo},
}

@article{PAN2025110175,
    title = {Enhanced feedback analysis of vertical load reliability parameters for airplane landing gear using an improved generative adversarial network and explainable artificial intelligence techniques},
    journal = {Engineering Applications of Artificial Intelligence},
    volume = {145},
    pages = {110175},
    year = {2025},
    issn = {0952-1976},
    doi = {https://doi.org/10.1016/j.engappai.2025.110175},
    author = {Weihuang Pan and Yunwen Feng and Cheng Lu and Jiaqi Liu and Jingcui Liang},
}

@article{BOOYSE2020106612,
    title = {Deep digital twins for detection, diagnostics and prognostics},
    journal = {Mechanical Systems and Signal Processing},
    volume = {140},
    pages = {106612},
    year = {2020},
    issn = {0888-3270},
    doi = {https://doi.org/10.1016/j.ymssp.2019.106612},
    author = {Wihan Booyse and Daniel N. Wilke and Stephan Heyns},
}

@article{NarayanaPichika,
    title = {Automatic Signal Denoising and Multi-Component Fault Classification Based on Deep Learning Using Integrated Condition Monitoring in a Wind Turbine Gearbox},
    journal = {Journal of Vibration Engineering \& Technologies},
    volume = {12},
    pages = {8623–8637},
    year = {2024},
    doi = {https://doi.org/10.1007/s42417-024-01380-6},
    author = {Narayana Pichika and Vamshi Kasam and Sabareesh Geetha Rajasekharan and Aruna Malapati},
}

@article{YANG2024117689,
    title = {Detection of wind turbine blade abnormalities through a deep learning model integrating VAE and neural ODE},
    journal = {Ocean Engineering},
    volume = {302},
    pages = {117689},
    year = {2024},
    issn = {0029-8018},
    doi = {https://doi.org/10.1016/j.oceaneng.2024.117689},
    author = {Zeyun Yang and Mingqiang Xu and Shuqing Wang and Jun Li and Zhen Peng and Fei Jin and Yuan Yang},
}

@ARTICLE{10153965,
    author={Wang, Lei and He, Yigang and Shao, Kaixuan and Xing, Zhikai and Zhou, Yazhong},
    journal={IEEE Transactions on Instrumentation and Measurement}, 
    title={An Unsupervised Approach to Wind Turbine Blade Icing Detection Based on Beta Variational Graph Attention Autoencoder}, 
    year={2024},
    volume={73},
    number={},
    pages={1-12},
    doi={10.1109/TIM.2023.3286011}
}

@article{Liu_2023,
    doi = {10.1088/1361-6501/aca496},
    year = {2022},
    month = {dec},
    publisher = {IOP Publishing},
    volume = {34},
    number = {3},
    pages = {035902},
    author = {Liu, Jiarui and Yang, Guotian and Li, Xinli and Hao, Shumin and Guan, Yingming and Li, Yaqi},
    title = {A deep generative model based on CNN-CVAE for wind turbine condition monitoring},
    journal = {Measurement Science and Technology},
}

@article{TAO2025112127,
    title = {LLM-based framework for bearing fault diagnosis},
    journal = {Mechanical Systems and Signal Processing},
    volume = {224},
    pages = {112127},
    year = {2025},
    issn = {0888-3270},
    doi = {https://doi.org/10.1016/j.ymssp.2024.112127},
    author = {Laifa Tao and Haifei Liu and Guoao Ning and Wenyan Cao and Bohao Huang and Chen Lu},
}

@article{SOLEIMANIBABAKAMALI2023110404,
    title = {Zero-shot transfer learning for structural health monitoring using generative adversarial networks and spectral mapping},
    journal = {Mechanical Systems and Signal Processing},
    volume = {198},
    pages = {110404},
    year = {2023},
    issn = {0888-3270},
    doi = {https://doi.org/10.1016/j.ymssp.2023.110404},
    author = {Mohammad Hesam Soleimani-Babakamali and Roksana Soleimani-Babakamali and Kourosh Nasrollahzadeh and Onur Avci and Serkan Kiranyaz and Ertugrul Taciroglu},
}

@Misc{peft,
    title = {PEFT: State-of-the-art Parameter-Efficient Fine-Tuning methods},
    author = {Sourab Mangrulkar and Sylvain Gugger and Lysandre Debut and Younes Belkada and Sayak Paul and Benjamin Bossan},
    howpublished = {\url{https://github.com/huggingface/peft}},
    year = {2022}
}

@article{GAO2023105277,
    title = {Enhanced data imputation framework for bridge health monitoring using Wasserstein generative adversarial networks with gradient penalty},
    journal = {Structures},
    volume = {57},
    pages = {105277},
    year = {2023},
    issn = {2352-0124},
    doi = {https://doi.org/10.1016/j.istruc.2023.105277},
    author = {Shuai Gao and Chunfeng Wan and Zhenwei Zhou and Jiale Hou and Liyu Xie and Songtao Xue},
}

@article{LULECI2023110370,
    title = {CycleGAN for undamaged-to-damaged domain translation for structural health monitoring and damage detection},
    journal = {Mechanical Systems and Signal Processing},
    volume = {197},
    pages = {110370},
    year = {2023},
    issn = {0888-3270},
    doi = {https://doi.org/10.1016/j.ymssp.2023.110370},
    author = {Furkan Luleci and F. {Necati Catbas} and Onur Avci},
}

@article{anaissi2023multi,
    title={Multi-objective variational autoencoder: an application for smart infrastructure maintenance},
    author={Anaissi, Ali and Zandavi, Seid Miad and Suleiman, Basem and Naji, Mohamad and Braytee, Ali},
    journal={Applied Intelligence},
    volume={53},
    number={10},
    pages={12047--12062},
    year={2023},
    publisher={Springer}
}

@article{CORACA2023109025,
    title = {An unsupervised structural health monitoring framework based on Variational Autoencoders and Hidden Markov Models},
    journal = {Reliability Engineering \& System Safety},
    volume = {231},
    pages = {109025},
    year = {2023},
    issn = {0951-8320},
    doi = {https://doi.org/10.1016/j.ress.2022.109025},
    author = {Eduardo M. Coraça and Janito V. Ferreira and Eurípedes G.O. Nóbrega},
}

@article{LULECI2023106146,
    title = {Improved undamaged-to-damaged acceleration response translation for Structural Health Monitoring},
    journal = {Engineering Applications of Artificial Intelligence},
    volume = {122},
    pages = {106146},
    year = {2023},
    issn = {0952-1976},
    doi = {https://doi.org/10.1016/j.engappai.2023.106146},
    author = {Furkan Luleci and Onur Avci and F. Necati Catbas},
}

@article{JANSSEN2020115483,
    title = {Data processing and augmentation of acoustic array signals for fault detection with machine learning},
    journal = {Journal of Sound and Vibration},
    volume = {483},
    pages = {115483},
    year = {2020},
    issn = {0022-460X},
    doi = {https://doi.org/10.1016/j.jsv.2020.115483},
    author = {L.A.L. Janssen and I. {Lopez Arteaga}},
}

@article{Fu_2023,
    doi = {10.1088/1361-6501/acabdb},
    year = {2023},
    month = {jan},
    publisher = {IOP Publishing},
    volume = {34},
    number = {4},
    pages = {045005},
    author = {Fu, Wenlong and Jiang, Xiaohui and Li, Bailin and Tan, Chao and Chen, Baojia and Chen, Xiaoyue},
    title = {Rolling bearing fault diagnosis based on 2D time-frequency images and data augmentation technique},
    journal = {Measurement Science and Technology},
}

@article{YE2024,
    title = {A spatiotemporal recurrent neural network for missing data imputation in tunnel monitoring},
    journal = {Journal of Rock Mechanics and Geotechnical Engineering},
    year = {2024},
    issn = {1674-7755},
    doi = {https://doi.org/10.1016/j.jrmge.2024.11.019},
    author = {Junchen Ye and Yuhao Mao and Ke Cheng and Xuyan Tan and Bowen Du and Weizhong Chen},
}

@article{ZHANG2025107031,
    title = {Domain-guided conditional diffusion model for unsupervised domain adaptation},
    journal = {Neural Networks},
    volume = {184},
    pages = {107031},
    year = {2025},
    issn = {0893-6080},
    doi = {https://doi.org/10.1016/j.neunet.2024.107031},
    author = {Yulong Zhang and Shuhao Chen and Weisen Jiang and Yu Zhang and Jiangang Lu and James T. Kwok},
}

@inproceedings{NEURIPS2024_cb1ba6a4,
    author = {Chen, Zhichao and Li, Haoxuan and Wang, Fangyikang and Zhang, Odin and Xu, Hu and Jiang, Xiaoyu and Song, Zhihuan and Wang, Hao},
    booktitle = {Advances in Neural Information Processing Systems},
    editor = {A. Globerson and L. Mackey and D. Belgrave and A. Fan and U. Paquet and J. Tomczak and C. Zhang},
    pages = {112050--112103},
    publisher = {Curran Associates, Inc.},
    title = {Rethinking the Diffusion Models for Missing Data Imputation: A Gradient Flow Perspective},
    volume = {37},
    year = {2024}
}

@article{WANG2025110973,
    title = {Time-frequency informed stacked long short-term memory-based generative adversarial network for missing data imputation in sensor networks},
    journal = {Engineering Applications of Artificial Intelligence},
    volume = {155},
    pages = {110973},
    year = {2025},
    issn = {0952-1976},
    doi = {https://doi.org/10.1016/j.engappai.2025.110973},
    author = {Zixin Wang and Malleswari Kachireddy and Tarutal Ghosh Mondal and Wen Tang and Mohammad R. Jahanshahi},
}

@article{LI2021109377,
    title = {A large-scale sensor missing data imputation framework for dams using deep learning and transfer learning strategy},
    journal = {Measurement},
    volume = {178},
    pages = {109377},
    year = {2021},
    issn = {0263-2241},
    doi = {https://doi.org/10.1016/j.measurement.2021.109377},
    author = {Yangtao Li and Tengfei Bao and Hao Chen and Kang Zhang and Xiaosong Shu and Zexun Chen and Yuhan Hu},
}

@article{REN2021107734,
    title = {Incremental Bayesian matrix/tensor learning for structural monitoring data imputation and response forecasting},
    journal = {Mechanical Systems and Signal Processing},
    volume = {158},
    pages = {107734},
    year = {2021},
    issn = {0888-3270},
    doi = {https://doi.org/10.1016/j.ymssp.2021.107734},
    author = {Pu Ren and Xinyu Chen and Lijun Sun and Hao Sun},
}

@article{QiAngWang2022,
    author = {Qi-Ang Wang and Chang-Bao Wang and Zhan-Guo Ma and Wei Chen and Yi-Qing Ni and Chu-Fan Wang and Bing-Gang Yan and Pei-Xuan Guan},
    title ={Bayesian dynamic linear model framework for structural health monitoring data forecasting and missing data imputation during typhoon events},
    journal = {Structural Health Monitoring},
    volume = {21},
    number = {6},
    pages = {2933-2950},
    year = {2022},
    doi = {10.1177/14759217221079529},
}

@article{chirici2016meta,
    title={A meta-analysis and review of the literature on the k-Nearest Neighbors technique for forestry applications that use remotely sensed data},
    author={Chirici, Gherardo and Mura, Matteo and McInerney, Daniel and Py, Nicolas and Tomppo, Erkki O and Waser, Lars T and Travaglini, Davide and McRoberts, Ronald E},
    journal={Remote Sensing of Environment},
    volume={176},
    pages={282--294},
    year={2016},
    publisher={Elsevier}
}

@article{van2015time,
    title={Time-series analysis of GPS monitoring data from a long-span bridge considering the global deformation due to air temperature changes},
    author={Van Le, Hien and Nishio, Mayuko},
    journal={Journal of Civil Structural Health Monitoring},
    volume={5},
    number={4},
    pages={415--425},
    year={2015},
    publisher={Springer}
}

@article{Huachen2022,
    author = {Huachen Jiang and Chunfeng Wan and Kang Yang and Youliang Ding and Songtao Xue},
    title ={Continuous missing data imputation with incomplete dataset by generative adversarial networks–based unsupervised learning for long-term bridge health monitoring},
    journal = {Structural Health Monitoring},
    volume = {21},
    number = {3},
    pages = {1093-1109},
    year = {2022},
    doi = {10.1177/14759217211021942},
}

@article{GAO2022112095,
    title = {Missing data imputation framework for bridge structural health monitoring based on slim generative adversarial networks},
    journal = {Measurement},
    volume = {204},
    pages = {112095},
    year = {2022},
    issn = {0263-2241},
    doi = {https://doi.org/10.1016/j.measurement.2022.112095},
    author = {Shuai Gao and Wenlong Zhao and Chunfeng Wan and Huachen Jiang and Youliang Ding and Songtao Xue},
}

@ARTICLE{9851940,
    author={Sim, Yeon-Sub and Hwang, Jae-Sang and Mun, Sung-Duk and Kim, Tae-Joon and Chang, Seung Jin},
    journal={IEEE Access}, 
    title={Missing Data Imputation Algorithm for Transmission Systems Based on Multivariate Imputation With Principal Component Analysis}, 
    year={2022},
    volume={10},
    number={},
    pages={83195-83203},
}

@ARTICLE{10239215,
    author={Huo, Jiuyuan and Qi, Chenbo and Li, Chaojie and Wang, Na},
    journal={IEEE Transactions on Instrumentation and Measurement}, 
    title={Data Augmentation Fault Diagnosis Method Based on Residual Mixed Self-Attention for Rolling Bearings Under Imbalanced Samples}, 
    year={2023},
    volume={72},
    number={},
    pages={1-14},
    doi={10.1109/TIM.2023.3311062}
}

@article{elreedy2024theoretical,
    title={A theoretical distribution analysis of synthetic minority oversampling technique (SMOTE) for imbalanced learning},
    author={Elreedy, Dina and Atiya, Amir F and Kamalov, Firuz},
    journal={Machine Learning},
    volume={113},
    number={7},
    pages={4903--4923},
    year={2024},
    publisher={Springer}
}

@ARTICLE{10136604,
    author={Wang, Dongdong and Dong, Yining and Wang, Han and Tang, Gang},
    journal={IEEE Sensors Journal}, 
    title={Limited Fault Data Augmentation With Compressed Sensing for Bearing Fault Diagnosis}, 
    year={2023},
    volume={23},
    number={13},
    pages={14499-14511},
    doi={10.1109/JSEN.2023.3277563}
}

@article{OH202072,
    title = {Data augmentation for bearing fault detection with a light weight CNN},
    journal = {Procedia Computer Science},
    volume = {175},
    pages = {72-79},
    year = {2020},
    issn = {1877-0509},
    doi = {https://doi.org/10.1016/j.procs.2020.07.013},
    author = {Jin Woo Oh and Jongpil Jeong},
}

@ARTICLE{8737894,
    author={Meng, Zong and Guo, Xiaolin and Pan, Zuozhou and Sun, Dengyun and Liu, Shuang},
    journal={IEEE Access}, 
    title={Data Segmentation and Augmentation Methods Based on Raw Data Using Deep Neural Networks Approach for Rotating Machinery Fault Diagnosis}, 
    year={2019},
    volume={7},
    number={},
    pages={79510-79522},
    doi={10.1109/ACCESS.2019.2923417}
}

@ARTICLE{9673114,
    author={You, Zhichao and Gao, Hongli and Li, Shichao and Guo, Liang and Liu, Yuekai and Li, Jingbo},
    journal={IEEE Transactions on Industrial Electronics}, 
    title={Multiple Activation Functions and Data Augmentation-Based Lightweight Network for In Situ Tool Condition Monitoring}, 
    year={2022},
    volume={69},
    number={12},
    pages={13656-13664},
}

@article{NGUYEN2022114172,
    title = {Deep learning-based autonomous damage-sensitive feature extraction for impedance-based prestress monitoring},
    journal = {Engineering Structures},
    volume = {259},
    pages = {114172},
    year = {2022},
    issn = {0141-0296},
    doi = {https://doi.org/10.1016/j.engstruct.2022.114172},
    author = {Thanh-Truong Nguyen and Thi {Tuong Vy Phan} and Duc-Duy Ho and Ananta {Man Singh Pradhan} and Thanh-Canh Huynh},
}

@ARTICLE{9259076,
    author={Dang, Hung V. and Tran-Ngoc, Hoa and Nguyen, Tung V. and Bui-Tien, T. and De Roeck, Guido and Nguyen, Huan X.},
    journal={IEEE Transactions on Automation Science and Engineering}, 
    title={Data-Driven Structural Health Monitoring Using Feature Fusion and Hybrid Deep Learning}, 
    year={2021},
    volume={18},
    number={4},
    pages={2087-2103},
    doi={10.1109/TASE.2020.3034401}
}

@article{li2023machine,
    title={A machine learning-based data augmentation strategy for structural damage classification in civil infrastructure system},
    author={Li, Lechen and Betti, Raimondo},
    journal={Journal of Civil Structural Health Monitoring},
    volume={13},
    number={6},
    pages={1265--1285},
    year={2023},
    publisher={Springer}
}

@ARTICLE{8815712,
    author={Hu, Tianhao and Tang, Tang and Chen, Ming},
    journal={IEEE Access}, 
    title={Data Simulation by Resampling—A Practical Data Augmentation Algorithm for Periodical Signal Analysis-Based Fault Diagnosis}, 
    year={2019},
    volume={7},
    number={},
    pages={125133-125145},
    doi={10.1109/ACCESS.2019.2937838}
}

@article{HOU2022111206,
    title = {Deep learning and data augmentation based data imputation for structural health monitoring system in multi-sensor damaged state},
    journal = {Measurement},
    volume = {196},
    pages = {111206},
    year = {2022},
    issn = {0263-2241},
    doi = {https://doi.org/10.1016/j.measurement.2022.111206},
    author = {Jiale Hou and Huachen Jiang and Chunfeng Wan and Letian Yi and Shuai Gao and Youliang Ding and Songtao Xue},
}

@ARTICLE{10697302,
    author={Jia, Xiaodong and Chen, Xiao},
    journal={IEEE Transactions on Industrial Informatics}, 
    title={Unsupervised Wind Turbine Blade Damage Detection With Memory-Aided Denoising Reconstruction}, 
    year={2025},
    volume={21},
    number={1},
    pages={762-770},
    doi={10.1109/TII.2024.3459612}
}

@article{CHEN2023116063,
    title = {Reconstruction of long-term strain data for structural health monitoring with a hybrid deep-learning and autoregressive model considering thermal effects},
    journal = {Engineering Structures},
    volume = {285},
    pages = {116063},
    year = {2023},
    issn = {0141-0296},
    doi = {https://doi.org/10.1016/j.engstruct.2023.116063},
    author = {Chengbin Chen and Liqun Tang and Yonghui Lu and Yong Wang and Zejia Liu and Yiping Liu and Licheng Zhou and Zhenyu Jiang and Bao Yang},
}

@article{LI2025112062,
    title = {A Dual-Discriminator Conditional GAN for rapid compressive sensing of structural vibration data},
    journal = {Mechanical Systems and Signal Processing},
    volume = {224},
    pages = {112062},
    year = {2025},
    issn = {0888-3270},
    doi = {https://doi.org/10.1016/j.ymssp.2024.112062},
    author = {Tao Li and Zhongyu Zhang and Rui Hou and Bo Liu and Kangkang Zheng and Dongwei Ren},
}

@article{ZHANG2025124378,
    title = {Domain-specific large language models for fault diagnosis of heating, ventilation, and air conditioning systems by labeled-data-supervised fine-tuning},
    journal = {Applied Energy},
    volume = {377},
    pages = {124378},
    year = {2025},
    issn = {0306-2619},
    doi = {https://doi.org/10.1016/j.apenergy.2024.124378},
    author = {Jian Zhang and Chaobo Zhang and Jie Lu and Yang Zhao},
}

@article{AliMardanshahi2025,
    title = {Sensing Techniques for Structural Health Monitoring: A State-of-the-Art Review on Performance Criteria and New-Generation Technologies},
    journal = {Sensors},
    issue = {5},
    volume = {25},
    pages = {1424},
    year = {2025},
    doi = {https://doi.org/10.3390/s25051424},
    author = {Ali Mardanshahi and Abhilash Sreekumar and Xin Yang and Swarup Kumar Barman and Dimitrios Chronopoulos},
}

@article{SHU2025111783,
    title = {DF-CDM: Conditional diffusion model with data fusion for structural dynamic response reconstruction},
    journal = {Mechanical Systems and Signal Processing},
    volume = {222},
    pages = {111783},
    year = {2025},
    issn = {0888-3270},
    doi = {https://doi.org/10.1016/j.ymssp.2024.111783},
    author = {Jiangpeng Shu and Hongchuan Yu and Gaoyang Liu and Yuanfeng Duan and Hao Hu and He Zhang},
}

@misc{hu2024evaluationimprovement,
    title={Evaluation and Improvement of Fault Detection for Large Language Models}, 
    author={Qiang Hu and Jin Wen and Maxime Cordy and Yuheng Huang and Wei Ma and Xiaofei Xie and Lei Ma},
    year={2024},
    eprint={2404.14419},
    archivePrefix={arXiv},
    primaryClass={cs.SE},
    url={https://arxiv.org/abs/2404.14419}, 
}

@article{chen2025large,
    title={Large Models for Machine Monitoring and Fault Diagnostics: Opportunities, Challenges, and Future Direction},
    author={Chen, Xuefeng and Lei, Yaguo and Li, Yanfu and Parkinson, Simon and Li, Xiang and Liu, Jinxin and Lu, Fan and Wang, Huan and Wang, Zisheng and Yang, Bin and others},
    journal={Journal of Dynamics, Monitoring and Diagnostics},
    volume={4},
    number={2},
    pages={76--90},
    year={2025}
}

@article{LIN2025103208,
    title = {FD-LLM: Large language model for fault diagnosis of complex equipment},
    journal = {Advanced Engineering Informatics},
    volume = {65},
    pages = {103208},
    year = {2025},
    issn = {1474-0346},
    doi = {https://doi.org/10.1016/j.aei.2025.103208},
    author = {Lin Lin and Sihao Zhang and Song Fu and Yikun Liu},
}

@INPROCEEDINGS{10774330,
  author={Pang, Zhendong and Zhang, Hao and Li, Teng},
    booktitle={2024 IEEE 22nd International Conference on Industrial Informatics (INDIN)}, 
    title={Hybrid Fine-Tuning in Large Language Model Learning for Machinery Fault Diagnosis}, 
    year={2024},
    volume={},
    number={},
    pages={1-6},
    doi={10.1109/INDIN58382.2024.10774330}
}

@misc{j2024finetuningllmenterprise,
    title={Fine Tuning LLM for Enterprise: Practical Guidelines and Recommendations}, 
    author={Mathav Raj J and Kushala VM and Harikrishna Warrier and Yogesh Gupta},
    year={2024},
    eprint={2404.10779},
    archivePrefix={arXiv},
    primaryClass={cs.SE},
    url={https://arxiv.org/abs/2404.10779}, 
}

@article{WANG2025110760,
    title = {An enhanced generative adversarial network for longer vibration time data generation under variable operating conditions for imbalanced bearing fault diagnosis},
    journal = {Engineering Applications of Artificial Intelligence},
    volume = {151},
    pages = {110760},
    year = {2025},
    issn = {0952-1976},
    doi = {https://doi.org/10.1016/j.engappai.2025.110760},
    author = {Teng Wang and Zhi Chao Ong and Shin Yee Khoo and Pei Yi Siow and Jinlai Zhang and Tao Wang},
}

@Article{app13053136,
    AUTHOR = {Wang, Yu and Vinogradov, Alexey},
    TITLE = {Improving the Performance of Convolutional GAN Using History-State Ensemble for Unsupervised Early Fault Detection with Acoustic Emission Signals},
    JOURNAL = {Applied Sciences},
    VOLUME = {13},
    YEAR = {2023},
    NUMBER = {5},
    ARTICLE-NUMBER = {3136},
    URL = {https://www.mdpi.com/2076-3417/13/5/3136},
    ISSN = {2076-3417},
    DOI = {10.3390/app13053136}
}

@inproceedings{asami2022data,
    title={Data Augmentation with Synthesized Damaged Roof Images Generated by GAN.},
    author={Asami, Koki and Fujita, Shono and Hiroi, Kei and Hatayama, Michinori},
    booktitle={ISCRAM},
    pages={256--265},
    year={2022}
}

@ARTICLE{10577252,
    author={Li, Huipeng and Wang, Congqing and Liu, Yang},
    journal={IEEE Transactions on Instrumentation and Measurement}, 
    title={A High-Quality GAN With Data Augmentation for Aircraft Skin Defect Detection Under Limited Data}, 
    year={2024},
    volume={73},
    number={},
    pages={1-12},
    doi={10.1109/TIM.2024.3420361}
}

@article{CHEN2024110101,
    title = {Acoustic signal recovering for rubbing in a dual-rotor system based on diffusion probabilistic models},
    journal = {Applied Acoustics},
    volume = {224},
    pages = {110101},
    year = {2024},
    issn = {0003-682X},
    doi = {https://doi.org/10.1016/j.apacoust.2024.110101},
    author = {Ao Chen and Zhi-Yuan Wu and Dong-Wu Li and Dong Wang and Wen-Ming Zhang},
}

@ARTICLE{10892083,
    author={Wu, Peng and Yu, Gongye and Han, Yongming and Ma, Bo},
    journal={IEEE Transactions on Reliability}, 
    title={Cross-Domain Acoustic Diagnosis Method of Rotating Machinery Based on Vibration and Acoustic Migration}, 
    year={2025},
    volume={74},
    number={3},
    doi={10.1109/TR.2024.3521335}
}

@ARTICLE{10018473,
    author={Shao, Haidong and Li, Wei and Cai, Baoping and Wan, Jiafu and Xiao, Yiming and Yan, Shen},
    journal={IEEE Transactions on Industrial Informatics}, 
    title={Dual-Threshold Attention-Guided GAN and Limited Infrared Thermal Images for Rotating Machinery Fault Diagnosis Under Speed Fluctuation}, 
    year={2023},
    volume={19},
    number={9},
    pages={9933-9942},
    doi={10.1109/TII.2022.3232766}
}

@article{SINGHRAGHAV2023104020,
    title = {Crack propagation and fatigue life estimation of spur gear with and without spalling failure},
    journal = {Theoretical and Applied Fracture Mechanics},
    volume = {127},
    pages = {104020},
    year = {2023},
    issn = {0167-8442},
    doi = {https://doi.org/10.1016/j.tafmec.2023.104020},
    author = {Mahendra {Singh Raghav} and Shivdayal Patel},
}

@article{LI20242958,
    title = {Fault diagnosis of nuclear power plant sliding bearing-rotor systems using deep convolutional generative adversarial networks},
    journal = {Nuclear Engineering and Technology},
    volume = {56},
    number = {8},
    pages = {2958-2973},
    year = {2024},
    issn = {1738-5733},
    doi = {https://doi.org/10.1016/j.net.2024.02.056},
    author = {Qi Li and Weiwei Zhang and Feiyu Chen and Guobing Huang and Xiaojing Wang and Weimin Yuan and Xin Xiong},
}

@article{SHEN2024109159,
    title = {Piston aero-engine fault cross-domain diagnosis based on unpaired generative transfer learning},
    journal = {Engineering Applications of Artificial Intelligence},
    volume = {137},
    pages = {109159},
    year = {2024},
    issn = {0952-1976},
    doi = {https://doi.org/10.1016/j.engappai.2024.109159},
    author = {Pengfei Shen and Fengrong Bi and Xiaoyang Bi and Mingzhi Guo and Yunyi Lu},
}

@article{WEN2024111663,
    title = {{GRU-AE-wiener}: A generative adversarial network assisted hybrid gated recurrent unit with Wiener model for bearing remaining useful life estimation},
    journal = {Mechanical Systems and Signal Processing},
    volume = {220},
    pages = {111663},
    year = {2024},
    issn = {0888-3270},
    doi = {https://doi.org/10.1016/j.ymssp.2024.111663},
    author = {Long Wen and Shaoquan Su and Xinyu Li and Weiping Ding and Ke Feng},
}

@article{YANG2025112996,
    title = {Damage imaging in structural health monitoring with fine-tuned conditional diffusion model},
    journal = {Mechanical Systems and Signal Processing},
    volume = {236},
    pages = {112996},
    year = {2025},
    issn = {0888-3270},
    doi = {https://doi.org/10.1016/j.ymssp.2025.112996},
    author = {Xin Yang and Sergio Cantero-Chinchilla and Morteza Moradi and Panagiotis Komninos and Chen Fang and Yunlai Liao and Pradeep Kundu and Dimitrios Zarouchas and Dimitrios Chronopoulos},
}

@article{XinY101049,
    author = {Xin Yang  and Pradeep Kundu  and Dimitrios Chronopoulos },
    title = {Fine-tuned diffusion model for high-resolution sparse-array ultrasonic guided wave imaging},
    journal = {IET Conference Proceedings},
    volume = {2025},
    issue = {10},
    pages = {148-151},
    year = {2025},
    doi = {10.1049/icp.2025.2347},
}

@article{hibat2024framework,
    title={A framework for demonstrating practical quantum advantage: comparing quantum against classical generative models},
    author={Hibat-Allah, Mohamed and Mauri, Marta and Carrasquilla, Juan and Perdomo-Ortiz, Alejandro},
    journal={Communications Physics},
    volume={7},
    number={1},
    pages={68},
    year={2024},
    publisher={Nature Publishing Group UK London}
}

@article{Zhu_2022,
    title={Earthquake Phase Association Using a Bayesian Gaussian Mixture Model},
    volume={127},
    ISSN={2169-9356},
    DOI={10.1029/2021jb023249},
    number={5},
    journal={Journal of Geophysical Research: Solid Earth},
    publisher={American Geophysical Union (AGU)},
    author={Zhu, Weiqiang and McBrearty, Ian W. and Mousavi, S. Mostafa and Ellsworth, William L. and Beroza, Gregory C.},
    year={2022},
}

@inproceedings{NEURIPS2024_9ad996b5,
    author = {Li, Senmao and Hu, Taihang and van de Weijer, Joost and Khan, Fahad Shahbaz and Liu, Tao and Li, Linxuan and Yang, Shiqi and Wang, Yaxing and Cheng, Ming-Ming and Yang, Jian},
    booktitle = {Advances in Neural Information Processing Systems},
    editor = {A. Globerson and L. Mackey and D. Belgrave and A. Fan and U. Paquet and J. Tomczak and C. Zhang},
    pages = {85203--85240},
    publisher = {Curran Associates, Inc.},
    title = {Faster Diffusion: Rethinking the Role of the Encoder for Diffusion Model Inference},
    volume = {37},
    year = {2024}
}

@article{3408318,
    author = {Zhang, Yi and Li, Miaomiao and Wang, Siwei and Dai, Sisi and Luo, Lei and Zhu, En and Xu, Huiying and Zhu, Xinzhong and Yao, Chaoyun and Zhou, Haoran},
    title = {Gaussian mixture model clustering with incomplete data},
    year = {2021},
    volume = {17},
    number = {1s},
    issn = {1551-6857},
    doi = {10.1145/3408318},
    journal = {ACM Transactions on Multimedia Computing, Communications, and Applications},
    numpages = {14},
}

@misc{zhang2024scalingmeetsllmfinetuning,
    title={When Scaling Meets LLM Finetuning: The Effect of Data, Model and Finetuning Method}, 
    author={Biao Zhang and Zhongtao Liu and Colin Cherry and Orhan Firat},
    year={2024},
    eprint={2402.17193},
    archivePrefix={arXiv},
    primaryClass={cs.CL},
    url={https://arxiv.org/abs/2402.17193}, 
}

@Article{rs17203398,
    AUTHOR = {Zhang, Jiyuan and Wang, Zhonggen and Chen, Jing and Wang, Fei and Gao, Lyuzhou},
    TITLE = {An Automated Framework for Abnormal Target Segmentation in Levee Scenarios Using Fusion of {UAV}-Based Infrared and Visible Imagery},
    JOURNAL = {Remote Sensing},
    VOLUME = {17},
    YEAR = {2025},
    NUMBER = {20},
    ARTICLE-NUMBER = {3398},
    ISSN = {2072-4292},
    DOI = {10.3390/rs17203398}
}

@misc{naveed2024comprehensiveoverviewlargelanguage,
    title={A Comprehensive Overview of Large Language Models}, 
    author={Humza Naveed and Asad Ullah Khan and Shi Qiu and Muhammad Saqib and Saeed Anwar and Muhammad Usman and Naveed Akhtar and Nick Barnes and Ajmal Mian},
    year={2024},
    eprint={2307.06435},
    archivePrefix={arXiv},
    primaryClass={cs.CL},
    url={https://arxiv.org/abs/2307.06435}, 
}

@article{ZHAO2025109520,
    title = {Denoising diffusion probabilistic model-enabled data augmentation method for intelligent machine fault diagnosis},
    journal = {Engineering Applications of Artificial Intelligence},
    volume = {139},
    pages = {109520},
    year = {2025},
    issn = {0952-1976},
    doi = {https://doi.org/10.1016/j.engappai.2024.109520},
    author = {Pengcheng Zhao and Wei Zhang and Xiaoshan Cao and Xiang Li},
}

@article{WANG2023106872,
    title = {A dynamic spectrum loss generative adversarial network for intelligent fault diagnosis with imbalanced data},
    journal = {Engineering Applications of Artificial Intelligence},
    volume = {126},
    pages = {106872},
    year = {2023},
    issn = {0952-1976},
    doi = {https://doi.org/10.1016/j.engappai.2023.106872},
    author = {Xin Wang and Hongkai Jiang and Yunpeng Liu and Shaowei Liu and Qiao Yang},
}

@article{LIU2022110888,
    title = {A conditional variational autoencoding generative adversarial networks with self-modulation for rolling bearing fault diagnosis},
    journal = {Measurement},
    volume = {192},
    pages = {110888},
    year = {2022},
    issn = {0263-2241},
    doi = {https://doi.org/10.1016/j.measurement.2022.110888},
    author = {Yunpeng Liu and Hongkai Jiang and Yanfeng Wang and Zhenghong Wu and Shaowei Liu},
}

@article{SHEN2024109299,
    title = {Image augmentation for nondestructive testing in engineering structures based on denoising diffusion probabilistic model},
    journal = {Journal of Building Engineering},
    volume = {89},
    pages = {109299},
    year = {2024},
    issn = {2352-7102},
    doi = {https://doi.org/10.1016/j.jobe.2024.109299},
    author = {Wei Shen and Dongyang Zeng and Yang Zhang and Xi Tian and Ziqi Li},
}

@article{GUO2025110854,
    title = {SDCGAN: A CycleGAN-based single-domain generalization method for mechanical fault diagnosis},
    journal = {Reliability Engineering \& System Safety},
    volume = {258},
    pages = {110854},
    year = {2025},
    issn = {0951-8320},
    doi = {https://doi.org/10.1016/j.ress.2025.110854},
    author = {Yu Guo and Xiangyu Li and Jundong Zhang and Ziyi Cheng},
}

@article{YANG2024111597,
    title = {A decision-level sensor fusion scheme integrating ultrasonic guided wave and vibration measurements for damage identification},
    journal = {Mechanical Systems and Signal Processing},
    volume = {219},
    pages = {111597},
    year = {2024},
    issn = {0888-3270},
    doi = {https://doi.org/10.1016/j.ymssp.2024.111597},
    author = {Xin Yang and Chen Fang and Pradeep Kundu and Jian Yang and Dimitrios Chronopoulos},
}

@article{DENG2025116595,
    title = {Unsupervised PG-DDPM-augmented mixed dataset for training an accurate concrete bridge crack detection model under small samples},
    journal = {Measurement},
    volume = {245},
    pages = {116595},
    year = {2025},
    issn = {0263-2241},
    doi = {https://doi.org/10.1016/j.measurement.2024.116595},
    author = {Jianghua Deng and Linxin Hua and Ye Lu and Chenyang Wang and Jiao Che},
}

@article{NajafiDataset,
    title={Thermal image of equipment (Induction Motor)},
    author={Najafi, Mohamad and Baleghi, Yasser and Mirimani, Seyyed Mehdi},
    Journal={Mendeley Data},
    year = {2020},
    doi = {10.17632/m4sbt8hbvk.1},
}

@article{ZHANG2024114795,
    title = {CBAM-CRLSGAN: A novel fault diagnosis method for planetary transmission systems under small samples scenarios},
    journal = {Measurement},
    volume = {234},
    pages = {114795},
    year = {2024},
    issn = {0263-2241},
    doi = {https://doi.org/10.1016/j.measurement.2024.114795},
    author = {Jie Zhang and Yun Kong and Zhuyun Chen and Te Han and Qinkai Han and Mingming Dong and Fulei Chu},
}

@article{WAN2025109614,
    title = {A novel hybrid data-driven domain generalization approach with dual-perspective feature fusion for intelligent fault diagnosis},
    journal = {Engineering Applications of Artificial Intelligence},
    volume = {139},
    pages = {109614},
    year = {2025},
    issn = {0952-1976},
    doi = {https://doi.org/10.1016/j.engappai.2024.109614},
    author = {Lanjun Wan and Jian Zhou and Jiaen Ning and Yuanyuan Li and Changyun Li},
}

@misc{urain2024dgmsrobotics,
    title={Deep Generative Models in Robotics: A Survey on Learning from Multimodal Demonstrations}, 
    author={Julen Urain and Ajay Mandlekar and Yilun Du and Mahi Shafiullah and Danfei Xu and Katerina Fragkiadaki and Georgia Chalvatzaki and Jan Peters},
    year={2024},
    eprint={2408.04380},
    archivePrefix={arXiv},
    primaryClass={cs.RO},
    url={https://arxiv.org/abs/2408.04380}, 
}

@article{LI2025110570,
    title = {A novel reinforcement learning agent for rotating machinery fault diagnosis with data augmentation},
    journal = {Reliability Engineering \& System Safety},
    volume = {253},
    pages = {110570},
    year = {2025},
    issn = {0951-8320},
    doi = {https://doi.org/10.1016/j.ress.2024.110570},
    author = {Zhenning Li and Hongkai Jiang and Xin Wang},
}

@article{OGUNFOWORA2023244,
    title = {Reinforcement and deep reinforcement learning-based solutions for machine maintenance planning, scheduling policies, and optimization},
    journal = {Journal of Manufacturing Systems},
    volume = {70},
    pages = {244-263},
    year = {2023},
    issn = {0278-6125},
    doi = {https://doi.org/10.1016/j.jmsy.2023.07.014},
    author = {Oluwaseyi Ogunfowora and Homayoun Najjaran},
}

@misc{raissi2017PINN,
    title={Physics Informed Deep Learning (Part I): Data-driven Solutions of Nonlinear Partial Differential Equations}, 
    author={Maziar Raissi and Paris Perdikaris and George Em Karniadakis},
    year={2017},
    eprint={1711.10561},
    archivePrefix={arXiv},
    primaryClass={cs.AI},
    url={https://arxiv.org/abs/1711.10561}, 
}

@misc{he2024llmsmeetmultimodalgeneration,
    title={LLMs Meet Multimodal Generation and Editing: A Survey}, 
    author={Yingqing He and Zhaoyang Liu and Jingye Chen and Zeyue Tian and Hongyu Liu and Xiaowei Chi and Runtao Liu and Ruibin Yuan and Yazhou Xing and Wenhai Wang and Jifeng Dai and Yong Zhang and Wei Xue and Qifeng Liu and Yike Guo and Qifeng Chen},
    year={2024},
    eprint={2405.19334},
    archivePrefix={arXiv},
    primaryClass={cs.AI},
    url={https://arxiv.org/abs/2405.19334}, 
}

@article{YANG2024115387,
    title = {Transfer learning-based Gaussian process classification for lattice structure damage detection},
    journal = {Measurement},
    volume = {238},
    pages = {115387},
    year = {2024},
    issn = {0263-2241},
    doi = {https://doi.org/10.1016/j.measurement.2024.115387},
    author = {Xin Yang and Amin Farrokhabadi and Ali Rauf and Yongcheng Liu and Reza Talemi and Pradeep Kundu and Dimitrios Chronopoulos},
}

@ARTICLE{10440027,
    author={Gawde, Shreyas and Patil, Shruti and Kumar, Satish and Kamat, Pooja and Kotecha, Ketan and Alfarhood, Sultan},
    journal={IEEE Access}, 
    title={Explainable predictive maintenance of rotating machines using {IME}, {SHAP}, {PDP}, {ICE}}, 
    year={2024},
    volume={12},
    number={},
    pages={29345-29361},
    doi={10.1109/ACCESS.2024.3367110}
}

@misc{huang2024trustllm,
    title={TrustLLM: Trustworthiness in Large Language Models}, 
    author={Yue Huang and Lichao Sun and Haoran Wang and Siyuan Wu and Qihui Zhang and Yuan Li and Chujie Gao and Yixin Huang and Wenhan Lyu and Yixuan Zhang and Xiner Li and Zhengliang Liu and Yixin Liu and Yijue Wang and Zhikun Zhang and Bertie Vidgen and Bhavya Kailkhura and Caiming Xiong and Chaowei Xiao and Chunyuan Li and Eric Xing and Furong Huang and Hao Liu and Heng Ji and Hongyi Wang and Huan Zhang and Huaxiu Yao and Manolis Kellis and Marinka Zitnik and Meng Jiang and Mohit Bansal and James Zou and Jian Pei and Jian Liu and Jianfeng Gao and Jiawei Han and Jieyu Zhao and Jiliang Tang and Jindong Wang and Joaquin Vanschoren and John Mitchell and Kai Shu and Kaidi Xu and Kai-Wei Chang and Lifang He and Lifu Huang and Michael Backes and Neil Zhenqiang Gong and Philip S. Yu and Pin-Yu Chen and Quanquan Gu and Ran Xu and Rex Ying and Shuiwang Ji and Suman Jana and Tianlong Chen and Tianming Liu and Tianyi Zhou and William Wang and Xiang Li and Xiangliang Zhang and Xiao Wang and Xing Xie and Xun Chen and Xuyu Wang and Yan Liu and Yanfang Ye and Yinzhi Cao and Yong Chen and Yue Zhao},
    year={2024},
    eprint={2401.05561},
    archivePrefix={arXiv},
    primaryClass={cs.CL},
    url={https://arxiv.org/abs/2401.05561}, 
}

@article{Yanfei2020,
    author = {Kang, Yanfei and Hyndman, Rob J. and Li, Feng},
    title = {GRATIS: GeneRAting TIme Series with diverse and controllable characteristics},
    journal = {Statistical Analysis and Data Mining: An ASA Data Science Journal},
    volume = {13},
    number = {4},
    pages = {354-376},
    doi = {https://doi.org/10.1002/sam.11461},
    year = {2020}
}

@misc{liu2024trustworthyllms,
    title={Trustworthy LLMs: a Survey and Guideline for Evaluating Large Language Models' Alignment}, 
    author={Yang Liu and Yuanshun Yao and Jean-Francois Ton and Xiaoying Zhang and Ruocheng Guo and Hao Cheng and Yegor Klochkov and Muhammad Faaiz Taufiq and Hang Li},
    year={2024},
    eprint={2308.05374},
    archivePrefix={arXiv},
    primaryClass={cs.AI},
    url={https://arxiv.org/abs/2308.05374}, 
}

@inproceedings{2022_Yang,
    author = {Xin Yang and Fengxiang Chen},
    title = {Deep Uncertainty Quantification of Prognostic Techniques for Proton Exchange Membrane Fuel Cell},
    year = {2022},
    pages = {2022-01-7001},
    month = {feb},
    publisher = {SAE International},
    doi={10.4271/2022-01-7001},
}

@article{Qing03042022,
    author = {Xinlin Qing and Yunlai Liao and Yihan Wang and Binqiang Chen and Fanghong Zhang and Yishou Wang and},
    title = {Machine Learning Based Quantitative Damage Monitoring of Composite Structure},
    journal = {International Journal of Smart and Nano Materials},
    volume = {13},
    number = {2},
    pages = {167--202},
    year = {2022},
    publisher = {Taylor \& Francis},
    doi = {10.1080/19475411.2022.2054878},
}

@article{Yaowen2021,
    author = {Ou, Yaowen and Tatsis, Konstantinos E. and Dertimanis, Vasilis K. and Spiridonakos, Minas D. and Chatzi, Eleni N.},
    title = {Vibration-based monitoring of a small-scale wind turbine blade under varying climate conditions. {P}art {I}: {A}n experimental benchmark},
    journal = {Structural Control and Health Monitoring},
    volume = {28},
    number = {6},
    pages = {e2660},
    doi = {https://doi.org/10.1002/stc.2660},
    year = {2021},
}

@article{Ju_2024,
    doi = {10.1088/1361-6501/ad7a97},
    year = {2024},
    month = {sep},
    publisher = {IOP Publishing},
    volume = {35},
    number = {12},
    pages = {122004},
    author = {Ju, Zedong and Chen, Yinsheng and Qiang, Yukang and Chen, Xinyi and Ju, Chao and Yang, Jingli},
    title = {A systematic review of data augmentation methods for intelligent fault diagnosis of rotating machinery under limited data conditions},
    journal = {Measurement Science and Technology},
}

@misc{ren2025aigcindustrial,
    title={{AIGC} for Industrial Time Series: From Deep Generative Models to Large Generative Models}, 
    author={Lei Ren and Haiteng Wang and Jinwang Li and Yang Tang and Chunhua Yang},
    year={2025},
    eprint={2407.11480},
    archivePrefix={arXiv},
    primaryClass={cs.LG},
    url={https://arxiv.org/abs/2407.11480}, 
}

@misc{ye2025autoregression,
    title={Beyond Autoregression: Discrete Diffusion for Complex Reasoning and Planning}, 
    author={Jiacheng Ye and Jiahui Gao and Shansan Gong and Lin Zheng and Xin Jiang and Zhenguo Li and Lingpeng Kong},
    year={2025},
    eprint={2410.14157},
    archivePrefix={arXiv},
    primaryClass={cs.CL},
    url={https://arxiv.org/abs/2410.14157}, 
}

@article{huxiaosong2021,
    author = {Hu, Xiaosong and Yang, Xin and Feng, Fei and Liu, Kailong and Lin, Xianke},
    title = {A Particle Filter and Long Short-Term Memory Fusion Technique for Lithium-Ion Battery Remaining Useful Life Prediction},
    journal = {Journal of Dynamic Systems, Measurement, and Control},
    volume = {143},
    number = {6},
    pages = {061001},
    year = {2021},
    month = {01},
    issn = {0022-0434},
    doi = {10.1115/1.4049234},
}

@article{CHEN2020106683,
    title = {A deep learning method for bearing fault diagnosis based on Cyclic Spectral Coherence and Convolutional Neural Networks},
    journal = {Mechanical Systems and Signal Processing},
    volume = {140},
    pages = {106683},
    year = {2020},
    issn = {0888-3270},
    doi = {https://doi.org/10.1016/j.ymssp.2020.106683},
    author = {Zhuyun Chen and Alexandre Mauricio and Weihua Li and Konstantinos Gryllias},
}

@article{VANDREVEN2024132711,
    title = {A systematic approach for data generation for intelligent fault detection and diagnosis in District Heating},
    journal = {Energy},
    volume = {307},
    pages = {132711},
    year = {2024},
    issn = {0360-5442},
    doi = {https://doi.org/10.1016/j.energy.2024.132711},
    author = {Jonne {van Dreven} and Veselka Boeva and Shahrooz Abghari and Håkan Grahn and Jad {Al Koussa}},
}

@article{ZHAO2024122807,
    title = {A comparison review of transfer learning and self-supervised learning: Definitions, applications, advantages and limitations},
    journal = {Expert Systems with Applications},
    volume = {242},
    pages = {122807},
    year = {2024},
    issn = {0957-4174},
    author = {Zehui Zhao and Laith Alzubaidi and Jinglan Zhang and Ye Duan and Yuantong Gu},
}

@article{WU2023439,
    title = {A transformer-based approach for novel fault detection and fault classification/diagnosis in manufacturing: A rotary system application},
    journal = {Journal of Manufacturing Systems},
    volume = {67},
    pages = {439-452},
    year = {2023},
    issn = {0278-6125},
    doi = {https://doi.org/10.1016/j.jmsy.2023.02.018},
    author = {Haiyue Wu and Matthew J. Triebe and John W. Sutherland},
}

@article{CHEN2019106272,
    title = {Mechanical fault diagnosis using Convolutional Neural Networks and Extreme Learning Machine},
    journal = {Mechanical Systems and Signal Processing},
    volume = {133},
    pages = {106272},
    year = {2019},
    issn = {0888-3270},
    doi = {https://doi.org/10.1016/j.ymssp.2019.106272},
    author = {Zhuyun Chen and Konstantinos Gryllias and Weihua Li},
}

@article{AHMEDMURTAZA2024102935,
    title = {Paradigm shift for predictive maintenance and condition monitoring from Industry 4.0 to Industry 5.0: A systematic review, challenges and case study},
    journal = {Results in Engineering},
    volume = {24},
    pages = {102935},
    year = {2024},
    issn = {2590-1230},
    doi = {https://doi.org/10.1016/j.rineng.2024.102935},
    author = {Aitzaz {Ahmed Murtaza} and Amina Saher and Muhammad {Hamza Zafar} and Syed {Kumayl Raza Moosavi} and Muhammad {Faisal Aftab} and Filippo Sanfilippo},
}

@article{LI2025110027,
    title = {Causality-guided fault diagnosis under visual interference in fused deposition modeling},
    journal = {Engineering Applications of Artificial Intelligence},
    volume = {143},
    pages = {110027},
    year = {2025},
    issn = {0952-1976},
    doi = {https://doi.org/10.1016/j.engappai.2025.110027},
    author = {Qian Li and Tingting Huang and Jie Liu and Shanggang Wang},
}

@article{YANG2025110312,
    title = {A global information-guided denoising diffusion probabilistic model for fault diagnosis with imbalanced data},
    journal = {Engineering Applications of Artificial Intelligence},
    volume = {147},
    pages = {110312},
    year = {2025},
    issn = {0952-1976},
    doi = {https://doi.org/10.1016/j.engappai.2025.110312},
    author = {Han Yang and Yan Song and Daichao Wang and Jinghao Xing and Yibin Li},
}

@article{XU2023107063,
    title = {Intelligent fault diagnosis of bearings under small samples: A mechanism-data fusion approach},
    journal = {Engineering Applications of Artificial Intelligence},
    volume = {126},
    pages = {107063},
    year = {2023},
    issn = {0952-1976},
    doi = {https://doi.org/10.1016/j.engappai.2023.107063},
    author = {Kun Xu and Xianguang Kong and Qibin Wang and Bing Han and Liqiang Sun},
}

@article{Xiaoming2021,
    author = {Xiaoming Lei and Limin Sun and Ye Xia},
    title ={Lost data reconstruction for structural health monitoring using deep convolutional generative adversarial networks},
    journal = {Structural Health Monitoring},
    volume = {20},
    number = {4},
    pages = {2069-2087},
    year = {2021},
    doi = {10.1177/1475921720959226},
}

@misc{xiao2022tacklinggenerative,
    title={Tackling the Generative Learning Trilemma with Denoising Diffusion GANs}, 
    author={Zhisheng Xiao and Karsten Kreis and Arash Vahdat},
    year={2022},
    eprint={2112.07804},
    archivePrefix={arXiv},
    primaryClass={cs.LG},
    url={https://arxiv.org/abs/2112.07804}, 
}

@misc{finn2017modelagnosticmetalearningfastadaptation,
    title={Model-Agnostic Meta-Learning for Fast Adaptation of Deep Networks}, 
    author={Chelsea Finn and Pieter Abbeel and Sergey Levine},
    year={2017},
    eprint={1703.03400},
    archivePrefix={arXiv},
    primaryClass={cs.LG},
    url={https://arxiv.org/abs/1703.03400}, 
}

@article{LIU2025116989,
    title = {Frequency domain guided latent diffusion model for domain generalization in cross-machine fault diagnosis},
    journal = {Measurement},
    volume = {249},
    pages = {116989},
    year = {2025},
    issn = {0263-2241},
    doi = {https://doi.org/10.1016/j.measurement.2025.116989},
    author = {Xiaolin Liu and Fuzheng Liu and Xiangyi Geng and Longqing Fan and Mingshun Jiang and Faye Zhang},
}

@ARTICLE{10688397,
    author={Meng, Yuquan and Dong, Zhiqiao and Lu, Kuan-Chieh and Li, Shichen and Shao, Chenhui},
    journal={IEEE Transactions on Industrial Informatics}, 
    title={Meta-Learning-Based Domain Generalization for Cost-Effective Tool Condition Monitoring in Ultrasonic Metal Welding}, 
    year={2025},
    volume={21},
    number={1},
    pages={653-662},
    doi={10.1109/TII.2024.3456671}
}

@ARTICLE{10091197,
    author={Ren, Lei and Mo, Tingyu and Cheng, Xuejun},
    journal={IEEE Transactions on Industrial Informatics}, 
    title={Meta-learning Based Domain Generalization Framework for Fault Diagnosis With Gradient Aligning and Semantic Matching}, 
    year={2024},
    volume={20},
    number={1},
    pages={754-764},
    doi={10.1109/TII.2023.3264111}
}

@ARTICLE{Zhuang9759507,
    author={Zhuang, Jichao and Jia, Minping and Ding, Yifei and Zhao, Xiaoli},
    journal={IEEE/ASME Transactions on Mechatronics}, 
    title={Health Assessment of Rotating Equipment With Unseen Conditions Using Adversarial Domain Generalization Toward Self-Supervised Regularization Learning}, 
    year={2022},
    volume={27},
    number={6},
    pages={4675-4685},
    doi={10.1109/TMECH.2022.3163289}}

@Article{ir202307,
    AUTHOR = {Chun Liu and Shaojie Li and Hongtian Chen and Xianchao Xiu and Chen Peng},
    TITLE = {Semi-supervised joint adaptation transfer network with conditional adversarial learning for rotary machine fault diagnosis},
    JOURNAL = {Intelligence \& Robotics},
    VOLUME = {3},
    YEAR = {2023},
    NUMBER = {2},
    ARTICLE-NUMBER = {131-43},
    ISSN = {2770-3541},
    DOI = {10.20517/ir.2023.07}
}

@article{Kouw_2021,
    title={A Review of Domain Adaptation without Target Labels},
    volume={43},
    ISSN={1939-3539},
    DOI={10.1109/tpami.2019.2945942},
    number={3},
    journal={IEEE Transactions on Pattern Analysis and Machine Intelligence},
    publisher={Institute of Electrical and Electronics Engineers (IEEE)},
    author={Kouw, Wouter M. and Loog, Marco},
    year={2021},
    month=mar, 
    pages={766–785} 
}

@ARTICLE{Lou9789138,
    author={Lou, Yunxia and Kumar, Anil and Xiang, Jiawei},
    journal={IEEE Transactions on Instrumentation and Measurement}, 
    title={Machinery Fault Diagnosis Based on Domain Adaptation to Bridge the Gap Between Simulation and Measured Signals}, 
    year={2022},
    volume={71},
    number={},
    pages={1-9},
    doi={10.1109/TIM.2022.3180416}
}

@article{CACERESCASTELLANOS20237746,
    title = {Condition Monitoring using Domain-Adversarial Networks with Convolutional Kernel Features},
    journal = {IFAC-PapersOnLine},
    volume = {56},
    number = {2},
    pages = {7746-7752},
    year = {2023},
    note = {22nd IFAC World Congress},
    issn = {2405-8963},
    doi = {https://doi.org/10.1016/j.ifacol.2023.10.1180},
    author = {Cesar Caceres-Castellanos and Moritz Fehsenfeld and Karl-Philipp Kortmann},
}

@article{WANG2025110662,
    title = {A dynamic collaborative adversarial domain adaptation network for unsupervised rotating machinery fault diagnosis},
    journal = {Reliability Engineering \& System Safety},
    volume = {255},
    pages = {110662},
    year = {2025},
    issn = {0951-8320},
    doi = {https://doi.org/10.1016/j.ress.2024.110662},
    author = {Xin Wang and Hongkai Jiang and Mingzhe Mu and Yutong Dong},
}

@ARTICLE{Sun10504387,
    author={Sun, Shilong and Ding, Hao and Huang, Haodong and Zhao, Zida and Wang, Dong and Xu, Wenfu},
    journal={IEEE Transactions on Instrumentation and Measurement}, 
    title={A Novel Cross-Domain Data Augmentation and Bearing Fault Diagnosis Method Based on an Enhanced Generative Model}, 
    year={2024},
    volume={73},
    number={},
    pages={1-9},
    doi={10.1109/TIM.2024.3390242}
}

@article{WANG2023455,
    title = {Partial adversarial domain adaptation by dual-domain alignment for fault diagnosis of rotating machines},
    journal = {ISA Transactions},
    volume = {136},
    pages = {455-467},
    year = {2023},
    issn = {0019-0578},
    doi = {https://doi.org/10.1016/j.isatra.2022.11.021},
    author = {Xuan Wang and Bo She and Zhangsong Shi and Shiyan Sun and Fenqi Qin},
}

@article{XU2024112396,
    title = {Multi-source domain adaptation using diffusion denoising for bearing fault diagnosis under variable working conditions},
    journal = {Knowledge-Based Systems},
    volume = {302},
    pages = {112396},
    year = {2024},
    issn = {0950-7051},
    doi = {https://doi.org/10.1016/j.knosys.2024.112396},
    author = {Xuefang Xu and Xu Yang and Zijian Qiao and Pengfei Liang and Changbo He and Peiming Shi},
}

@article{GUO2025115951,
    title = {Bearing fault diagnostic framework under unknown working conditions based on condition-guided diffusion model},
    journal = {Measurement},
    volume = {242},
    pages = {115951},
    year = {2025},
    issn = {0263-2241},
    doi = {https://doi.org/10.1016/j.measurement.2024.115951},
    author = {Zhibin Guo and Lefei Xu and Yuhao Zheng and Jingsong Xie and Tiantian Wang},
}

@misc{radford2021learning,
    title={Learning Transferable Visual Models From Natural Language Supervision}, 
    author={Alec Radford and Jong Wook Kim and Chris Hallacy and Aditya Ramesh and Gabriel Goh and Sandhini Agarwal and Girish Sastry and Amanda Askell and Pamela Mishkin and Jack Clark and Gretchen Krueger and Ilya Sutskever},
    year={2021},
    eprint={2103.00020},
    archivePrefix={arXiv},
    primaryClass={cs.CV},
    url={https://arxiv.org/abs/2103.00020}, 
}

@misc{wang2022generalizing,
    title={Generalizing to Unseen Domains: A Survey on Domain Generalization}, 
    author={Jindong Wang and Cuiling Lan and Chang Liu and Yidong Ouyang and Tao Qin and Wang Lu and Yiqiang Chen and Wenjun Zeng and Philip S. Yu},
    year={2022},
    eprint={2103.03097},
    archivePrefix={arXiv},
    primaryClass={cs.LG},
    url={https://arxiv.org/abs/2103.03097}, 
}

@misc{ho2020denoisingdiffusion,
    title={Denoising Diffusion Probabilistic Models}, 
    author={Jonathan Ho and Ajay Jain and Pieter Abbeel},
    year={2020},
    eprint={2006.11239},
    archivePrefix={arXiv},
    primaryClass={cs.LG},
    url={https://arxiv.org/abs/2006.11239}, 
}

@ARTICLE{10103569,
    author={Chen, Peng and Xu, Chaojun and Ma, Zhigang and Jin, Yaqiang},
    journal={IEEE Transactions on Instrumentation and Measurement}, 
    title={A Mixed Samples-Driven Methodology Based on Denoising Diffusion Probabilistic Model for Identifying Damage in Carbon Fiber Composite Structures}, 
    year={2023},
    volume={72},
    number={},
    pages={1-11},
}

@article{CHEN2025111813,
    title = {Automated structural resilience evaluation based on a multi-scale Transformer network using field monitoring data},
    journal = {Mechanical Systems and Signal Processing},
    volume = {222},
    pages = {111813},
    year = {2025},
    issn = {0888-3270},
    doi = {https://doi.org/10.1016/j.ymssp.2024.111813},
    author = {Zepeng Chen and Qitian Liu and Zhenghao Ding and Feng Liu},
}

@ARTICLE{9625031,
    author={Nath, Aneesh G. and Udmale, Sandeep S. and Raghuwanshi, Divyanshu and Singh, Sanjay Kumar},
    journal={IEEE Sensors Journal}, 
    title={Structural Rotor Fault Diagnosis Using Attention-Based Sensor Fusion and Transformers}, 
    year={2022},
    volume={22},
    number={1},
    pages={707-719},
    doi={10.1109/JSEN.2021.3130183}
}

@article{li2022structural,
    title={Structural health monitoring data anomaly detection by transformer enhanced densely connected neural networks},
    author={Li, Jun and Chen, Wupeng and Fan, Gao},
    journal={Smart Structures and Systems},
    volume={30},
    number={6},
    pages={613--626},
    year={2022}
}

@article{Jichen2025,
    author = {Jichen Tian and Yanling Li and Yonghua Luo and Han Zhang and Xiang Lu},
    title ={Multisource information fusion model for deformation safety monitoring of earth and rock dams based on deep graph feature fusion},
    journal = {Structural Health Monitoring},
    volume = {24},
    number = {2},
    pages = {925-940},
    year = {2025},
    doi = {10.1177/14759217241244549},
}

@article{RUAN2023101877,
    title = {CNN parameter design based on fault signal analysis and its application in bearing fault diagnosis},
    journal = {Advanced Engineering Informatics},
    volume = {55},
    pages = {101877},
    year = {2023},
    issn = {1474-0346},
    doi = {https://doi.org/10.1016/j.aei.2023.101877},
    author = {Diwang Ruan and Jin Wang and Jianping Yan and Clemens Gühmann},
}

@article{JANA2022108723,
    title = {CNN and Convolutional Autoencoder (CAE) based real-time sensor fault detection, localization, and correction},
    journal = {Mechanical Systems and Signal Processing},
    volume = {169},
    pages = {108723},
    year = {2022},
    issn = {0888-3270},
    doi = {https://doi.org/10.1016/j.ymssp.2021.108723},
    author = {Debasish Jana and Jayant Patil and Sudheendra Herkal and Satish Nagarajaiah and Leonardo Duenas-Osorio},
}

@article{liu2022data,
    title={Data anomaly detection for structural health monitoring using a combination network of GANomaly and CNN},
    author={Liu, Gaoyang and Niu, Yanbo and Zhao, Weijian and Duan, Yuanfeng and Shu, Jiangpeng},
    journal={Smart Structures and Systems},
    volume={29},
    number={1},
    pages={53-62},
    year={2022}
}

@article{JIN2022379,
    title = {A Time Series Transformer based method for the rotating machinery fault diagnosis},
    journal = {Neurocomputing},
    volume = {494},
    pages = {379-395},
    year = {2022},
    issn = {0925-2312},
    doi = {https://doi.org/10.1016/j.neucom.2022.04.111},
    author = {Yuhong Jin and Lei Hou and Yushu Chen},
}

@article{WANG2024111067,
    title = {A small sample piezoelectric impedance-based structural damage identification using Signal Reshaping-based Enhance Attention Transformer},
    journal = {Mechanical Systems and Signal Processing},
    volume = {208},
    pages = {111067},
    year = {2024},
    issn = {0888-3270},
    doi = {https://doi.org/10.1016/j.ymssp.2023.111067},
    author = {Xian Wang and Zhuo Chen and Wenjun Sun and Nan Shao and Zengying You and Jiawen Xu and Ruqiang Yan},
}

@article{YingHan2025,
    author = {Ying Han and Tanmiao Liu and Kun Li},
    title ={Fault diagnosis of wind turbine blades under wide-weather multi-operating conditions based on multi-modal information fusion and deep learning},
    journal = {Structural Health Monitoring},
    volume = {0},
    number = {0},
    pages = {14759217251333761},
    year = {2025},
    doi = {10.1177/14759217251333761},
}

@article{FALCHI2024111382,
    title = {Deep learning and structural health monitoring: Temporal Fusion Transformers for anomaly detection in masonry towers},
    journal = {Mechanical Systems and Signal Processing},
    volume = {215},
    pages = {111382},
    year = {2024},
    issn = {0888-3270},
    doi = {https://doi.org/10.1016/j.ymssp.2024.111382},
    author = {Fabrizio Falchi and Maria Girardi and Gianmarco Gurioli and Nicola Messina and Cristina Padovani and Daniele Pellegrini},
}

@ARTICLE{10922758,
    author={Li, Haiyang and Wang, Jing and Wu, Shuguang and Gao, Yehemin and Wang, Yawei and Nie, Guigen},
    journal={IEEE Transactions on Instrumentation and Measurement}, 
    title={Transforming Structural Health Monitoring: Leveraging Multisource Data Fusion With Two-Stage Encoder Transformer for Bridge Deformation Prediction}, 
    year={2025},
    volume={74},
    number={},
    pages={1-13},
    doi={10.1109/TIM.2025.3550241}
}

@article{SHEN2023100442,
    title = {Multi-route fusion method of GNSS and accelerometer for structural health monitoring},
    journal = {Journal of Industrial Information Integration},
    volume = {32},
    pages = {100442},
    year = {2023},
    issn = {2452-414X},
    doi = {https://doi.org/10.1016/j.jii.2023.100442},
    url = {https://www.sciencedirect.com/science/article/pii/S2452414X23000158},
    author = {Nan Shen and Liang Chen and Ruizhi Chen},
}

@article{ZHU2023116573,
    title = {Multi-rate Kalman filtering for structural dynamic response reconstruction by fusing multi-type sensor data with different sampling frequencies},
    journal = {Engineering Structures},
    volume = {293},
    pages = {116573},
    year = {2023},
    issn = {0141-0296},
    doi = {https://doi.org/10.1016/j.engstruct.2023.116573},
    author = {Zimo Zhu and Jubin Lu and Songye Zhu},
}

@article{PAL2024111482,
    title = {Data fusion based on short-term memory Kalman filtering using intermittent-displacement and acceleration signal with a time-varying bias},
    journal = {Mechanical Systems and Signal Processing},
    volume = {216},
    pages = {111482},
    year = {2024},
    issn = {0888-3270},
    doi = {https://doi.org/10.1016/j.ymssp.2024.111482},
    author = {Ashish Pal and Satish Nagarajaiah},
}

@article{Wang0309714,
    author = {Wang, Qiushi AND Sun, Zhicheng AND Zhu, Yueming AND Li, Dong AND Ma, Yunbin},
    journal = {PLOS ONE},
    publisher = {Public Library of Science},
    title = {A fault diagnosis method based on an improved diffusion model under limited sample conditions},
    year = {2024},
    month = {09},
    volume = {19},
    doi = {https://doi.org/10.1371/journal.pone.0309714},
    pages = {1-19},
    number = {9},
}

@article{YAO2024111397,
    title = {A novel stochastic process diffusion model for wind turbines condition monitoring and fault identification with multi-parameter information fusion},
    journal = {Mechanical Systems and Signal Processing},
    volume = {214},
    pages = {111397},
    year = {2024},
    issn = {0888-3270},
    doi = {https://doi.org/10.1016/j.ymssp.2024.111397},
    author = {Qingtao Yao and Hankun Bing and Guopeng Zhu and Ling Xiang and Aijun Hu},
}

@article{HerTerng,
    author = {Her-Terng Yau and Ping-Huan Kuo and Shang-Yi Yu},
    title ={Bearing fault detection system based on a deep diffusion model},
    journal = {Structural Health Monitoring},
    volume = {0},
    number = {0},
    pages = {14759217241274335},
    year = {2024},
    doi = {10.1177/14759217241274335},
}

@ARTICLE{10750268,
    author={Zhang, Zhenyu and Pang, Jian and Ding, Jianxi and Yuan, Xin’an and Liu, Weifeng and Lu, Xiaoping and Chen, Honglong},
    journal={IEEE Sensors Journal}, 
    title={Conditional Denoising Diffusion Model-Based Defect Reconstruction for Alternating Current Field Measurement}, 
    year={2025},
    volume={25},
    number={1},
    pages={919-928},
    doi={10.1109/JSEN.2024.3491186}
}

@article{wei2025generation,
    title={Generation of Fault Data from Multiple Types of Bridge Monitoring Sensors Based on Time Series Diffusion Models Fusing Control Conditions and Pseudo Prompt Enhancement},
    author={Wei, Shenglin and others},
    journal={Academic Journal of Computing \& Information Science},
    volume={8},
    number={3},
    pages={1--9},
    doi = {https://doi.org/10.25236/AJCIS.2025.080301},
    year={2025},
    publisher={Francis Academic Press}
}

@article{YI2024111481,
    title = {Time series diffusion method: A denoising diffusion probabilistic model for vibration signal generation},
    journal = {Mechanical Systems and Signal Processing},
    volume = {216},
    pages = {111481},
    year = {2024},
    issn = {0888-3270},
    doi = {https://doi.org/10.1016/j.ymssp.2024.111481},
    author = {Haiming Yi and Lei Hou and Yuhong Jin and Nasser A. Saeed and Ali Kandil and Hao Duan},
}

@article{WANG2022234,
    title = {Mix-VAEs: A novel multisensor information fusion model for intelligent fault diagnosis},
    journal = {Neurocomputing},
    volume = {492},
    pages = {234-244},
    year = {2022},
    issn = {0925-2312},
    doi = {https://doi.org/10.1016/j.neucom.2022.04.044},
    author = {Cunjun Wang and Cun Xin and Zili Xu and Manqing Qin and Mengfu He},
}

@article{LEI2013108,
    title = {A review on empirical mode decomposition in fault diagnosis of rotating machinery},
    journal = {Mechanical Systems and Signal Processing},
    volume = {35},
    number = {1},
    pages = {108-126},
    year = {2013},
    issn = {0888-3270},
    doi = {https://doi.org/10.1016/j.ymssp.2012.09.015},
    author = {Yaguo Lei and Jing Lin and Zhengjia He and Ming J. Zuo},
}

@Article{electronics13244912,
    AUTHOR = {Alsaif, Khalid M. and Albeshri, Aiiad A. and Khemakhem, Maher A. and Eassa, Fathy E.},
    TITLE = {Multimodal Large Language Model-Based Fault Detection and Diagnosis in Context of Industry 4.0},
    JOURNAL = {Electronics},
    VOLUME = {13},
    YEAR = {2024},
    NUMBER = {24},
    ARTICLE-NUMBER = {4912},
    ISSN = {2079-9292},
    DOI = {10.3390/electronics13244912}
}

@misc{qaid2024fdllmlargelanguagemodel,
    title={FD-LLM: Large Language Model for Fault Diagnosis of Machines}, 
    author={Hamzah A. A. M. Qaid and Bo Zhang and Dan Li and See-Kiong Ng and Wei Li},
    year={2024},
    eprint={2412.01218},
    archivePrefix={arXiv},
    primaryClass={cs.AI},
    url={https://arxiv.org/abs/2412.01218}, 
}

@article{ZHENG2024110382,
    title = {Empirical study on fine-tuning pre-trained large language models for fault diagnosis of complex systems},
    journal = {Reliability Engineering \& System Safety},
    volume = {252},
    pages = {110382},
    year = {2024},
    issn = {0951-8320},
    doi = {https://doi.org/10.1016/j.ress.2024.110382},
    author = {Shuwen Zheng and Kai Pan and Jie Liu and Yunxia Chen},
}

@article{schneider2024GenXAI,
    title={Explainable Generative AI (GenXAI): a survey, conceptualization, and research agenda},
    author={Schneider, Johannes},
    journal={Artificial Intelligence Review},
    volume={57},
    number={11},
    pages={289},
    year={2024},
    publisher={Springer},
    doi={https://doi.org/10.1007/s10462-024-10916-x}, 
}

@misc{manduchi2025chall,
    title={On the {C}hallenges and {O}pportunities in {G}enerative {AI}}, 
    author={Laura Manduchi and Kushagra Pandey and Clara Meister and Robert Bamler and Ryan Cotterell and Sina Däubener and Sophie Fellenz and Asja Fischer and Thomas Gärtner and Matthias Kirchler and Marius Kloft and Yingzhen Li and Christoph Lippert and Gerard de Melo and Eric Nalisnick and Björn Ommer and Rajesh Ranganath and Maja Rudolph and Karen Ullrich and Guy Van den Broeck and Julia E Vogt and Yixin Wang and Florian Wenzel and Frank Wood and Stephan Mandt and Vincent Fortuin},
    year={2025},
    eprint={2403.00025},
    archivePrefix={arXiv},
    primaryClass={cs.LG},
    url={https://arxiv.org/abs/2403.00025}, 
}

@article{Kingma_2019,
    title={An Introduction to Variational Autoencoders},
    volume={12},
    ISSN={1935-8245},
    DOI={10.1561/2200000056},
    number={4},
    journal={Foundations and Trends® in Machine Learning},
    publisher={Now Publishers},
    author={Kingma, Diederik P. and Welling, Max},
    year={2019},
    pages={307–392} 
}

@misc{xu2023qalora,
    title={QA-LoRA: Quantization-Aware Low-Rank Adaptation of Large Language Models}, 
    author={Yuhui Xu and Lingxi Xie and Xiaotao Gu and Xin Chen and Heng Chang and Hengheng Zhang and Zhengsu Chen and Xiaopeng Zhang and Qi Tian},
    year={2023},
    eprint={2309.14717},
    archivePrefix={arXiv},
    primaryClass={cs.LG},
    url={https://arxiv.org/abs/2309.14717}, 
}

@article{PANG2024110312,
    title = {ParInfoGPT: An LLM-based two-stage framework for reliability assessment of rotating machine under partial information},
    journal = {Reliability Engineering \& System Safety},
    volume = {250},
    pages = {110312},
    year = {2024},
    issn = {0951-8320},
    doi = {https://doi.org/10.1016/j.ress.2024.110312},
    author = {Zhendong Pang and Yingxin Luan and Jiahong Chen and Teng Li},
}

@article{XU2024102650,
    title = {Few-shot learning for structural health diagnosis of civil infrastructure},
    journal = {Advanced Engineering Informatics},
    volume = {62},
    pages = {102650},
    year = {2024},
    issn = {1474-0346},
    doi = {https://doi.org/10.1016/j.aei.2024.102650},
    author = {Yang XU and Yunlei FAN and Yuequan BAO and Hui LI},
}

@Inbook{Tomczak2022,
    author={Tomczak, Jakub M.},
    title={Why Deep Generative Modeling?},
    bookTitle={Deep Generative Modeling},
    year={2022},
    publisher={Springer International Publishing},
    address={Cham},
    pages={1--12},
    isbn={978-3-030-93158-2},
    doi={10.1007/978-3-030-93158-2_1},
}

@article{GM2020100285,
    title = {A comprehensive survey and analysis of generative models in machine learning},
    journal = {Computer Science Review},
    volume = {38},
    pages = {100285},
    year = {2020},
    issn = {1574-0137},
    doi = {https://doi.org/10.1016/j.cosrev.2020.100285},
    author = {Harshvardhan GM and Mahendra Kumar Gourisaria and Manjusha Pandey and Siddharth Swarup Rautaray},
}

@article{ZHANG2025125059,
    title = {Deep generative models in energy system applications: Review, challenges, and future directions},
    journal = {Applied Energy},
    volume = {380},
    pages = {125059},
    year = {2025},
    issn = {0306-2619},
    doi = {https://doi.org/10.1016/j.apenergy.2024.125059},
    author = {Xiangyu Zhang and Andrew Glaws and Alexandre Cortiella and Patrick Emami and Ryan N. King},
}

@INPROCEEDINGS{9934291,
    author={Kossale, Youssef and Airaj, Mohammed and Darouichi, Aziz},
    booktitle={2022 8th International Conference on Optimization and Applications (ICOA)}, 
    title={Mode Collapse in Generative Adversarial Networks: An Overview}, 
    year={2022},
    volume={},
    number={},
    pages={1-6},
    doi={10.1109/ICOA55659.2022.9934291}
}

@inproceedings{Lucas2019UnderstandingPC,
    title={Understanding Posterior Collapse in Generative Latent Variable Models},
    author={James Lucas and G. Tucker and Roger Baker Grosse and Mohammad Norouzi},
    booktitle={DGS@ICLR},
    year={2019},
    url={https://api.semanticscholar.org/CorpusID:156054774}
}

@article{Bond_Taylor_2022,
    title={Deep Generative Modelling: A Comparative Review of VAEs, GANs, Normalizing Flows, Energy-Based and Autoregressive Models},
    volume={44},
    ISSN={1939-3539},
    DOI={10.1109/tpami.2021.3116668},
    number={11},
    journal={IEEE Transactions on Pattern Analysis and Machine Intelligence},
    publisher={Institute of Electrical and Electronics Engineers (IEEE)},
    author={Bond-Taylor, Sam and Leach, Adam and Long, Yang and Willcocks, Chris G.},
    year={2022},
    month=nov, pages={7327–7347} 
}

@ARTICLE{Chen10081265,
    author={Chen, Yuejian and Rao, Meng and Feng, Ke and Niu, Gang},
    journal={IEEE Transactions on Instrumentation and Measurement}, 
    title={Modified Varying Index Coefficient Autoregression Model for Representation of the Nonstationary Vibration From a Planetary Gearbox}, 
    year={2023},
    volume={72},
    number={},
    pages={1-12},
    doi={10.1109/TIM.2023.3259048}
}

@article{ZHANG2022109175,
    title = {Vibration feature extraction using signal processing techniques for structural health monitoring: A review},
    journal = {Mechanical Systems and Signal Processing},
    volume = {177},
    pages = {109175},
    year = {2022},
    issn = {0888-3270},
    doi = {https://doi.org/10.1016/j.ymssp.2022.109175},
    author = {Chunwei Zhang and Asma A. Mousavi and Sami F. Masri and Gholamreza Gholipour and Kai Yan and Xiuling Li},
}

@article{HU2025113122,
    title = {A health monitoring and early fault detection method of rotating machines based on latent variables of diffusion model},
    journal = {Mechanical Systems and Signal Processing},
    volume = {237},
    pages = {113122},
    year = {2025},
    issn = {0888-3270},
    doi = {https://doi.org/10.1016/j.ymssp.2025.113122},
    author = {Wenyang Hu and Qi Li and Tianyang Wang and Fulei Chu},
}

@article{djemili2024wind,
    title={A wind turbine bearing fault detection method based on improved CEEMDAN and AR-MEDA},
    author={Djemili, Ilyes and Medoued, Ammar and Soufi, Youcef},
    journal={Journal of Vibration Engineering \& Technologies},
    volume={12},
    number={3},
    pages={4225--4246},
    year={2024},
    publisher={Springer}
}

@article{ZHUANG2024111186,
    title = {Health prognosis of bearings based on transferable autoregressive recurrent adaptation with few-shot learning},
    journal = {Mechanical Systems and Signal Processing},
    volume = {211},
    pages = {111186},
    year = {2024},
    issn = {0888-3270},
    doi = {https://doi.org/10.1016/j.ymssp.2024.111186},
    author = {Jichao Zhuang and Minping Jia and Cheng-Geng Huang and Michael Beer and Ke Feng},
}

@inproceedings{NEURIPS2019_378a063b,
    author = {Du, Yilun and Mordatch, Igor},
    booktitle = {Advances in Neural Information Processing Systems},
    editor = {H. Wallach and H. Larochelle and A. Beygelzimer and F. d\textquotesingle Alch\'{e}-Buc and E. Fox and R. Garnett},
    pages = {},
    publisher = {Curran Associates, Inc.},
    title = {Implicit Generation and Modeling with Energy Based Models},
    url = {https://proceedings.neurips.cc/paper_files/paper/2019/file/378a063b8fdb1db941e34f4bde584c7d-Paper.pdf},
    volume = {32},
    year = {2019}
}

@misc{moran2025interpretableDGMs,
    title={Towards Interpretable Deep Generative Models via Causal Representation Learning}, 
    author={Gemma E. Moran and Bryon Aragam},
    year={2025},
    eprint={2504.11609},
    archivePrefix={arXiv},
    primaryClass={stat.ML},
    url={https://arxiv.org/abs/2504.11609}, 
}

@article{RETZLAFF2024101243,
    title = {Post-hoc vs ante-hoc explanations: xAI design guidelines for data scientists},
    journal = {Cognitive Systems Research},
    volume = {86},
    pages = {101243},
    year = {2024},
    issn = {1389-0417},
    doi = {https://doi.org/10.1016/j.cogsys.2024.101243},
    author = {Carl O. Retzlaff and Alessa Angerschmid and Anna Saranti and David Schneeberger and Richard Röttger and Heimo Müller and Andreas Holzinger},
}

@article{XU2025102721,
    title = {Interpretability research of deep learning: A literature survey},
    journal = {Information Fusion},
    volume = {115},
    pages = {102721},
    year = {2025},
    issn = {1566-2535},
    doi = {https://doi.org/10.1016/j.inffus.2024.102721},
    author = {Biao Xu and Guanci Yang},
}

@misc{song2022ddim,
    title={Denoising Diffusion Implicit Models}, 
    author={Jiaming Song and Chenlin Meng and Stefano Ermon},
    year={2022},
    eprint={2010.02502},
    archivePrefix={arXiv},
    primaryClass={cs.LG},
    url={https://arxiv.org/abs/2010.02502}, 
}

@article{Kobyzev_2021,
    title={Normalizing {F}lows: An Introduction and Review of Current Methods},
    volume={43},
    ISSN={1939-3539},
    DOI={10.1109/tpami.2020.2992934},
    number={11},
    journal={IEEE Transactions on Pattern Analysis and Machine Intelligence},
    publisher={Institute of Electrical and Electronics Engineers (IEEE)},
    author={Kobyzev, Ivan and Prince, Simon J.D. and Brubaker, Marcus A.},
    year={2021},
    month=nov, pages={3964–3979} 
}

@misc{song2021scorebasedgenerative,
    title={Score-Based Generative Modeling through Stochastic Differential Equations}, 
    author={Yang Song and Jascha Sohl-Dickstein and Diederik P. Kingma and Abhishek Kumar and Stefano Ermon and Ben Poole},
    year={2021},
    eprint={2011.13456},
    archivePrefix={arXiv},
    primaryClass={cs.LG},
    url={https://arxiv.org/abs/2011.13456}, 
}

@misc{ruthotto2021introduction,
    title={An Introduction to Deep Generative Modeling}, 
    author={Lars Ruthotto and Eldad Haber},
    year={2021},
    eprint={2103.05180},
    archivePrefix={arXiv},
    primaryClass={cs.LG},
    url={https://arxiv.org/abs/2103.05180}, 
}

@article{CHEN2024128167,
    title = {A comprehensive survey for generative data augmentation},
    journal = {Neurocomputing},
    volume = {600},
    pages = {128167},
    year = {2024},
    issn = {0925-2312},
    doi = {https://doi.org/10.1016/j.neucom.2024.128167},
    author = {Yunhao Chen and Zihui Yan and Yunjie Zhu},
}

@misc{ma2021identifiable,
    title={Identifiable Generative Models for Missing Not at Random Data Imputation}, 
    author={Chao Ma and Cheng Zhang},
    year={2021},
    eprint={2110.14708},
    archivePrefix={arXiv},
    primaryClass={cs.LG},
    url={https://arxiv.org/abs/2110.14708}, 
}

@ARTICLE{10058512,
    author={Branikas, Efstathios and Murray, Paul and West, Graeme},
    journal={IEEE Access}, 
    title={A Novel Data Augmentation Method for Improved Visual Crack Detection Using Generative Adversarial Networks}, 
    year={2023},
    volume={11},
    number={},
    pages={22051-22059},
    doi={10.1109/ACCESS.2023.3251988}
}

@article{LIU2021107488,
    title = {Intelligent fault diagnosis under small sample size conditions via Bidirectional InfoMax GAN with unsupervised representation learning},
    journal = {Knowledge-Based Systems},
    volume = {232},
    pages = {107488},
    year = {2021},
    issn = {0950-7051},
    doi = {https://doi.org/10.1016/j.knosys.2021.107488},
    author = {Shen Liu and Jinglong Chen and Shuilong He and Enyong Xu and Haixin Lv and Zitong Zhou},
}

@article{LIU2024124511,
    title = {Generative artificial intelligence and data augmentation for prognostic and health management: Taxonomy, progress, and prospects},
    journal = {Expert Systems with Applications},
    volume = {255},
    pages = {124511},
    year = {2024},
    issn = {0957-4174},
    doi = {https://doi.org/10.1016/j.eswa.2024.124511},
    author = {Shen Liu and Jinglong Chen and Yong Feng and Zongliang Xie and Tongyang Pan and Jingsong Xie},
}

@article{WANG2024110394,
    title = {Data augmentation based on diffusion probabilistic model for remaining useful life estimation of aero-engines},
    journal = {Reliability Engineering \& System Safety},
    volume = {252},
    pages = {110394},
    year = {2024},
    issn = {0951-8320},
    doi = {https://doi.org/10.1016/j.ress.2024.110394},
    author = {Wei Wang and Honghao Song and Shubin Si and Wenhao Lu and Zhiqiang Cai},
}

@article{LI2022101552,
    title = {Multi-mode data augmentation and fault diagnosis of rotating machinery using modified ACGAN designed with new framework},
    journal = {Advanced Engineering Informatics},
    volume = {52},
    pages = {101552},
    year = {2022},
    issn = {1474-0346},
    doi = {https://doi.org/10.1016/j.aei.2022.101552},
    author = {Wei Li and Xiang Zhong and Haidong Shao and Baoping Cai and Xingkai Yang},
}

@article{QU2020106610,
    title = {A novel wind turbine data imputation method with multiple optimizations based on GANs},
    journal = {Mechanical Systems and Signal Processing},
    volume = {139},
    pages = {106610},
    year = {2020},
    issn = {0888-3270},
    doi = {https://doi.org/10.1016/j.ymssp.2019.106610},
    author = {Fuming Qu and Jinhai Liu and Yanjuan Ma and Dong Zang and Mingrui Fu},
}

@misc{singh2024rethinking,
    title={Rethinking Interpretability in the Era of Large Language Models}, 
    author={Chandan Singh and Jeevana Priya Inala and Michel Galley and Rich Caruana and Jianfeng Gao},
    year={2024},
    eprint={2402.01761},
    archivePrefix={arXiv},
    primaryClass={cs.CL},
    url={https://arxiv.org/abs/2402.01761}, 
}

@article{van2010software,
    title={Software survey: {VOS}viewer, a computer program for bibliometric mapping},
    author={Van Eck, Nees and Waltman, Ludo},
    journal={Scientometrics},
    volume={84},
    number={2},
    pages={523--538},
    year={2010},
    doi = {https://doi.org/10.1007/s11192-009-0146-3},
}

@article{ZHANG2025110663,
    title = {Multimodal data imputation and fusion for trustworthy fault diagnosis of mechanical systems},
    journal = {Engineering Applications of Artificial Intelligence},
    volume = {150},
    pages = {110663},
    year = {2025},
    issn = {0952-1976},
    doi = {https://doi.org/10.1016/j.engappai.2025.110663},
    author = {Jie Zhang and Yun Kong and Qinkai Han and Tianyang Wang and Mingming Dong and Hui Liu and Fulei Chu},
}

@misc{hu2021lora,
      title={LoRA: Low-Rank Adaptation of Large Language Models}, 
      author={Edward J. Hu and Yelong Shen and Phillip Wallis and Zeyuan Allen-Zhu and Yuanzhi Li and Shean Wang and Lu Wang and Weizhu Chen},
      year={2021},
      eprint={2106.09685},
      archivePrefix={arXiv},
      primaryClass={cs.CL},
      url={https://arxiv.org/abs/2106.09685}, 
}

@misc{hu2023llmadapters,
      title={LLM-Adapters: An Adapter Family for Parameter-Efficient Fine-Tuning of Large Language Models}, 
      author={Zhiqiang Hu and Lei Wang and Yihuai Lan and Wanyu Xu and Ee-Peng Lim and Lidong Bing and Xing Xu and Soujanya Poria and Roy Ka-Wei Lee},
      year={2023},
      eprint={2304.01933},
      archivePrefix={arXiv},
      primaryClass={cs.CL},
      url={https://arxiv.org/abs/2304.01933}, 
}

@misc{li2021prefixtuning,
      title={Prefix-Tuning: Optimizing Continuous Prompts for Generation}, 
      author={Xiang Lisa Li and Percy Liang},
      year={2021},
      eprint={2101.00190},
      archivePrefix={arXiv},
      primaryClass={cs.CL},
      url={https://arxiv.org/abs/2101.00190}, 
}

@article{Zhao3649447,
    author = {Zhao, Fei and Zhang, Chengcui and Geng, Baocheng},
    title = {Deep Multimodal Data Fusion},
    year = {2024},
    publisher = {Association for Computing Machinery},
    volume = {56},
    number = {9},
    issn = {0360-0300},
    doi = {https://doi.org/10.1145/3649447},
    journal = {ACM computing surveys},
}

@misc{houlsby2019parameterefficient,
      title={Parameter-Efficient Transfer Learning for NLP}, 
      author={Neil Houlsby and Andrei Giurgiu and Stanislaw Jastrzebski and Bruna Morrone and Quentin de Laroussilhe and Andrea Gesmundo and Mona Attariyan and Sylvain Gelly},
      year={2019},
      eprint={1902.00751},
      archivePrefix={arXiv},
      primaryClass={cs.LG},
      url={https://arxiv.org/abs/1902.00751}, 
}

@article{SHI2023101940,
    title = {Intelligent layout generation based on deep generative models: A comprehensive survey},
    journal = {Information Fusion},
    volume = {100},
    pages = {101940},
    year = {2023},
    issn = {1566-2535},
    doi = {https://doi.org/10.1016/j.inffus.2023.101940},
    author = {Yong Shi and Mengyu Shang and Zhiquan Qi},
}

@misc{dettmers2023qlora,
    title={QLoRA: Efficient Finetuning of Quantized LLMs}, 
    author={Tim Dettmers and Artidoro Pagnoni and Ari Holtzman and Luke Zettlemoyer},
    year={2023},
    eprint={2305.14314},
    archivePrefix={arXiv},
    primaryClass={cs.LG},
    url={https://arxiv.org/abs/2305.14314}, 
}

@misc{vaswani2023attentionneed,
      title={Attention Is All You Need}, 
      author={Ashish Vaswani and Noam Shazeer and Niki Parmar and Jakob Uszkoreit and Llion Jones and Aidan N. Gomez and Lukasz Kaiser and Illia Polosukhin},
      year={2023},
      eprint={1706.03762},
      archivePrefix={arXiv},
      primaryClass={cs.CL},
      url={https://arxiv.org/abs/1706.03762}, 
}

@misc{li2024genqquantizationlowdata,
      title={GenQ: Quantization in Low Data Regimes with Generative Synthetic Data}, 
      author={Yuhang Li and Youngeun Kim and Donghyun Lee and Souvik Kundu and Priyadarshini Panda},
      year={2024},
      eprint={2312.05272},
      archivePrefix={arXiv},
      primaryClass={cs.CV},
      url={https://arxiv.org/abs/2312.05272}, 
}

@misc{rombach2022highresolution,
      title={High-Resolution Image Synthesis with Latent Diffusion Models}, 
      author={Robin Rombach and Andreas Blattmann and Dominik Lorenz and Patrick Esser and Björn Ommer},
      year={2022},
      eprint={2112.10752},
      archivePrefix={arXiv},
      primaryClass={cs.CV},
      url={https://arxiv.org/abs/2112.10752}, 
}

\end{document}